\documentclass[12pt]{article}
\pdfoutput=1

\usepackage[utf8]{inputenc}

\usepackage{draft}
\usepackage{putex}
 
 \usepackage[table]{xcolor}
 \usepackage{dsfont}
\usepackage[all]{xy}

\usepackage{stackrel,amssymb}

\usepackage{graphicx}
\usepackage{caption}
\usepackage{amsmath,bm}
\usepackage{array}
\usepackage{makecell}
\usepackage{subcaption}
\usepackage{epstopdf}
\usepackage{enumerate}
\usepackage{mathtools}
\usepackage{cite}
\usepackage{tensor}
\usepackage{slashed}
\usepackage{braket}
\usepackage[utf8]{inputenc}
\usepackage{rotating}
\usepackage{hhline}
\usepackage{bigfoot}

\usepackage[
      colorlinks=true,
      linkcolor=blue,
      urlcolor=blue,
      filecolor=black,
      citecolor=red,
      ]{hyperref}

\usepackage{tikz}
\usetikzlibrary{intersections, calc,arrows.meta,bending, positioning,decorations.pathreplacing,decorations.pathmorphing, decorations.markings}

\tikzset{
  midarrow/.style={
    postaction={decorate},
    decoration={
      markings,
      mark=at position 0.5 with {\arrow[scale=1.5]{Stealth}}
    }
  }
}

\def\XXint#1#2#3{{\setbox0=\hbox{$#1{#2#3}{\int}$ }
		\vcenter{\hbox{$#2#3$ }}\kern-.6\wd0}}

  \usepackage{adjustbox}
\usepackage{multirow}
\usepackage{tikz-cd}

\numberwithin{equation}{section}

\newcommand{\cI}{{\cal I}}

\def\<{\langle}
\def\>{\rangle}

\def\pa{\partial}

\def\wD{\widehat{\Delta}}
\def\d{\text{d}}
\def\i{\text{i}}
\def\h{\hat{h}}

\newcommand{\leftrarrows}{\mathrel{\raise.75ex\hbox{\oalign{%
				$\scriptstyle\leftarrow$\cr
				\vrule width0pt height.5ex$\hfil\scriptstyle\relbar$\cr}}}}
\newcommand{\lrightarrows}{\mathrel{\raise.75ex\hbox{\oalign{%
				$\scriptstyle\relbar$\hfil\cr
				$\scriptstyle\vrule width0pt height.5ex\smash\rightarrow$\cr}}}}
\newcommand{\Rrelbar}{\mathrel{\raise.75ex\hbox{\oalign{%
				$\scriptstyle\relbar$\cr
				\vrule width0pt height.5ex$\scriptstyle\relbar$}}}}

\makeatletter
\def\leftrightarrowsfill@{\arrowfill@\leftrarrows\Rrelbar\lrightarrows}
\newcommand{\xleftrightarrows}[2][]{\ext@arrow 3399\leftrightarrowsfill@{#1}{#2}}
\makeatother

\begin{document}
\preprint{OU-HET 1293}

\title{\makebox[\textwidth][c]{\fontsize{28pt}{24pt}\selectfont
Subdimensional Disorder and Logarithmic Defect}}

\authors{{\Large Soichiro Shimamori\worksat{\Osaka} and Yifan Wang\worksat{\NYU}}\vspace{10mm}}

\institution{Osaka}{Department of Physics, The University of Osaka,
Machikaneyama-Cho 1-1, Toyonaka 560-0043, Japan}

\institution{NYU}{Center for Cosmology and Particle Physics, New York University, New York, NY 10003, USA}

\abstract{We study quenched disorder localized on a $p$-dimensional subspacetime in a $d$-dimensional conformal field theory. Motivated by the logarithmic behavior often associated with disorder, we introduce a defect setup in which bulk local operators transform in ordinary conformal representations, while defect local operators assemble into logarithmic multiplets. We refer to such objects as logarithmic defects and investigate their model-independent properties dictated solely by conformal symmetry and its representation theory, including correlation functions, logarithmic defect operator expansions, and conformal blocks. As a concrete example, we analyze the free scalar theory with a generalized pinning defect subject to random coupling fluctuations, and we identify a half-line of fixed points describing the corresponding logarithmic conformal defects. Along the way, we propose a candidate monotone governing defect renormalization group flows induced by subdimensional disorder, and demonstrate its consistency with such flows in the free theory across various dimensions. We further comment on generalizations and outline a broader program for bootstrapping logarithmic defects.
}
\date{}

\maketitle

\tableofcontents

\section{Introduction and Summary}
\label{sec:intro}
Random systems have long played a central role in condensed matter and statistical physics, as the introduction of disorder often gives rise to rich and fascinating phenomena. One of the earliest and most influential studies in this context is Anderson's seminal work on localization \cite{Anderson:1958vr}, which demonstrated that quenched disorder can fundamentally alter the transport properties of waves including electromagnetic waves and spin waves on lattice, leading to the absence of diffusion in low-dimensional systems. Since then, a wide variety of disordered models have been explored, revealing phenomena, e.g. Griffiths phases \cite{Griffiths:1969} and spin glass phases \cite{Edwards:1975aas, Sherrington:1975zz} (see e.g. \cite{Fytas_2018, Vojta_2019, Altieri_2024} for the review of recent developments of these topics).

In recent years, the study of disorder has gained renewed interest in the context of holography and strongly correlated quantum systems. In particular, the Sachdev-Ye-Kitaev (SYK) model \cite{Maldacena:2016hyu, Rosenhaus:2018dtp, Chowdhury:2021qpy} and its variants, namely one-dimensional systems of fermions with randomly distributed coupling constants, have been extensively investigated. These models exhibit maximally chaotic behavior \cite{Maldacena:2015waa, Maldacena:2016hyu} and nontrivial conformal symmetry in the infrared \cite{Maldacena:2016hyu, Gross:2017vhb, Nayak:2019khe}, providing a solvable playground to explore the interplay between randomness and quantum dynamics.
Moreover, the SYK model is believed to be holographically dual to the nearly $\text{AdS}_2$ gravity \cite{Maldacena:2016hyu, Rosenhaus:2018dtp}, providing one of the simplest models for quantum gravity. This duality highlights the essential role of ensemble averaging in establishing a consistent gravitational description, where Euclidean wormholes naturally compute averaged boundary observables (see e.g. \cite{Saad:2018bqo, Maldacena:2019ufo, Garcia-Garcia:2020ttf, Cotler:2022rud}).

In quantum computing and quantum information, quenched disorder also provides a unifying framework for understanding crucial aspects such as quantum error correction \cite{Dennis:2001nw,Wang:2002ph,Li:2021dsr} and measurement-induced phase transitions \cite{Bao:2019qah,Jian:2019mny,Fisher:2022qey}. In particular, there is often a mapping such that the effects of quantum noise and dephasing can be captured by classical quenched disorder in an effective statistical system.\footnote{An important difference is that, unlike ordinary quenched disorder which is uncorrelated, the disorder distribution obtained from the quantum-to-classical mapping follows Born rule and thus is highly correlated \cite{Bao:2019qah,Jian:2019mny,Fisher:2022qey}. In the replica approach to quenched disorder with $n$ replica copies, the former amounts to taking the limit $n\to 0$ whereas the latter corresponds to $n\to 1$, as is expected to produce von Neumann entropies.  
} 
For example, the
study of error recovery using quantum error-correcting codes can be mapped onto classical disordered spin models, such as the random-bond Ising model (see e.g. \cite{HNishimori_1980,Nishimori:1981ajd,Gruzberg_2001,Komargodski:2016auf}) with binary distribution, where the accuracy threshold (which measures the efficacy of the code)
coincides with the critical point on the Nishimori line \cite{HNishimori_1980,Nishimori:1981ajd},\footnote{Unlike the well-known Ising fixed points (in 2d and 3d) described by unitary CFTs \cite{Belavin:1984vu,El-Showk:2012cjh}, the Nishimori critical point \cite{HNishimori_1980,Nishimori:1981ajd} is believed to be logarithmic \cite{Gurarie:2004ce} and 
has not been solved.  
See recent work \cite{Delfino:2024atu,Putz:2025cxk} and references therein for recent progress.} defined by matching error probabilities and effective thermal fluctuations \cite{Dennis:2001nw,Wang:2002ph}. Likewise, in monitored quantum circuits, averaging over measurement outcomes generates effective spacetime disorder in the corresponding statistical mechanics representations of entanglement dynamics (see e.g. \cite{Nahum:2020hec,Zhu:2022bpk,Wang:2025xwk}). Intuitively, via this quantum-to-classical mapping, the interplay between unitary entanglement growth and measurement-induced disentangling mirrors the competition between ordering and disorder in classical systems.

\begin{figure}[!htb]
    \centering
    \begin{tikzpicture}[>=stealth]

\draw[thick,midarrow] (0,0) .. controls (2,-1) and (2,-3) .. (4,-4);
\fill[blue!80] (0,0) circle (3pt);
\node[left] at (-0.3,0) {UV CFT};
\node[right] at (0.5,0.2) {$\langle\mathcal{O}\mathcal{O}\rangle$ : power law};

\fill[red!80] (4,-4) circle (3pt);
\node[left] at (3.5,-4) {IR defect CFT};
\node[right] at (4,-3.5) {$\langle\mathcal{O}\mathcal{O}\rangle$ : logarithmic contributions};
\node[right] at (-1.5,-1.8) {Subdimensional};
\node[right] at (-1,-2.3) {disorder};

\end{tikzpicture}
    \caption{Illustration of the defect setup studied in this paper. In the UV CFT, the correlation functions show usual power law, whereas in the IR defect CFT they develop logarithmic behavior due to subdimensional disorder.}
    \label{fig:rg_flow}
\end{figure}
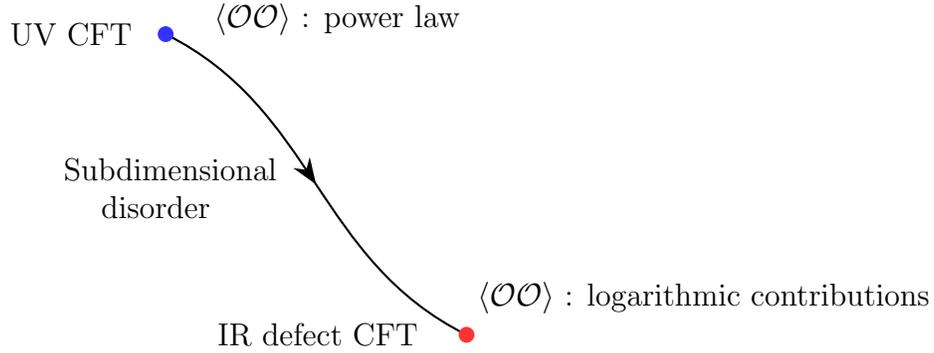

While these developments underscore the significant role of disorder across various branches of physics, it is worth noting that most previous studies have focused on cases where the random couplings are distributed throughout the entire space. In such cases, the disorder averaging effectively acts as a global modification of the theory, and the resulting averaged theory generally differs drastically from the original one. In this work, we propose to study a different but closely related scenario: instead of distributing disorder across the whole space, we consider the quenched disorder {\it localized along a subspacetime}.\footnote{Throughout this paper, we focus on  CFT in Euclidean spacetime $\mR^d$. Without loss of generality, we assume the disorder coupling is turned on in the subspacetime including the Euclidean time and a subspace in the Euclidean space. 
Here we study classical disorder (see e.g. \cite{Aharony:2015aea, Piazza:2024wll}), where the random couplings vary on this subspacetime. In principle, one could also study quantum disorder, where the couplings fluctuate only in space. The latter case leads to novel phenomena, for example the appearance of Lifshitz scaling (see e.g. \cite{Aharony:2018mjm}). We leave the exploration of quantum disorder in subdimensional spacetime for future work.} In other words, the coupling constants fluctuate randomly only on a lower-dimensional subsystem, while the rest of the system remains unaffected.
This problem was first investigated in the 2d Ising model with a random boundary magnetic field \cite{Cardy:1991kp,Igloi_1991,DeMartino:1997km,Pleimling_2004} (see also the review \cite{Pleimling_2004rev}). More recent developments on boundary disorder can be found in \cite{Jeng:2001dy,Gupta:2021plc,Ge:2024ubs}. In the present work, we extend the discussion to general codimensions. Concretely, we consider a $d$-dimensional Euclidean spacetime $\mathbb{R}^d$, where the UV fixed point is described by a conformal field theory (CFT) with action $S_{\text{CFT}}$.
We deform this UV theory by adding the following localized term on a $p$-dimensional subspacetime $\mathbb{R}^p$ $(p<d)$:
\begin{align}
S=S_{\text{CFT}}+\int_{\mathbb{R}^{p}}\d^p \hat{x}\, {g}(\hat{x}) O(\hat{x})\,, 
\label{eq:gendisord}\end{align}
where $\hat{x}$ denotes the coordinates on $\mathbb{R}^p$, and $O$ is a scalar primary operator in the UV CFT. Throughout this paper, we assume that the coupling ${g}$ is a random variable distributed according to some probability functional $P[g]$.
A natural and intriguing question then arises: what is the fate of this localized deformation in the infrared, particularly after performing the ensemble average over the random couplings? More specifically, under what conditions does such a spatially localized disorder flow to a {\it conformal defect} (Figure~\ref{fig:rg_flow})? Furthermore, what are the special properties of such a conformal defect?

To begin, the relevance of conventional disorder in the infrared is governed by the Harris criterion \cite{Harris:1974zz} (see also \cite{Aharony:2015aea} for a review). As we explain in Section~\ref{subsec:confandfree}, this criterion admits a direct analog for subdimensional disorder, which depends on the conformal dimension $\Delta_O$ of the disorder operator (in the original theory without disorder, i.e. at $c=0$) in \eqref{eq:gendisord},\footnote{We emphasize that the Harris criterion \eqref{Harrisdef} applies only to Gaussian disorder, which we focus on in this paper. For other quenched disorders such as long-range correlated disorder, the Harris criterion is modified \cite{Weinrib:1983zz}.}
\ie 
{\rm Relevant:}~\Delta_O<{p\over 2}\,,\quad 
{\rm Irrelevant:}~\Delta_O>{p\over 2}\,,\quad 
{\rm Marginal:}~\Delta_O={p\over 2}\,.
\label{Harrisdef}
\fe 
This implies, for unitary CFTs, new defect fixed points from subdimensional disorder is only possible for $p\geq d-2$.

To better understand the defect fixed points arising from ensemble averaging, we elucidate  the systematics of conformal defects with logarithmic behavior on their worldvolume in Section~\ref{sec:systematics}, under the assumption that localized disorder flows to such a defect in the IR. The meaning of these logarithmic features will be reviewed below.

Recall that in a $d$-dimensional Euclidean spacetime, the conformal symmetry is described by the group $\mathrm{SO}(1,d+1)$. The introduction of a $p$-dimensional conformal defect breaks this symmetry as
\begin{align}
\mathrm{SO}(1,d+1) \longrightarrow \mathrm{SO}(1,p+1)\times \mathrm{SO}(d-p)\,,
\end{align}
where $\mathrm{SO}(1,p+1)$ corresponds to the residual conformal symmetry along the defect worldvolume, while $\mathrm{SO}(d-p)$ represents the rotational symmetry in the transverse directions \cite{McAvity:1993ue, McAvity:1995zd, Kapustin:2005py, Liendo:2012hy, Billo:2016cpy}. Consequently, there exist two types of local operators: bulk local operators $\mathcal{O}(x)$ in the ambient spacetime and defect local operators $\widehat{\mathcal{O}}(\hat{x})$ confined to the subspacetime. In conventional unitary defect CFT, bulk operators are classified by primary operators under $\text{SO}(1, d+1)$, while defect operators are classified by primaries under $\text{SO}(1, p+1)\times \text{SO}(d-p)$. In particular, the dilatation operator $D$ acts {\it diagonally} on these primaries. For instance, when two primaries $\mathcal{O}_{1}$ and $\mathcal{O}_{2}$ have the same conformal dimension $\Delta$, the dilatation operator acts as 
\begin{align}
    \begin{aligned}
    \text{ordinary CFT : }
        D\begin{pmatrix}
            \mathcal{O}_{1} \\
            \mathcal{O}_{2}
        \end{pmatrix} = 
        \begin{pmatrix}
            \Delta & 0 \\
            0 & \Delta 
        \end{pmatrix}
        \begin{pmatrix}
            \mathcal{O}_{1} \\
            \mathcal{O}_{2}
        \end{pmatrix} \,. 
    \end{aligned}
\end{align}
This diagonal structure makes correlation functions exhibit standard power-law behavior.

However, conformal defects generated by ensemble averaging over localized disorder are expected to fall outside this standard framework. Indeed, quenched disorder is often observed to induce logarithmic behavior in correlation functions \cite{cardy1999logarithmiccorrelationsquenchedrandom,cardy2001stresstensorquenchedrandom,Kogan:2002mg,Cardy:2013rqg, Aharony:2015aea}. This is one of the hallmarks of logarithmic CFTs (logCFTs), in which the dilatation operator $D$ acts {\it non-diagonally} on operators (see e.g. \cite{Gurarie:1993xq,   Flohr:2001zs,Gaberdiel:2001tr,Kawai:2002fu,Creutzig:2013hma,Hogervorst:2016itc} for reviews on logCFTs). For example, when two operators $\mathcal{O}_{1}$ and $\mathcal{O}_{2}$ share the same conformal dimension $\Delta$, the dilatation operator may act as
\begin{align}
    \begin{aligned}
    \text{logarithmic CFT : }
        D\begin{pmatrix}
            \mathcal{O}_{1} \\
            \mathcal{O}_{2}
        \end{pmatrix} = 
        \begin{pmatrix}
            \Delta & 1 \\
            0 & \Delta 
        \end{pmatrix}
        \begin{pmatrix}
            \mathcal{O}_{1} \\
            \mathcal{O}_{2}
        \end{pmatrix} \,,
    \end{aligned}
\end{align}
up to a change of basis. Clearly, this matrix is a Jordan cell and cannot be diagonalized. As reviewed in the main text, this non-diagonal structure gives rise to logarithmic terms in correlators. We refer to the lowest conformal weight operators $(\mathcal{O}_1 , \mathcal{O}_2 )$ satisfying the above property as the \textit{logarithmic primaries} of a \textit{logarithmic multiplet} with rank two and conformal dimension $\Delta$. For the above reason, in the present work we expect that the conformal defects arising from localized ensemble averaging exhibit the following structure of operator spectrum: bulk local operators behave as ordinary $\text{SO}(1, d+1)$ primaries, while defect local operators form logarithmic representations of $\text{SO}(1, p+1)\times \text{SO}(d-p)$. Henceforth we will refer to such defects as {\it logarithmic conformal defects}. One of our main goals is to establish model-independent systematics for such logarithmic conformal defects. In particular, we obtain general results for the form of correlation functions, the conformal block decomposition, and the expansion of bulk operators in terms of defect operators, namely the logarithmic defect-operator-expansion (logDOE), using only symmetry considerations and representation theory.

As a concrete realization of the outlined systematic analysis, we consider a $d$-dimensional free scalar field theory as the UV CFT, deformed by a localized disorder as below,
\begin{align}\label{eq:free_theory}
        S=\frac{1}{2}\int_{\mathbb{R}^{d}}\d^d x(\partial \sigma)^2+\int_{\mathbb{R}^{p}}\d^p \hat{x}\, {\h}(\hat{x}) \sigma(\hat{x})\,. 
\end{align}
For specificity, we assume that the coupling $\h(\hat{x})$ is sampled from a Gaussian distribution with variance $c$ which measures the strength of the disorder, which is marginal according to the Harris criterion \eqref{Harrisdef} and turns out to be exactly marginal (see Section~\ref{subsec:line_fixed_point}).

It is instructive to contrast our setup with the localized magnetic (or pinning) defects studied in \cite{Allais:2014fqa, Cuomo:2021kfm}, where the defect coupling is uniform rather than randomly fluctuating. In that case, for $d<4$ and $p=1$, the coupling grows indefinitely under the RG flow, leading to runaway behavior without an IR-stable defect CFT description \cite{Cuomo:2021kfm}. There, the instability can be remedied by introducing bulk interactions. By contrast, in our setup the instability is avoided through disorder averaging. In fact, as we will show in the main text, using the replica method, the disorder-averaged theory can exhibit an IR-stable fixed line parametrized by the disorder strength $c$ even though the bulk theory remains free. This is also to be contrasted with the earlier studies of boundary disorder in the 2d Ising model \cite{Cardy:1991kp,Igloi_1991,DeMartino:1997km} where the disorder is marginally irrelevant and does not produce new boundary CFT.

The correlation functions for the disordered system in \eqref{eq:free_theory} can be computed explicitly for general $p$, allowing a direct comparison with the model-independent systematics derived in Section~\ref{sec:systematics}. This comparison provides an explicit diagnostic for whether the disorder produces a nontrivial conformal defect or not, in this free theory. Interestingly, we find that infrared behavior matches that of a logarithmic conformal defect only when the defect codimension is two, i.e. $p=d-2$. Moreover, combining the bulk equations of motion, we find that various defect operators organize into logarithmic multiplets. Notably, the displacement operator, which encodes the response to local deformations of the defect's position, also forms a part of the logarithmic multiplet. 

We further extend this analysis to the natural generalization of a single scalar field: an $\text{O}(N)$ free scalar theory, deformed by
\begin{align}\label{eq:freeON}
        S=\frac{1}{2}\int_{\mathbb{R}^{d}}\d^d x(\partial \vec{\sigma})^2+\int_{\mathbb{R}^{p}}\d^p \hat{x}\, {h}(\hat{x}) \vec{n}\cdot\vec{\sigma}(\hat{x})\,. 
\end{align}
Here, $\vec{n}$ is the fixed vector, and the bulk $\text{O}(N)$ symmetry is broken to $\text{O}(N-1)$ around the vector $\vec{n}$. This setup also perfectly matches the expected properties of a logarithmic conformal defect only when $p=d-2$. In particular, the tilt operator, which arises as a universal operator due to the breaking of the bulk $\text{O}(N)$ symmetry, is found to form a logarithmic multiplet as well.

The rest of the paper is structured as follows.In Section~\ref{sec:Subdimensional Quenched Disorder}, we introduce the notion of subdimensional quenched disorder and specify the conditions under which it gives rise to conformal defects. As a concrete example, we study random-field defects in free scalar field theories, which are conformal if and only if $p=d-2$. We then analyze the associated defect RG flow, establishing the existence of a line of disordered codimension-two fixed points in the free theory and highlighting the emergence of logarithmic behavior. In Section~\ref{sec:systematics}, we develop the model-independent systematics of logarithmic defect CFTs: beginning with a brief review of logarithmic conformal representation theory, we derive the logarithmic defect operator-product-expansion and discuss general constraints from bulk two-point functions and bootstrap considerations. Section~\ref{sec:back_to_free} returns to the free-theory setting, where we provide explicit realizations of logarithmic conformal defects. There we analyze the structure of logarithmic multiplets, extract detailed CFT data, and confirm consistency with the general framework developed earlier. Finally, in Section~\ref{sec:disc}, we outline extensions of this work and promising directions for future study. Technical details are collected in the Appendices.

\section{Subdimensional Quenched Disorder}\label{sec:Subdimensional Quenched Disorder}
In this section, we introduce the notion of subdimensional quenched disorder, and analyze the concrete models described in \hyperref[sec:intro]{Introduction}. 
In Section~\ref{subsec:general}, we set up our notation and formulate the subdimensional disorder in its most general form. In Section~\ref{subsec:confandfree}, we formulate the Harris-type criterion for conformality in the presence of subdimensional disorder. To illustrate this criterion, we consider the disordered theory \eqref{eq:free_theory} with bulk dynamics governed by a free scalar field. We show that subdimensional disorder produces a conformal defect only when $p=d-2$, as evidenced by the behavior of certain bulk correlators. In Section~\ref{subsec:line_fixed_point}, we analyze the fixed-point structure using the replica method. In particular, we derive the beta function for the disorder strength $c$ and find that it is exactly marginal, implying that the disordered theory defines a defect conformal manifold parametrized by $c$. Moreover, through explicit computations of disorder-averaged correlation functions, we uncover logarithmic behavior, which serves as a characteristic signature of disordered fixed points.

\subsection{Defect from Subdimensional Quenched Disorder}\label{subsec:general}
We now set up the general framework of subdimensional quenched disorder in $d$-dimensional Euclidean spacetime $\mathbb{R}^d$. For later convenience, we split the coordinate for $\mathbb{R}^d$ as 
\begin{align}
 x^{\mu} = (\hat{x}^{a},\, x_{\perp}^{i}) \,, \qquad 
 a=1,2,\dots,p \,,\quad i=p+1,\dots,d \,,
\end{align}
where $\hat{x}^{a}$ parameterize the $p$-dimensional subspace $\mathbb{R}^p$ on which the disorder is taken, while $x_{\perp}^{i}$ denote the remaining transverse directions. As mentioned in \hyperref[sec:intro]{Introduction}, we consider the disordered system defined by the following action,
\begin{align}
    S[X, g]\equiv S_{\text{CFT}}+\int_{\mathbb{R}^{p}}\d^p \hat{x}\, g(\hat{x}) O(\hat{x})\,, 
    \label{disorderact}
\end{align}
where $S_{\text{CFT}}$ is the UV CFT action, $O(\hat{x})$ is some primary operator in the UV CFT and $X$ represents the bulk fields collectively. Whenever the transverse coordinates are suppressed, they are taken to be at $x^i_\perp=0$, which is the location of the disorder. 

We assume that the coupling $g(\hat{x})$ is randomly fluctuating over $\mathbb{R}^p$, and the probability distribution functional $P[g]$ satisfies the following normalization condition:
\begin{align}\label{eq:normalization}
    \int \mathcal{D}g \, P[g]=1\,. 
\end{align}
For a fixed configuration of the subdimensional coupling $g$, one can define the connected correlation function as
\begin{align}
    \langle\, \cdots\, \rangle_{g}^{\text{conn}}\equiv \frac{1}{Z_{g}}\int \mathcal{D}X (\cdots)e^{-S[X, g]}\,, \quad Z_{g}\equiv \int \mathcal{D}X e^{-S[X, g]}\,, 
\end{align}
where $\cdots$ means some inserted operators, and the subscript $g$ is attached just to emphasize that $g$ is fixed. In terms of this, we introduce the subdimensional disorder-averaged correlator by\footnote{Let us emphasize that the disorder coupling should not be regarded as a dynamical field. A dynamical field $X$ is integrated over in the path integral for a given configuration of defect couplings $\{g(\hat{x})\}_{\hat{x}\in \mathbb{R}^p}$. The physical observables are then obtained by averaging the correlation functions $\langle\cdots\rangle_{g}^{\mathrm{conn}}$ over the probability distribution of $g$. Importantly, throughout this paper, the order of these operations cannot be interchanged: one must first perform the path integral over the dynamical fields for a fixed $g$, and only afterwards take the disorder average.}
\begin{align}\label{eq:def_desorder_corre}
    \overline{\langle\, \cdots\, \rangle}\equiv \int \mathcal{D}g \, P[g] \langle\, \cdots\, \rangle_{g}^{\text{conn}}\,. 
\end{align}
In principle, one may consider various probability distributions (see relevant discussions in Section~\ref{sec:disc}). In this paper, for simplicity, we exclusively consider the Gaussian distribution defined by
 \begin{align}\label{eq:gauss}
     P[g]\propto \exp{\left[-\frac{1}{2c^2}\int_{\mathbb{R}^{p}}\d^p \hat{x} \, \left({g}(\hat{x})\right)^2 \right]} \,, 
 \end{align}
 where $c$ is the width of the Gaussian function which characterizes the strength of the disorder. The omitted overall factor above can be fixed from the normalization condition \eqref{eq:normalization}. In this case, the one and two-point functions of subdimensional coupling $g$ is written as
 \begin{align}
     \overline{\langle\, g(\hat{x})\, \rangle}=0 \,, \quad \overline{\langle\, g(\hat{x})g(\hat{y})\, \rangle}=c^2 \delta^{p}(\hat{x}-\hat{y})\,.
 \end{align}
 One can deduce all other correlation functions from this two-point function via the Wick's theorem. For instance, the four-point function reads
 \begin{align}
     \begin{aligned}
         &\overline{\langle\, g(\hat{x}_1 )g(\hat{x}_2 )g(\hat{x}_3 )g(\hat{x}_4 )\, \rangle} \\
         &\quad =c^4 \left[\delta^{p}(\hat{x}_1-\hat{x}_2)\delta^{p}(\hat{x}_3-\hat{x}_4)+\delta^{p}(\hat{x}_1-\hat{x}_3)\delta^{p}(\hat{x}_2-\hat{x}_4)+\delta^{p}(\hat{x}_1-\hat{x}_4)\delta^{p}(\hat{x}_2-\hat{x}_3)\right]\,. 
     \end{aligned}
 \end{align}

\subsection{Conformality from Disorder and  Free Theory Example}\label{subsec:confandfree}

A natural way to assess the relevance of quenched disorder is to adapt the Harris criterion \cite{Harris:1974zz} to the defect setting (see also \cite{Aharony:2015aea} for a review). 
The strength of disorder is characterized by a dimensionless variance parameter $c$, which requires the random coupling $g(\hat{x})$ to have dimension $p/2$, as dictated by \eqref{eq:gauss}.
It then follows that the subdimensional disorder in \eqref{disorderact} is relevant if $\Delta_O < p/2$, marginal if $\Delta_O = p/2$, and irrelevant if $\Delta_O > p/2$, as summarized in \eqref{Harrisdef}.
We therefore expect disorder to generate nontrivial conformal defects only when $\Delta_O \leq p/2$.\footnote{See \cite{Aharony:2015aea} for a discussion of ``dangerously irrelevant’’ disorder, which can still affect infrared correlation functions through contact terms.}
For a unitary CFT, however, this Harris bound is in tension with the unitarity bound for scalar operators, which imposes $p \geq d-2$. The borderline case is realized by the free scalar theory, which we now turn to.

As a concrete illustration of conformal defects arising from subdimensional disorder, let us consider the simplest bulk theory: a free massless scalar field in $d$ dimensions.
We introduce quenched disorder by coupling the scalar to a localized random magnetic field $\h(\hat{x})$ (corresponding to $g(\hat{x})$ in \eqref{disorderact}) along a $p$-dimensional subspace:
\begin{align}\label{eq:free_theory2}
        S=\frac{1}{2}\int_{\mathbb{R}^{d}}\d^d x(\partial \sigma)^2+\int_{\mathbb{R}^{p}}\d^p \hat{x}\, {\h}(\hat{x}) \sigma(\hat{x})\,,
\end{align}
where $\h(\hat{x})$ is drawn from a Gaussian distribution.
This model can be regarded as a disordered generalization of a pinning defect, where the scalar couples to a random local magnetic field.
We will show that scale invariance is preserved only when the defect has codimension two, i.e. $p = d-2$.
From the Harris criterion discussed above, this corresponds precisely to the marginal case, since the free scalar has scaling dimension $\Delta_\sigma = \tfrac{d-2}{2}$.
This provides the first indication that the codimension-two case is distinguished and potentially admits a conformal fixed point.

To make this relation between scale invariance and the dimensional constraint $p=d-2$ more concrete, let us examine the bulk two-point function of $\sigma$ explicitly. For a given coupling $\h$, we obtain
\begin{align}
    {\langle\, \sigma(x) \sigma(y)\, \rangle_{\h}}=\frac{C_{d}}{|x-y|^{d-2}}+C_{d}^{2}\int \d^{p}\hat{z}\,\d^{p}\hat{w}\,\frac{\h(\hat{z}){\h}(\hat{w})}{|x-\hat{z}|^{d-2}|y-\hat{w}|^{d-2}}\,,
\end{align}
where 
\begin{align}\label{Cddef}
    C_{d}\equiv \frac{\Gamma\left(d/2-1\right)}{4\pi^{d/2}}\,.
\end{align}
Then, the averaging over the coupling ${h}$ gives rise to
\begin{align}\label{eq:two_pt_defect}
    \begin{aligned}
        \overline{\langle\, \sigma(x) \sigma(y)\, \rangle}
        &=\frac{C_{d}}{|x-y|^{d-2}}+{c}^{2}C_{d}^{2}\int \d^{p}\hat{z}\,\frac{1}{|x-\hat{z}|^{d-2}|y-\hat{z}|^{d-2}}\,. 
    \end{aligned}
\end{align}
Note that to avoid IR singularities, we must require $1\leq p\leq \min(2d-4, d-1)$.
Suppose that the scale invariance does hold even with the disorder, the two-point function must be subject to the following Ward-Takahashi identity: 
\begin{align}
    \overline{\langle\, \sigma(\lambda x) \sigma(\lambda y)\, \rangle}=\lambda^{-(d-2)}\overline{\langle\, \sigma(x) \sigma(y)\, \rangle}\,. 
\end{align}
However the second term in the RHS of \eqref{eq:two_pt_defect} shows the different scaling response in general
\begin{align}
    \overline{\langle\, \sigma(\lambda x) \sigma(\lambda y)\, \rangle_{\hat{h}}}=\lambda^{-(d-2)}\frac{C_{d}}{|x-y|^{d-2}}+\lambda^{p-2d+4}\, \hat{c}^{2}C_{d}^{2}\int \d^{p}\hat{z}\,\frac{1}{|x-\hat{z}|^{d-2}|y-\hat{z}|^{d-2}}\,.
\end{align}
This implies that the scale invariance for subdimensional disorder in the free theory requires that the bulk and defect dimensions must be related as $p=d-2$.  

\subsection{Half-line of Logarithmic Fixed Points in Free Theory}\label{subsec:line_fixed_point}
In the previous section, we saw that conformality of subdimensional disorder in the free scalar theory can only occur at codimension two (i.e. $p=d-2$). In this section, we turn to a fixed-point analysis for this case. Under this dimensional constraint, not only the localized magnetic field but also a localized mass term becomes marginal. It is therefore necessary to analyze the fixed points of the theory including both contributions. To this end, let us consider the following action:
\begin{align}\label{eq:free_theory_mass}
    S=S_{\text{bulk}}[\sigma]+S_{\text{D}}[\sigma, \h]\,, 
\end{align}
where $S_{\text{bulk}}$ and $S_{\text{D}}$ are defined by
\ie\label{eq:action}
        S_{\text{bulk}}\equiv \frac{1}{2}\int_{\mathbb{R}^{d}}\d^d x(\partial \sigma)^2\,,\quad 
        S_{\text{D}} \equiv \int_{\mathbb{R}^{p}}\d^p \hat{x}\, {\h}(\hat{x}) \sigma(\hat{x})+\frac{\gamma}{2}\int_{\mathbb{R}^{p}} \d^p\hat{x} (\sigma(\hat{x}))^{2}\,. 
\fe 
Here, $\gamma$ is the defect coupling constant (to the mass term) and together with the disorder strength $c$, they are the deformation parameters of the theory. In what follows, we compute the corresponding beta functions, and discuss the stability of the fixed points. 
We proceed by employing the replica method as in \cite{Ludwig:1987gs,Dotsenko:1994sy,Dotsenko:1994im,Cardy_1996, Dotsenko_2000, Shimada:2009dm}, to reformulate the disorder-averaged theory in terms of a conventional field theory, which allows us to define beta functions in the usual sense and study the RG flow of $c$ and $\gamma$. To be more concrete, let us consider the disorder-averaged correlation function of a local operator $\mathcal{O}(\sigma)$ made up of $\sigma$, and deform this as follows:
\begin{align}
    \begin{aligned}
        \overline{\langle \mathcal{O}(\sigma) \rangle}&=\overline{\langle\mathcal{O}(\sigma)\rangle \langle\mathbf{1}\rangle^{n-1}} \\
        &=\int \mathcal{D}\h \frac{P[\h]}{Z_{\h}^{n}}\int \mathcal{D}\vec{\sigma}\, \mathcal{O}(\sigma_{1})e^{-\sum_{A=1}^{n}\left( S_{\text{bulk}}[\sigma_{A}]-S_{\text{D}}[\sigma_{A}, \h]\right)}\,, 
    \end{aligned}
\end{align}
where $\vec{\sigma}\equiv (\sigma_{1}, \sigma_{2}, \cdots, \sigma_{n})$. In the first line, we have simply inserted one, and in the second line, we have used the definition of the disorder-averaged correlator \eqref{eq:def_desorder_corre}. Applying the replica trick for the quenched disorder amounts to taking the limit $n\to 0$, in which $Z_{\h}^{n}\to 1$. Since the distribution $P[\h]$ is Gaussian, we can straightforwardly integrate out the disorder field $\h$, and obtain the effective replica action $S_{\text{rep}}$ describing interactions among different replicas:
\begin{align}\label{eq:replica_action}
    S_{\text{rep}}[\vec{\sigma}]\equiv \sum_{A=1}^{n}S_{\text{bulk}}[\sigma_{A}]+\sum_{A, B}\int \d^{p}\hat{x}\, m_{AB} \sigma_{A}(\hat{x})\sigma_{B}(\hat{x})\,, 
\end{align}
where $m_{AB}$ is defined by 
\begin{align}
    m_{AB}\equiv -\frac{c^2}{2}+\gamma\, \delta_{AB}\,. 
\end{align}
With the replica action \eqref{eq:replica_action} at hand, the beta functions for defect couplings $c^2$ and $\gamma$ can now be defined in the standard QFT sense as follows:
\begin{align}
    \beta_{c^2, n}\equiv \frac{\d c^2}{\d \log\mu}\,, \quad \beta_{\gamma, n}\equiv \frac{\d \gamma}{\d \log\mu}\,,
\end{align}
where the subscript $n$ signals that they  are defined in the $n$-times replicated theory.
These beta functions are determined by solving the Callan–Symanzik equation for the bulk one-point function $\langle\, \sigma_{A}\sigma_B (x) \, \rangle_{\text{rep}}$. The explicit evaluation of $\beta_{c^2, n}$ and $\beta_{\gamma, n}$ is presented in Appendix \ref{app:beta}, and the results are summarized here,\footnote{In the absence of disorder, i.e.\ at $c=0$, the $\sigma^2$ deformation in codimension two is known to be marginally irrelevant (for $\C>0$) \cite{Lauria:2020emq}, as follows from the beta function for $\gamma$ in \eqref{eq:beta_function}.}
\begin{align}\label{eq:beta_function}
    \beta_{c^2, n}=C_{d}S_{p-1}\left(2\gamma c^2 -n c^4 \right)+\cdots\,, \quad \beta_{\gamma, n}=C_{d}S_{p-1}\gamma^{2}+\cdots\,, 
\end{align}
where $S_{p-1}\equiv 2\pi^{p/2}/\Gamma(\frac{p}{2})$ is the area of a unit $p$-sphere. In order to extract the beta functions $\beta_{c^2}$ and $\beta_{\gamma}$ for the original disordered system, we take the replica limit $n\to 0$,
\begin{align}
    \beta_{c^2, 0}=2C_{d}S_{p-1}\gamma c^2+\cdots\,, \quad \beta_{\gamma, 0}=C_{d}S_{p-1}\gamma^{2}+\cdots\,. 
\end{align} 

From these expressions, one can infer a half-line (due to the identification $c\to -c$) of fixed points given by
\begin{align}\label{eq:line_fixed_point}
\gamma^\ast = 0\,,\quad c: \ \text{arbitrary}\,.
\end{align}
In principle, one could verify this explicitly to higher order in the couplings by extending the analysis of Appendix~\ref{app:beta}. Instead, we confirm the exact marginality of $c$ by providing a fully non-perturbative solution for the conformal correlators as a function of $c$ in Section~\ref{sec:back_to_free}. Since the defect induced by subdimensional disorder is not a conventional local defect, we do not expect a corresponding exactly marginal local operator.

\subsection{Free Energy of Quenched Disorder Defect and RG Monotone}\label{subsec:defect_free_energy}
In the last subsection, we found the half-line of conformal fixed points \eqref{eq:line_fixed_point} by replica method. Since this conformal manifold includes the trivial point without any defect, one may expect that the averaged defect free energy, defined below, should vanish along the entire manifold, as what must happen when monotonicity theorems for defect RG are obeyed \cite{Affleck:1991tk,Friedan:2003yc,Jensen:2015swa,Casini:2018nym,Kobayashi:2018lil,Wang:2020xkc,Wang:2021mdq,Cuomo:2021rkm,Shachar:2022fqk,Casini:2023kyj}. In fact, this is expected from the replica approach (see below) to quenched disorder as we take the number of replica copies $n \to 0$ in the end. 

Indeed, for ordinary quenched disorder in 2d CFTs, this leads to the familiar vanishing central charge $c=0$ upon disordering (see e.g. \cite{cardy2001stresstensorquenchedrandom,Gurarie:2004ce}). To have a more interesting RG monotone in the disordered context, it was suggested in \cite{Gurarie:1999bp} (see also \cite{Patil:2025wqp}) to use the effective central charge $c_{\rm eff}\equiv \left.{\d c(n)\over \d n}\right|_{n=0}$ defined by a derivative of the central charge for the $n$-th replicated theory. Here we will also propose a direct analog for subdimensional disorder in general dimensions.

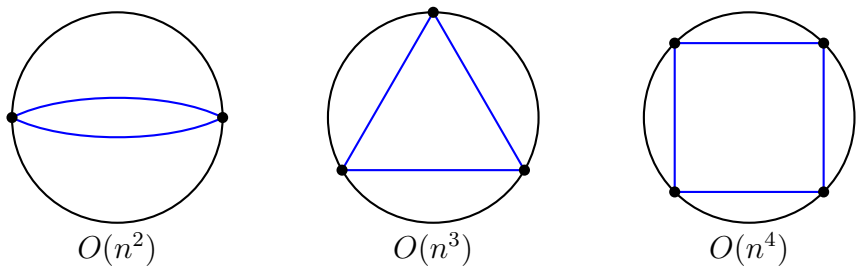
\begin{figure}[!htb]
    \centering
    \begin{tikzpicture}[scale=.7]

\begin{scope}[xshift=0cm]
  \draw[thick] (0,0) circle (2);
  \draw[blue, thick] (-2,0) .. controls (-1,0.5) and (1,0.5) .. (2,0);
  \draw[blue, thick] (-2,0) .. controls (-1,-0.5) and (1,-0.5) .. (2,0);
  \fill (-2,0) circle (3pt);
  \fill (2,0) circle (3pt);
  \node at (0,-2.5) {$O(n^2)$};
\end{scope}

\begin{scope}[xshift=6cm]
  \draw[thick] (0,0) circle (2);
  \draw[blue, thick] (90:2) -- (210:2) -- (-30:2) -- cycle;
  \fill (90:2) circle (3pt);
  \fill (210:2) circle (3pt);
  \fill (-30:2) circle (3pt);
  \node at (0,-2.5) {$O(n^3)$};
\end{scope}

\begin{scope}[xshift=12cm]
  \draw[thick] (0,0) circle (2);
  \draw[blue, thick] (45:2) -- (135:2) -- (225:2) -- (315:2) -- cycle;
  \fill (45:2) circle (3pt);
  \fill (135:2) circle (3pt);
  \fill (225:2) circle (3pt);
  \fill (315:2) circle (3pt);
  \node at (0,-2.5) {$O(n^4)$};
\end{scope}

\end{tikzpicture}
    \caption{Sample diagrams contributing to the averaged defect free energy. There are no contributions at order $O(n)$.}
    \label{fig:free_energy}
\end{figure}

We first confirm the vanishing of the averaged defect free energy $F_{D}$ defined below,
\begin{align}
    F_{D}\equiv-\overline{\log\frac{Z_{\h}}{Z_{\h=0}}}\, , 
    \label{averagedF}
\end{align}
where $Z_{\h}$ is the partition function in the presence of the spherical defect on $S^{p}$. 
Since the logarithm obstructs a direct evaluation of the above quantity, we employ the replica method as before. Namely, we introduce $n$ copies of the system and write
\begin{align}\label{Ffromreplica}
    F_{D}=-\lim_{n\to0}F_D(n)\,,\quad F_D(n)\equiv \frac{1}{n}\left[\overline{\left(\frac{Z_{\h}}{Z_{\h=0}}\right)^{n}}-1\right]\, . 
\end{align}
The averaged quantity defined above can be evaluated by first integrating over the disorder field $\h$, which yields
\begin{align}\label{replicaPF}
    \overline{\left(\frac{Z_{\h}}{Z_{\h=0}}\right)^{n}}=\left\langle e^{-\frac{c^2}{2}\sum_{A, B=1}^{n}\int_{S_{R}^p} \sigma_{A}\sigma_{B}}\right\rangle_{\text{rep}, c=0}\ , 
\end{align}
where $S_{R}^p$ is the spherical defect with radius $R$, and $\langle \cdots \rangle_{\text{rep}, c=0}$ denotes the partition function of the disorder-less replicated theory,
\begin{align}
    \langle \cdots \rangle_{\text{rep}, c=0}\equiv \frac{\int {\cal D} \vec{\sigma}(\cdots) e^{-\sum_{A=1}^{n}S_{\text{bulk}}[\sigma_{A}]}}{\int {\cal D} \vec{\sigma}e^{-\sum_{A=1}^{n}S_{\text{bulk}}[\sigma_{A}]}}\,,
\end{align}
with $S_{\text{bulk}}$ given in \eqref{eq:action}. This expression admits a straightforward perturbative evaluation. Note that by \eqref{Ffromreplica}, only the terms proportional to $n$ from \eqref{replicaPF} contribute nontrivially to the defect free energy in the replica limit $n\to 0$. However, the perturbative expansion admits no linear term in $n$, and the contributions starts from order $O(n^{2})$ (see Figure~\ref{fig:free_energy}), i.e.,
\begin{align}\label{eq:fd}
    F_{D}(n)=\frac{n (c^2 )^2 C_d^2}{4}\int_{S_{R}^p}\d^d \hat{x}\int_{S_{R}^p}\d^d \hat{y}\frac{1}{|\hat{x}-\hat{y}|^{2(d-2)}}+\mathcal{O}(n^2)\, .
\end{align}
Therefore, the universal part of the defect free energy always vanishes for our disordered free scalar, as is expected by the exactly marginality of the disorder strength. 

On the other hand, to have a more interesting RG monotone,  we define the effective free energy for the disordered defect by the derivative in $n$ at $n=0$,
\ie 
F_D^{\rm eff}\equiv \left.{\d F_D(n)\over \d n}\right|_{n=0}\,,
\label{effectiveF}
\fe
generalizing the work of \cite{Gurarie:1999bp,Patil:2025wqp} for bulk disorder. Then, by plugging the expression of \eqref{eq:fd} into this definition, we can obtain the \emph{exact} effective free energy,\footnote{Recall that $1\leq p \leq \min(2d-4, d-1)$ to avoid IR singularities of defect correlation functions. See also \eqref{eq:two_pt_defect} and one sentence below it.}
\begin{align}\label{eq:rg_monotone}
    F_D^{\rm eff}(R)=(c^2)^2 C_d^2 2^{p-2d+3}\pi^{p+\frac{1}{2}}\frac{\Gamma\left(\frac{p-2d+4}{2}\right)}{\Gamma(\frac{p+1}{2})\Gamma(p-d+2)}R^{2(p-d+2)}\ . 
\end{align}
Note that this implies that when $d=p-2$, the effective free energy vanishes exactly and there is no scale dependence. This is another manifestation of the exact marginality of disorder defect coupling $c$. We also give some comments on the divergence structures in the codimension-one disorder case $p=d-1$ with $d\geq 3$, where the effective free energy is given by
\begin{align}
    F_D^{\rm eff}(R)\overset{p=d-1}{=}(c^2)^2 C_d^2 2^{2-d}\pi^{d-\frac{1}{2}}\Gamma^{-1}\left(\frac{d}{2}\right)\Gamma\left(\frac{3-d}{2}\right)R^{2}\ .
\end{align}
The behavior differs depending on whether the bulk spacetime dimension $d$ is even or odd. When $d$ is odd, the effective free energy clearly diverges. However, this divergence can be removed by adding local counter terms on the defect. For instance, when $d=3$, we can add the cosmological counter term $\int_{S^{p}_R}\d^2 \hat{x}\sim R^2$ to the action, and get rid of UV divergence. 
On the other hand, when $d$ is even, there are no divergences in the effective free energy and the finite part is scheme-independent. This is because the defect we consider is invariant under transverse reflection and the possible form of local counter term on $p$-dimensional manifold is constrained to be $\mathcal{R}^{k}\mathcal{K}^{2m}$ where $\mathcal{R}$ and $\cal K$ are Ricci and extrinsic curvatures with indices suppressed.\footnote{In the case of boundary, additional logarithmic divergence also appear when $d$ is even \cite{Herzog:2015ioa, Fursaev:2015wpa}.}\footnote{Odd powers of the extrinsic curvatures $\mathcal{K}$ are allowed as local counter terms when the defect is equipped with a preferred orientation specified by the normal vectors. In such cases (e.g. for boundaries or oriented interfaces), the transverse reflection with respect to the defect is not a symmetry, and hence $\cal K \to -K$ is not required, allowing odd powers of $\cal K$.}

Finally, let us examine the behavior of the effective free energy for arbitrary bulk dimensions and specific defect dimensions $p=1,2$ in order. In the case of a line defect, i.e. $p=1$, it is natural, in light of known results \cite{Cuomo:2021rkm, Cuomo:2021kfm}, to consider the following quantity as a candidate for an RG monotone along the flow,
\begin{align}
    p=1\ : \ s_{D}^{\mathrm{eff}}(R)\equiv\left(1-R\frac{\partial}{\partial R}\right)F_D^{\rm eff}(R)\ , 
\end{align}
where $F_D^{\rm eff}(R)$ is given in \eqref{eq:rg_monotone}. This can be computed explicitly for our free theory example,
\begin{align}
    s_{D}^{\mathrm{eff}}(R)=(c^2)^2 C_d^2 2^{4-2d}\pi^{\frac{3}{2}}\frac{(2d-5)\Gamma\left(\frac{5-2d}{2}\right)}{\Gamma(3-d)}R^{6-2d}\ . 
\end{align}
To avoid IR singularities, the bulk dimensions should be $d\geq 3$.\footnote{Note that when $d=2$, the above quantity is proportional to $R^{2}$, which is another way to see the IR divergence.} For these integer bulk dimensions, the above defect entropy $s_{D}^{\mathrm{eff}}(R)$ vanishes, 
\begin{align}
    s_{D}^{\mathrm{eff}}(R) = 0 \qquad d\geq 3\ . 
\end{align}
This result is perfectly consistent with the Harris criterion \eqref{Harrisdef}, which says that the deformations should be marginal or irrelevant for $p=1$ and $d\geq 3$.

Next, we move on to the surface defect case $p=2$. In this case, guided by known results (see e.g. \cite{Shachar:2022fqk}), it is natural to consider the following quantity as a candidate for an RG monotone,
\begin{align}
    p=2\ : \ b_{D}^{\mathrm{eff}}(R)=\frac{3}{2}\left(R\frac{\partial}{\partial R} - R^2 \frac{\partial^2}{\partial R^2}\right)F_D^{\rm eff}(R)\ . 
\end{align}
By plugging \eqref{eq:rg_monotone} into the above, we obtain the following effective entropy, 
\begin{align}
    b_{D}^{\mathrm{eff}}(R)=3(c^2 )^2 C_{d}^{2}2^{7-2d}\pi (4-d)(d-3)\frac{\Gamma(3-d)}{\Gamma(4-d)}R^{8-2d}\ . 
\end{align}
We again require $d\geq 3$ to avoid IR singularities. This result is also perfectly consistent with the Harris criterion \eqref{Harrisdef}. Firstly, as described before, when $d=4$, the effective entropy $b_{D}^{\mathrm{eff}}(R)$ vanishes. When $d=3$, which corresponds to the relevant deformation, the effective entropy is given by
\begin{align}
    b_{D}^{\mathrm{eff}}(R)\overset{d=3}{=}-6\pi (c^2 )^2 C_{3}^{2} R^{2}\ , 
\end{align}
which is always negative except for UV fixed point ($R=0$) and consistent with the monotonicity theorem \cite{Jensen:2015swa}. Moreover, this implies that this relevant deformation does not terminate on non-trivial disorder defect fixed point since $\lim_{R\to \infty}b_{D}^{\mathrm{eff}}(R)=-\infty$ in the deep infrared. In some literature, this scenario is so-called \emph{run-away} \cite{Cuomo:2021kfm}. In this way, we have checked that our proposal for RG monotone for subdimensional disorder works well for bulk free scalar theories. It would be interesting to investigate whether \eqref{effectiveF} is indeed monotonic under other disordered defect RG flows.

\subsection{Symmetry Restoration by Subdimensional Disorder}\label{subsec:Z2_restoration}
It is often observed that a symmetry that is explicitly broken in the absence of disorder becomes restored once disorder is introduced.\footnote{See \cite{Antinucci:2023uzq} for discussions on this phenomenon from the view point of topological defects.} In this short section, we confirm this phenomenon in our pinning defect model \eqref{eq:free_theory2}. Before introducing the subdimensional disorder, the bulk spin-flip $\mathbb{Z}_2$ symmetry is explicitly broken on the defect. Indeed, if we compute the bulk one-point function of $\sigma$ in a fixed background configuration of the defect coupling $\h$, it turns out to be nonzero:
\begin{align}
    \langle\sigma(x)\rangle_{\h}=-C_{d}\int \d^{p}\hat{y}\frac{\h(\hat{y})}{|x-\hat{y}|^{p}}\,. 
\end{align}
However, once we average over $\h$ according to the Gaussian distribution, this one-point function vanishes, 
\begin{align}
    \overline{\langle\sigma(x)\rangle}=0\,. 
\end{align}
This clearly indicates that the $\mathbb{Z}_2$ symmetry is restored by the inclusion of the subdimensional disorder. Physically, this is reasonable because the Gaussian distribution is symmetric under the transformation $\h \to -\h$.
\subsection{First Pass: Logarithmic Behavior of Correlation Functions}\label{subsec:log_behaviour}
We now turn to the study of correlation functions on the fixed line described in \eqref{eq:line_fixed_point}. Since the fixed line occurs at $\gamma=0$, it is sufficient to analyze the original action \eqref{eq:free_theory2}. In this subsection, we focus exclusively on defect two-point functions, while correlation functions involving bulk local operators will be discussed in the next section. Our main goal here is to demonstrate explicitly that the disorder-averaged CFT correlators exhibit logarithmic dependence in contrast to the power-law behavior in usual CFTs. 

\subsubsection*{Two-point function: $\overline{\langle\,\sigma(\hat{x})\sigma(0) \, \rangle}$ }
We firstly investigate the disorder-averaged two-point function $\overline{\langle\,\sigma(\hat{x})\sigma(0) \, \rangle}$ on the defect, which can be deduced by setting $x_{\perp}\to 0$ and $y\to 0$ in \eqref{eq:two_pt_defect}:
\begin{align}\label{eq:defectscalar_two}
    \overline{\langle\, \sigma(\hat{x}) \sigma(0)\, \rangle}
    &=\frac{C_{d}}{|\hat{x}|^{p}}+{c}^{2}C_{d}^{2}\int \d^{p}\hat{z}\,\frac{1}{|\hat{x}-\hat{z}|^{p}|\hat{z}|^{p}}\,. 
\end{align}
The remaining integral clearly develops logarithmic singularities near the operator insertions at $\hat z=0$ and $\hat z=\hat{x}$. Indeed, upon evaluating the integral, one finds that it behaves as
\begin{align}
    \overline{\langle\, \sigma(\hat{x})\sigma(0)\, \rangle}=\frac{C_{d}}{|\hat{x}|^{p}}+\frac{C_{d} c^{2}}{2\pi}\frac{\log\mu^{2}|\hat{x}|^{2}}{|\hat{x}|^{p}}\,, 
\end{align}
where $\mu$ is an energy scale from regulating the integral. This logarithmic behavior is a distinctive feature of disorder averaging, and it clearly indicates that $\sigma(\hat{x})$ does not behave as an ordinary CFT primary operator.

\subsubsection*{Two-point function: $\overline{\langle\,\widetilde{D}_{i}(\hat{x})\widetilde{D}_{j}(0)) \, \rangle}$}
As a next step, we turn to a slightly more non-trivial example by considering a composite defect local operator $\widetilde{D}_{i}(\hat{x}) \equiv \sigma \partial_{i}\sigma(\hat{x})$ that carries transverse spin one. Its two-point function can be computed from
    \begin{align}
        \overline{\langle\, \widetilde{D}_{i}(\hat{x}) \widetilde{D}_{j}(\hat{y})\, \rangle}\equiv \frac{1}{4}\left.\frac{\partial}{\partial x_{\perp}^{i}}\frac{\partial}{\partial y_{\perp}^{j}} \overline{\langle\, \sigma^{2}(x)\sigma^{2}(y)\, \rangle}\right|_{x_{\perp}=y_{\perp}=0}\,. 
    \end{align}
After a straightforward calculation, one finds that the above two-point function can be expressed in terms of an integral of the form:
\begin{align}
      \overline{\langle\, \widetilde{D}_{i}(\hat{x}) \widetilde{D}_{j}(0)\, \rangle}=\frac{pC_{d}^{2}\delta_{ij}}{|\hat{x}|^{2p+2}}+\frac{p c^{2} C_{d}^{3} \delta_{ij}}{|\hat{x}|^{p+2}}\int \d^{p}\hat{z}\frac{1}{|\hat{x}-\hat{z}|^{p}|\hat{z}|^{p}}\,. 
\end{align}
This integral is the same as the one that we already encountered in the scalar two-point function \eqref{eq:defectscalar_two}. In particular, it exhibits a logarithmic divergence, confirming once again that the correlation functions do not follow a simple power-law behavior but instead display logarithmic corrections:
\begin{align}\label{eq:two_pt_Dtilde}
      \overline{\langle\, \widetilde{D}_{i}(\hat{x}) \widetilde{D}_{j}(0)\, \rangle}=pC_{d}^{2}\delta_{ij}\left(\frac{1}{|\hat{x}|^{2p+2}}+\frac{c^2}{2\pi}\frac{\log\mu^{2}|\hat{x}|^{2}}{|\hat{x}|^{2p+2}}\right)\,. 
\end{align}
In Section~\ref{sec:back_to_free}, we will see that these two-point functions are consistent with the structure of a logarithmic conformal defect, where defect local operators form nontrivial logarithmic multiplets. In particular, $\tilde D_i$ will be an operator in the logarithmic displacement multiplet.

\section{Systematics of Logarithmic Defect CFT}\label{sec:systematics}
From the defect two-point functions discussed in the previous sections and general discussion in \hyperref[sec:intro]{Introduction}, we expect defect local operators generated by subdimensional disorder to exhibit logarithmic behavior in their correlation functions. In particular, the appearance of logarithmic corrections indicates that these operators are naturally organized into logarithmic multiplets in logarithmic CFT (logCFT), rather than ordinary primaries and their descendants. Motivated by this perspective, we will refer to conformal defects whose local operator spectrum is built out of logarithmic multiplets as \emph{logarithmic defects}. The main goal of this section is to extract the model-independent features of such logarithmic conformal defects. In Section~\ref{subsec:rep_logCFT}, we review the representation theory of logCFT by focusing on primaries and the structures of correlation functions. In Section~\ref{subsec:logdefect_OPE}, we discuss how ordinary bulk CFT primary operators can be expanded in terms of logarithmic defect operators. Particularly, we determine the possible form of the defect-operator-expansion just from the representation theory. In Section~\ref{subsec:bulk_two_pt}, we also discuss the bootstrap equation that constrains the CFT data associated with logarithmic defects. 

\subsection{Review of Logarithmic Conformal Representation Theory}\label{subsec:rep_logCFT}
In this subsection, we present a brief review on the representation theory of the logCFTs, following \cite{Hogervorst:2016itc}. Readers who wish to explore this subject in more detail are referred to e.g. \cite{Hogervorst:2016itc, Cardy:2013rqg}. The conformal algebra $\mathfrak{so}(1, p+1)$ is generated by the dilatation $D$, translation $P^{a}$, special conformal transformation $K^a$, and rotation $M^{ab}=-M^{ba}$ in $p$ dimensions. In logCFTs, primary operators with conformal dimension $\widehat{\Delta}$ form a logarithmic multiplet of rank $N_{\widehat{\Delta}}\geq 1$, and such states are denoted as $\ket{\widehat{\mathcal{O}}_{I}}\, (I=1, 2, \cdots, N_{\widehat\Delta})$.\footnote{As our primary aim in this subsection is to arrange {\it defect} primaries into logarithmic multiplets, we adopt the convention of attaching a hat to most logarithmic variables.} They satisfy the following highest-weight condition,
\begin{align}
    K^{a}\ket{\widehat{\mathcal{O}}_{I}}=0\,. 
\end{align}
The descendant states are obtained by acting the translation operators $P^{a}$ on the primary states, just like in the standard CFT. The action of the dilatation operator $D$ on this logarithmic multiplet takes the following form,
\begin{align}\label{eq:dilatation}
    D\ket{\widehat{\mathcal{O}}_{I}}=-\text{i} \widehat{\bm{\Delta}}_{I}^{\, J}\ket{\widehat{\mathcal{O}}_{J}}\,,
\end{align}
where $\bm{\Delta}$ is a matrix of size $N_{\widehat{\Delta}}$. The crucial point is that this matrix cannot be diagonalized by any choice of basis. Instead, by fixing a basis, it can be brought into a Jordan block form
\begin{align}\label{eq:basis}
    \widehat{\bm{\Delta}} =  
    \begin{pmatrix} 
    \widehat{\Delta} & 1      & 0      & \cdots & 0 \\ 
    0      & \widehat{\Delta} & 1      & \cdots & 0 \\ 
    \vdots & \vdots & \vdots & \ddots & \vdots \\ 
    0      & 0      & 0      & \widehat{\Delta} & 1 \\ 
    0      & 0      & 0      & 0      & \widehat{\Delta} \end{pmatrix}\,.
\end{align}
It should be emphasized that, as will become evident later, the logarithmic behavior in logCFTs originates from the off–diagonal structure of the dilatation operator. By using the state-operator correspondence and translations, we can define the primary operator $\widehat{\mathcal{O}}_{I}(\hat{x})$, which transforms as
\begin{align}\label{eq:infinitesimal}
    \begin{aligned}
        \text{i}[P_{a}, \widehat{\mathcal{O}}_{I}(\hat{x})]&=\frac{\partial}{\partial \hat{x}^a }\widehat{\mathcal{O}}_{I}(\hat{x})\,, \\
        \text{i}[M_{ab}, \widehat{\mathcal{O}}_{I}(\hat{x})]&=\left(\Sigma_{ab}+\hat{x}_{a}\frac{\partial}{\partial \hat{x}^b}-\hat{x}_{b}\frac{\partial}{\partial \hat{x}^a }\right)\widehat{\mathcal{O}}_{I}(\hat{x})\,, \\
        \text{i}[D, \widehat{\mathcal{O}}_{I}(\hat{x})]&=\left(\bm{\widehat{\Delta}}_{I}^{\, J}+\delta_{I}^{J}\hat{x}\cdot \frac{\partial}{\partial \hat{x}}\right)\,\widehat{\mathcal{O}}_{J}(\hat{x})\,, \\
        \text{i}[K_{a}, \widehat{\mathcal{O}}_{I}(\hat{x})]&=-2\hat{x}_{a}\left(\bm{\widehat{\Delta}}_{I}^{\, J}+\delta_{I}^{J}\hat{x}\cdot \frac{\partial}{\partial \hat{x}}\right)\,\widehat{\mathcal{O}}_{J}(\hat{x})+|\hat{x}|^2 \frac{\partial}{\partial \hat{x}^a}\widehat{\mathcal{O}}_{I}(\hat{x}) +2\text{i}\hat{x}^{b}\Sigma_{ba}\widehat{\mathcal{O}}_{I}(\hat{x})\,, 
    \end{aligned}
\end{align}
where $\Sigma_{ab}$ is the matrix representation of the $p$-dimensional rotation group associated to the primary operator $\widehat{\mathcal{O}}_{I}(\hat{x})$. Notably, the action of $D$ and $K_{a}$ deviates from that in ordinary CFT, and the non-diagonal structure of $\bm{\Delta}$ implies mixing among the operators within a logarithmic multiplet. This mixing structure underlies the distinctive feature of logarithmic CFTs and will play a central role in the construction of correlation functions in the following sections. 

Finally, we present the finite form of the conformal transformation. From the above infinitesimal transformation rules \eqref{eq:infinitesimal}, one can deduce the finite transformation of logarithmic primaries. For a rank $N_{\widehat{\Delta}}$ logarithmic scalar primary operator $\widehat{\mathcal{O}}_{I}(\hat{x})$, the finite transformation under $\hat{x}\to \hat{x}'$ reads
\begin{align}\label{eq:conftrsf_log}
    \widehat{\mathcal{O}}'_{I}(\hat{x}')=\Omega(\hat{x})^{-\widehat{\Delta}/p}\Lambda_{I}^{\, J}\widehat{\mathcal{O}}_{J}(\hat{x})\,,
\end{align}
where $\Omega(\hat{x})$ is the Jacobian of the coordinate transformation,
\begin{align}
    \Omega(\hat{x})\equiv \left|\frac{\partial(\hat{x}')}{\partial(\hat{x})}\right|\,, 
\end{align}
and $\Lambda_{I}^{\, J}$ is an upper triangular matrix of the form
\begin{align}\label{eq:lambda}
    \Lambda_{I}^{\, J}=\frac{(-\ln \Omega)^{J-I}}{(J-I)!}\,, \quad J\geq I \,. 
\end{align}
For instance, in the case of $N_{\widehat{\Delta}}=2$, the above  transformation rule \eqref{eq:conftrsf_log} reads,
\begin{align}
    \begin{aligned}
        \widehat{\mathcal{O}}'_{1}(\hat{x}')&=\Omega(\hat{x})^{-\widehat{\Delta}/p}\left(\widehat{\mathcal{O}}_{1}(\hat{x})-\ln\Omega\, \widehat{\mathcal{O}}_{2}(\hat{x})\right)\,, \\
        \widehat{\mathcal{O}}'_{2}(\hat{x}')&=\Omega(\hat{x})^{-\widehat{\Delta}/p}\widehat{\mathcal{O}}_{2}(\hat{x})\,. 
    \end{aligned}
\end{align}

\subsection{Selected Correlation Functions}
In this subsection, we investigate how certain correlation functions are constrained by the logarithmic conformal representation theory developed in the previous section. As discussed in \hyperref[sec:intro]{Introduction}, we consider a setup in which ordinary CFT primaries $\mathcal{O}_{\Delta}(x)$ reside in the bulk, while logarithmic primary operators $\widehat{\mathcal{O}}^{I}_{\widehat{\Delta}}(\hat{x})$ are localized on a logarithmic defect. In general, correlation functions in such a defect CFT must satisfy the conformal Ward-Takahashi (WT) identities,\footnote{We assume the identity operator on the defect is in an ordinary representation (invariant under the residual conformal symmetry).}
\begin{align}\label{eq:CWTI}
    \langle\, [G, \mathcal{O}_{\Delta_1}(x_1)\mathcal{O}_{\Delta_2}(x_2)\cdots \widehat{\mathcal{O}}^{I_{1}}_{\widehat{\Delta}_{1}}(\hat{x}_{1})\widehat{\mathcal{O}}^{I_2}_{\widehat{\Delta}_{2}}(\hat{x}_{2})\cdots]\, \rangle=0\,, 
\end{align}
where $G$ is a generator of the residual conformal symmetry $\text{SO}(1, p+1)\times \text{SO}(d-p)$. To illustrate these constraints explicitly, we consider two representative cases in order. First, we study two-point functions of logarithmic primaries living on the defect, showing how their structure is fixed by the conformal Ward-Takahashi identities \eqref{eq:CWTI}. Second, we examine bulk–defect two-point functions and demonstrate how their form is similarly determined by these identities. This is without loss of generality since more general correlators are all determined by these basic correlators by operator-product-expansion (OPE).
Through these examples, we also highlight the characteristic features of logarithmic correlations and the interplay between bulk CFT primaries and defect logarithmic primaries.

\subsubsection{Defect Two-point Function}\label{subsubsec:defect_defect}
To begin with, we determine the defect scalar two-point function\footnote{The determination of the defect two-point function follows essentially the same logic as in the case of logarithmic CFT without a defect, see \cite[section 2.2]{Hogervorst:2016itc}. Therefore, readers who are already familiar with the analysis in ordinary homogeneous logCFT may safely skip to Section~\ref{subsubsec:bulk_defect}.}
\begin{align}
    \langle\, \widehat{\mathcal{O}}_{\widehat{\Delta}_1}^{I_1}(\hat{x})\widehat{\mathcal{O}}_{\widehat{\Delta}_2}^{I_2}(0)\, \rangle \,, \quad I_{i}=1, 2, \cdots, N_{i}\,, 
\end{align}
where $N_{i}\equiv N_{\widehat{\Delta}_{i}}$ is the rank of a logarithmic multiplet of conformal dimension $\widehat{\Delta}_{i}$. Firstly, the translation and rotational symmetries imply that the defect two-point function depends only on $s\equiv |\hat{x}|$: 
\begin{align}
    \langle\, \widehat{\mathcal{O}}_{\widehat{\Delta}_1}^{I_1}(\hat{x})\widehat{\mathcal{O}}_{\widehat{\Delta}_2}^{I_2}(0)\, \rangle=\frac{B_{I_{1}I_{2}}(s)}{s^{\widehat{\Delta}_1+\widehat{\Delta}_2}}\,. 
\end{align}
Here, we pull out the factor $s^{\widehat{\Delta}_1+\widehat{\Delta}_2}$ for convenience, and $B_{I_{1}I_{2}}$ is an $N_{1}\times N_{2}$ matrix.
Moreover, the conformal WT identities associated to the scale and special conformal transformations produce the following constraints,
\begin{align}
    s\frac{\d }{\d s}B_{I_{1}I_{2}}(s)&=-B_{(I_1 +1) I_2}(s)-B_{I_1(I_2 +1)}(s)\,, \\ 
    \left(s\frac{\d }{\d s} + \widehat{\Delta}_1-\widehat{\Delta}_2\right)B_{I_{1}I_{2}}(s)&=-2B_{(I_1 +1)I_2}(s)\,, \label{eq:WT_sct} 
\end{align}
where we have fixed the basis as in \eqref{eq:basis}. From these constraints, one can readily obtain
\begin{align}\label{eq:useful}
    (-\widehat{\Delta}_1 + \widehat{\Delta}_2 )B_{I_{1}I_{2}}(s)=B_{(I_1 +1)I_2}(s)-B_{I_1(I_2 +1)}(s)\,. 
\end{align}
This implies that, in order for $B_{I_{1}I_{2}}(s)$ to be non-trivial, the conformal dimensions for the two operators must be the same $\widehat{\Delta}\equiv \widehat{\Delta}_1 = \widehat{\Delta}_2$, thereby the ranks are also the same $N\equiv N_1 =N_2$. We thus obtain the result, familiar in the case of ordinary CFT, that two-point functions are only nontrivial for operators in the same conformal multiplet, which holds also in the logarithmic setting.
The constraint \eqref{eq:useful} further implies that $B_{I_{1}I_{2}}(s)$ only depends on $I_{1}+I_{2}$:
\begin{align}
     B_{I_{1}I_{2}}(s) \equiv \beta_{I_1 +I_2 -1}(s)\,. 
\end{align}
In terms of $\beta_n (s)$, the conformal WT identity \eqref{eq:WT_sct} takes the simpler form,
\begin{align}
    s\frac{\d }{\d s}\beta_n (s)=-2\beta_{n+1} (s)\,. 
\end{align}
By using $\beta_{N+1}=0$, we can iteratively solve this differential equation as
\begin{align}\label{betaexp}
    \beta_{n}(s)=\sum_{m=0}^{N-n}\frac{k_{n+m}}{m!}\left(-\log s^2 \right)^{m}\,. 
\end{align}
In summary, the defect scalar two-point function in logCFT can be fixed as
\begin{align}\label{eq:d2pt}
    \langle\, \widehat{\mathcal{O}}_{\widehat{\Delta}_1}^{I_1}(\hat{x})\widehat{\mathcal{O}}_{\widehat{\Delta}_2}^{I_2}(0)\, \rangle=\frac{\delta_{N_1,N_2}\delta_{\widehat{\Delta}_1 , \widehat{\Delta}_2}}{|\hat{x}|^{2\widehat{\Delta}_1}}\times
    \begin{cases}
        &\hspace{-3mm}\beta_{I_1 +I_2-1}(|\hat{x}|)\,, \quad 2\leq I_{1}+I_{2}\leq N_{1}+1\,, \\
        &\hspace{5mm}0 \quad\qquad\, , \quad \text{otherwise}\,. 
    \end{cases}
\end{align}
We now extend the above discussion to defect two-point functions of logarithmic primaries $\widehat{\mathcal{O}}_{\widehat{\Delta}_1 , i_{1}i_{2}, \cdots i_{s}}^{I_1}$ with transverse spin~$s$ associated to the rotation group $\mathrm{SO}(d-p)$.
For simplicity, we focus on operators transforming as symmetric traceless tensors under $\mathrm{SO}(d-p)$. This case is sufficient for the examples treated in the present paper. For defect operators, the rotational group $\text{SO}(d-p)$ behaves as a global symmetry, and the only invariant tensor is
\begin{align}
    \mathcal{I}_{i_{1}i_{2}, \cdots i_{s}}^{j_{1}j_{2}, \cdots j_{s}}\equiv \delta^{(j_{1}}_{i_1}\delta^{j_2}_{i_2}\cdots\delta^{j_{s})}_{i_s}\,, 
\end{align}
where the parenthesis is understood to impose the symmetric and traceless combination of indices. In terms of this invariant tensor, the spinning defect two-point function \eqref{eq:spin_defect} can be determined as
\begin{align}\label{eq:spin_defect_two}
    \langle\, \widehat{\mathcal{O}}_{\widehat{\Delta}_1 , i_{1}i_{2}, \cdots i_{s}}^{I_1}(\hat{x})\widehat{\mathcal{O}}_{\widehat{\Delta}_2}^{I_2, j_{1}j_{2}, \cdots j_{s}}(0)\, \rangle =\frac{\D_{N_1,N_2}\delta_{\widehat{\Delta}_1 , \widehat{\Delta}_2}}{|\hat{x}|^{2\widehat{\Delta}_1}}\,\mathcal{I}_{i_{1}i_{2}, \cdots i_{s}}^{j_{1}j_{2}, \cdots j_{s}}\times
    \begin{cases}
        &\hspace{-3mm}\beta_{I_1 +I_2-1}(|\hat{x}|)\,, \quad 2\leq I_{1}+I_{2}\leq N_1+1\,, \\
        &\hspace{5mm}0 \quad\qquad\, , \quad \text{otherwise}\,. 
    \end{cases}
\end{align}

\subsubsection{Bulk-defect Two-point Function}\label{subsubsec:bulk_defect}
We next determine the two-point function of a bulk scalar primary $\mathcal{O}_{\Delta}$ and a defect scalar logarithmic primary $\widehat{\mathcal{O}}^{I}_{\widehat{\Delta}}$.
Here, we set the position of defect local operator to the origin by using the translation symmetry along the defect. 
Firstly, recalling the form of the bulk-defect two-point function in normal defect CFT \cite{Billo:2016cpy}, it is convenient to write, 
\begin{align}\label{eq:b_d}
    \langle\,\mathcal{O}_{\Delta}(x)\widehat{\mathcal{O}}^{I}_{\widehat{\Delta}}(0) \, \rangle=\frac{C_{\Delta, \widehat{\Delta}}^{I}(x)}{|x_{\perp}|^{\Delta-\widehat{\Delta}}|x|^{2\widehat{\Delta}}}\,. 
\end{align}
Our remaining task is to fix the function $C_{\Delta, \widehat{\Delta}}^{I}(x)$ from the conformal WT identities \eqref{eq:CWTI}. Scale invariance forces this function to be subject to the following differential equation,
\begin{align}\label{eq:wt_scale}
    x\cdot \frac{\partial}{\partial x}C^{I}_{\Delta, \widehat{\Delta}}(x)=- C^{I+1}_{\Delta, \widehat{\Delta}}(x)\,. 
\end{align}
On the other hand, special conformal transformation invariance requires,
\begin{align}
    -2\hat{x}^{a}\left(\Delta + x\cdot \frac{\partial}{\partial x}\right)F^{a}(x) + x^{2} \frac{\partial}{\partial \hat{x}^{a}}F^{a}(x)=0\,,
\end{align}
which is equivalent to
\begin{align}\label{eq:wt_sct}
    -2\hat{x}^{a}x\cdot \frac{\partial}{\partial x}C^{I}_{\Delta, \widehat{\Delta}}(x)+|x|^{2} \frac{\partial}{\partial \hat{x}^{a}} C^{I}_{\Delta, \widehat{\Delta}}(x)=0\,. 
\end{align}
The residual Poincar\'e invariance constrains the function $C^{I}_{\Delta, \widehat{\Delta}}(x)$ to depend only on the two variables\footnote{Remark that $\log|\hat{x}|$ can be written in terms of $\tau$ and $\tilde{\tau}$.}
\begin{align}\label{eq:tau}
    \tau\equiv \log\frac{|x_{\perp}|}{|x|^{2}}\,, \qquad \tilde{\tau}\equiv \log |x_{\perp}|^{2}\,. 
\end{align}
Although at first sight the correlator seems to depend on both $\tau$ and $\tilde{\tau}$, it does depend only on $\tau$. Indeed, by using the following Leibniz rule,
\begin{align}
    \begin{aligned}
    \frac{\partial}{\partial x^{\mu}}
    &=\left({+}\frac{x_{\perp}^{i}\delta_{\mu}^{i}}{|x_{\perp}|^2}{-}\frac{2x_{\mu}}{|x|^{2}}\right)\frac{\partial}{\partial \tau}+2\frac{x_{\perp}^{i}\delta_{\mu}^{i}}{|x_{\perp}|^2}\frac{\partial}{\partial \tilde{\tau}}\,,
    \end{aligned}
\end{align}
the differential equation \eqref{eq:wt_sct} reduces to
\begin{align}
    \frac{\partial}{\partial \tilde{\tau}}C^{I}_{\Delta, \widehat{\Delta}}(\tau, \tilde{\tau})=0\,,
\end{align}
which clearly implies that the function $C^{I}_{\Delta, \widehat{\Delta}}$ depends only on $\tau$ and \eqref{eq:wt_scale} becomes,
\begin{align}
    \frac{\partial}{\partial \tau}C^{I}_{\Delta, \widehat{\Delta}}(\tau)=C^{I+1}_{\Delta, \widehat{\Delta}}(\tau)\,.  
\end{align}
This can be readily solved by the iterative method as before:
\begin{align}\label{eq:ci}
    C^{I}_{\Delta, \widehat{\Delta}}(\tau)=\sum_{k=0}^{N_{\widehat{\Delta}}-I}\frac{\lambda_{I+k}\tau^{k}}{k!}\,, 
\end{align}
where the coefficients $\lambda_{I+k}$ remain undetermined by residual conformal symmetry alone. We comment that this result is consistent with the method of images \cite{Nishioka:2022ook}, where the bulk-defect two-point function is \textit{kinematically} equivalent to the three-point function in the absence of a defect. More precisely, 
\begin{align}
    \langle\,\mathcal{O}_{\Delta}(x)\widehat{\mathcal{O}}^{I}_{\widehat{\Delta}}(0) \, \rangle_{\text{DCFT}}\cong \langle\,\mathcal{O}_{\Delta/2}(x)\mathcal{O}_{\Delta/2}(\bar{x})\mathcal{O}^{I}_{\widehat{\Delta}}(0) \, \rangle_{\text{CFT}}\,,
\end{align}
where $\bar{x}\equiv(\hat{x}^{a}, -x_{\perp}^{i})$.
The correlation function on the right hand side has been determined in \cite[equation (2.34)]{Hogervorst:2016itc}, and completely matches with our result \eqref{eq:b_d} and \eqref{eq:ci} including the $\tau$-dependence.

We can also extend the above analysis to the case where the defect operator carries spin. For simplicity, we keep the bulk primary to be a scalar. In general, a defect local primary can carry either parallel and transverse spins associated to the rotational groups $\text{SO}(p)$ and $\text{SO}(d-p)$, respectively. However, one can show that the bulk–defect two-point function involving a bulk scalar and a defect operator with parallel spin vanishes identically. Therefore, it suffices to consider defect primaries with transverse spin. Let $s$ be the transverse spin, the result takes the following form
\begin{align}\label{eq:b_d_spin}
    \langle\,\mathcal{O}_{\Delta}(x)\widehat{\mathcal{O}}^{I}_{\widehat{\Delta}, i_1 i_2 \cdots i_{s}}(0) \, \rangle=\frac{x_{\perp}^{(i_{1}}x_{\perp}^{i_{2}}\cdots x_{\perp}^{i_{s})}}{|x_{\perp}|^{\Delta-\widehat{\Delta}+s}|x|^{2\widehat{\Delta}}}C_{\Delta, \widehat{\Delta}}^{I}(x)\,,  
\end{align}
where $C^{I}_{\Delta, \widehat{\Delta}}$ has the same structure in \eqref{eq:ci}, and the parenthesis is to impose the symmetric and traceless combination of indices.

\subsection{Logarithmic Defect Operator Expansion}\label{subsec:logdefect_OPE}
In homogeneous CFTs, it is well known that when two local operators approach each other, their product can be systematically expanded into a sum of local operators through the operator-product-expansion (OPE). In the presence of a defect, this framework is enriched by the so-called defect-operator-expansion (DOE), in which a bulk local operator can be expanded in terms of local operators residing on the defect. The purpose of this subsection is to explore the logarithmic defect-operator-expansion (logDOE), which generalizes the standard DOE to the case where the defect supports logCFT local operators. The significance of this expansion lies in the fact that it bridges the representations of the bulk CFT with those of the defect, thereby providing a powerful tool to analyze the interplay between bulk and defect degrees of freedom. This feature will play an important role in later sections. 

For simplicity, in this work we focus on the logDOE of a bulk scalar primary ${\mathcal{O}}_{\Delta}$. From the analysis in the Section~\ref{subsubsec:bulk_defect}, it is clear that both scalar defect operators and spinning defect operators with transverse spin can contribute to the expansion.
Accordingly, we have the following general ansatz for the logDOE:
\begin{align}\label{eq:ansatz}
    \mathcal{O}_{\Delta}(x)=\sum_{\widehat{\Delta}, s}\sum_{I=1}^{N_{\widehat{\Delta}, s}}\frac{x_{\perp}^{i_{1}}x_{\perp}^{i_{2}}\cdots x_{\perp}^{i_{s}}}{|x_{\perp}|^{\Delta - \widehat{\Delta}+s}}B^{I}_{\Delta, \widehat{\Delta}}(|x_{\perp}|, \widehat{\partial})\,\widehat{\mathcal{O}}^{I}_{\widehat{\Delta}, i_{1}i_{2}\cdots i_s}(\hat{x})\,,
\end{align}
where the sum runs over defect logarithmic primaries (both scalar and spinning), and $N_{\widehat{\Delta}, s}$ represents the rank of a defect logarithmic primary with dimension $\widehat{\Delta}$ and transverse spin $s$. Also, the function $B^{I}_{\Delta, \widehat{\Delta}}(|x_{\perp}|, \widehat{\partial})$ encodes the dependence on the transverse distance as well as derivatives along the defect. For convenience, in the above expression we have explicitly extracted the overall power of $|x_{\perp}|$ together with the tensor structure associated with the transverse spin $s$.\footnote{This separation makes the covariance under the transverse rotation group $\text{SO}(d-p)$ manifest.}  

Our next task is to determine the coefficient functions $B^{I}_{\Delta, \widehat{\Delta}}(|x_{\perp}|, \widehat{\partial})$ in \eqref{eq:ansatz}. Since we have already factored out the tensor structures associated with the transverse spin, it suffices to focus only on the scalar contributions:
\begin{align}\label{eq:ansatz_scalar}
    \mathcal{O}_{\Delta}(x)\overset{\text{scalar}}{=}\sum_{\widehat{\Delta}}\sum_{I=1}^{N_{\widehat{\Delta}}}\frac{1}{|x_{\perp}|^{\Delta - \widehat{\Delta}}}B^{I}_{\Delta, \widehat{\Delta}}(|x_{\perp}|, \widehat{\partial})\,\widehat{\mathcal{O}}^{I}_{\widehat{\Delta}}(\hat{x})\,.
\end{align}
The essential requirement is that both sides of \eqref{eq:ansatz_scalar} must transform covariantly under the dilatation $x\to x'=\lambda x$,
\begin{align}
    \mathcal{O}_{\Delta}'(x')\overset{\text{scalar}}{=}\sum_{\widehat{\Delta}}\sum_{I=1}^{N_{\widehat{\Delta}}}\frac{1}{|x_{\perp}'|^{\Delta - \widehat{\Delta}}}B^{I}_{\Delta, \widehat{\Delta}}(|x_{\perp}'|, \widehat{\partial}')\,\widehat{\mathcal{O}}^{\prime\, I}_{\widehat{\Delta}}(\hat{x}')\,.
\end{align}
By recalling the transformation law \eqref{eq:conftrsf_log}, the above relation is reduced to
\begin{align}
    \mathcal{O}_{\Delta}(x)\overset{\text{scalar}}{=}\sum_{\widehat{\Delta}}\sum_{I, J=1}^{N_{\widehat{\Delta}}}\frac{1}{|x_{\perp}|^{\Delta - \widehat{\Delta}}}B^{I}_{\Delta, \widehat{\Delta}}(\lambda|x_{\perp}|, \lambda^{-1}\widehat{\partial})\,\Lambda^{I}_{\, J}\, \widehat{\mathcal{O}}^{J}_{\widehat{\Delta}}(\hat{x})\,,
\end{align}
where the matrix element $\Lambda^{I}_{\, J}$ is given in \eqref{eq:lambda}. By comparing this with \eqref{eq:ansatz_scalar}, we have
\begin{align}
    \sum_{J=1}^{N_{\widehat{\Delta}}}B^{I}_{\Delta, \widehat{\Delta}}(\lambda|x_{\perp}|, \lambda^{-1}\widehat{\partial})\,\Lambda^{I}_{\, J}=B^{J}_{\Delta, \widehat{\Delta}}(|x_{\perp}|, \widehat{\partial})\,. 
\end{align}
The most general solution is given by
\begin{align}
    B^{I}_{\Delta, \widehat{\Delta}}(|x_{\perp}|, \widehat{\partial})=\sum_{n=0}^{I-1}\frac{(\log|x_{\perp}|)^{n}}{n!}g_{I-n}(|x_{\perp}|^2 \Box_{\hat{x}})\,,
\end{align}
where $g_{I-n}(|x_{\perp}|^2 \Box_{\hat{x}})$ is non-singular under the limit $x_{\perp}\to 0$, and expanded as
\begin{align}\label{eq:a}
    g_{I-n}(|x_{\perp}|^2 \Box_{\hat{x}})=\sum_{k=0}^{\infty}A^{N_{\widehat{\Delta}}}_{I-n , k}\,|x_{\perp}|^{2k}\, \Box_{\hat{x}}^{k}\,. 
\end{align}
Here the index $k$ encodes the level of the descendant operators that enter in the DOE. 

What remains is to be fixed are the coefficients $A^{N_{\widehat{\Delta}}}_{I-n , k}$ in \eqref{eq:a}. In principle, imposing covariance of the logDOE under the full set of residual special conformal transformations parallel to the defect determines all of these coefficients except for $A^{N_{\widehat{\Delta}}}_{I-n , 0}$ for the primary. Rather than following this route here, we adopt an alternative and more direct method. In practice, this is carried out as follows. First, we evaluate the bulk–defect two-point function $\langle\mathcal{O}_{\Delta}(x)\widehat{\mathcal{O}}^{I}_{\widehat{\Delta}}(0)\rangle$
by employing the logDOE together with the defect two-point function \eqref{eq:d2pt} derived in the previous subsection. Second, we compare the resulting expression with the general form of the bulk–defect two-point function \eqref{eq:b_d} obtained earlier. This comparison uniquely fixes the coefficients $A^{N_{\widehat{\Delta}}}_{I-n , k}$ (up to $A^{N_{\widehat{\Delta}}}_{I-n , 0}$). 

By plugging the expression of logDOE \eqref{eq:ansatz_scalar} into the bulk-defect two-point function, we obtain
\begin{align}
    \langle\mathcal{O}_{\Delta}(x)\widehat{\mathcal{O}}^{I}_{\widehat{\Delta}}(0)\rangle 
    = \sum_{\widehat{\Delta}'}\sum_{J=1}^{N_{\widehat{\Delta}'}}\sum_{n=0}^{J-1}\frac{(\log|x_{\perp}|)^n}{n!}\frac{g_{J-n}(|x_{\perp}|^2 \Box_{\hat{x}})}{|x_{\perp}|^{\Delta-\widehat{\Delta}'}}\langle\widehat{\mathcal{O}}^{J}_{\widehat{\Delta}'}(\hat{x})\widehat{\mathcal{O}}^{I}_{\widehat{\Delta}}(0)\rangle\,. 
\end{align}
Moreover, by using the explicit form of the defect two-point function \eqref{eq:d2pt}, the above equation is rewritten as 
\begin{align}\label{eq:logdoe_general}
    \begin{aligned}
            \langle\mathcal{O}_{\Delta}(x)\widehat{\mathcal{O}}^{I}_{\widehat{\Delta}}(0)\rangle 
    &= \sum_{J=1}^{N_{\widehat{\Delta}}-I+1}\sum_{n=0}^{J-1}\sum_{m=0}^{N_{\widehat{\Delta}}-I-J+1}\sum_{\ell = 0}^{\infty}\frac{(-1)^{m}(\log|x_{\perp}|)^n}{n!\, m!}\frac{k_{I+J-1+m}A^{N_{\widehat{\Delta}}}_{J-n, \ell}}{|x_{\perp}|^{\Delta-\widehat{\Delta}-2\ell}}\,\Box_{\hat{x}}^{\ell}\frac{(\log |\hat{x}|^2)^m}{|\hat{x}|^{2\widehat{\Delta}}}\\
    &=\sum_{\ell = 0}^{\infty}\sum_{k=0}^{N_{\widehat{\Delta}}-I}\sum_{J=1}^{N_{\widehat{\Delta}}-I-k+1}\sum_{n=0}^{k}\frac{(\log|x_{\perp}|)^n}{n!\, (k-n)!}\frac{k_{I+J+k-1}A^{N_{\widehat{\Delta}}}_{J, \ell}}{|x_{\perp}|^{\Delta-\widehat{\Delta}-2\ell}}\,\Box_{\hat{x}}^{\ell}\frac{(-\log |\hat{x}|^2)^{k-n}}{|\hat{x}|^{2\widehat{\Delta}}}\,, 
    \end{aligned}
\end{align}
where in the second line, we have rearranged the sum by,
\begin{align}
    \sum_{J=1}^{N_{\widehat{\Delta}}-I+1}\sum_{n=0}^{J-1}\sum_{m=0}^{N_{\widehat{\Delta}}-I-J+1}f(J, n, m)=\sum_{k=0}^{N_{\widehat{\Delta}}-I}\sum_{J=1}^{N_{\widehat{\Delta}}-I-k+1}\sum_{n=0}^{k}f(J+n , n, k-n)\,.
\end{align}
In what follows, we first focus on the coefficients $A^{N_{\widehat{\Delta}}}_{J , 0}$, which are associated with the contributions of the defect primaries, and reorganize the above expression using the binomial expansion formula,
\begin{align}\label{eq:primary}
    \begin{aligned}
        &\langle\mathcal{O}_{\Delta}(x)\widehat{\mathcal{O}}^{I}_{\widehat{\Delta}}(0)\rangle 
        =\frac{1}{|x_{\perp}|^{\Delta-\widehat{\Delta}}|\hat{x}|^{2\widehat{\Delta}}}\sum_{k=0}^{N_{\widehat{\Delta}}-I}\sum_{J=1}^{N_{\widehat{\Delta}}-I-k+1}\frac{k_{I+J+k-1}A^{N_{\widehat{\Delta}}}_{J, 0}}{k!}\left(\log\frac{|x_{\perp}|}{|\hat{x}|^2}\right)^k +O\left(\frac{|x_{\perp}|^2}{|\hat{x}|^2}\right)\,. 
    \end{aligned}
\end{align}
On the other hand, 
by expanding \eqref{eq:b_d} and \eqref{eq:ci} around $|x_{\perp}|/|\hat{x}| = 0$, we obtain
\begin{align}
    \langle\mathcal{O}_{\Delta}(x)\widehat{\mathcal{O}}^{I}_{\widehat{\Delta}}(0)\rangle 
    =\frac{1}{|x_{\perp}|^{\Delta-\widehat{\Delta}}|\hat{x}|^{2\widehat{\Delta}}}\sum_{k=0}^{N_{\widehat{\Delta}}-I}\frac{\lambda_{I+k}}{k!}\left(\log\frac{|x_{\perp}|}{|\hat{x}|^2}\right)^k +O\left(\frac{|x_{\perp}|^2}{|\hat{x}|^2}\right)\,. 
\end{align}
By comparing this expression with \eqref{eq:primary}, we find the relation
\begin{align}\label{eq:primary_constraint}
    \lambda_{I}=\sum_{J=1}^{N_{\widehat{\Delta}}-I+1}k_{I+J-1}A_{J,0}^{N_{\widehat{\Delta}}}\,, \quad 1\leq I \leq N_{\widehat{\Delta}}\,. 
\end{align}

In the same manner, one can determine the coefficients $A_{J, \ell}^{N_{\widehat{\Delta}}}$ ($\ell\geq 1$) associated to the descendant operators. Since the derivation is rather involved and highly technical, we relegate the detailed computation to the Appendix \ref{app:logDOE} and only present the final results here,
\begin{align}
    A_{J, \ell}^{N_{\wD }}=\frac{(-4)^{-\ell}}{\ell! \left(\widehat{\Delta}+z-\frac{p-2}{2}\right)_{\ell}}\sum_{m=1}^{J}(-1)^{J+m}A_{m, 0}^{N_{\wD }}\sum_{0\leq k_{1}< \cdots < k_{J-m}\leq \ell -1}\prod_{r=1}^{J-m}\left(\wD -\frac{p-2}{2}+k_{r}\right)^{\hspace{-1mm}-1}\,. 
\end{align}
Note that, as is the case for the two-point functions of logarithmic primaries, the logDOE coefficients depend on the choice of basis among the multiple primaries in a logarithmic multiplet. In Section~\ref{subsec:bulk_two_pt}, we introduce a specific ``gauge choice'' following \cite{Hogervorst:2016itc} that simplifies various expressions and highlights the gauge-independent physical data encoded in the DOE. In Section~\ref{subsec:b_d_data}, we will give also explicit expressions for these logDOE coefficients for the logarithmic defect that emerges from subdimensional disorder in free theory.

\subsection{Bulk Two-point Function and Bootstrap Constraints}\label{subsec:bulk_two_pt}
In this subsection we consider the bulk scalar two-point function $\langle \mathcal{O}_{\Delta_1}(x_{1})\mathcal{O}_{\Delta_2}(x_{2}) \rangle$ in the presence of a logarithmic defect. The residual conformal invariance allows us to write the bulk two-point function as
\begin{align}\label{eq:bulk_2pt}
    \langle \mathcal{O}_{\Delta_1}(x_{1})\mathcal{O}_{\Delta_2}(x_{2}) \rangle =\frac{F(\xi, \phi)}{|x_{1\perp}|^{\Delta_1}|x_{2\perp}|^{\Delta_2}}\,, 
\end{align}
where $\xi$ and $\phi$ are cross-ratios defined by
\begin{align}
    \xi\equiv\frac{|x_{12}|^{2}}{|x_{1\perp}||x_{2\perp}|}\,, \quad \cos\phi\equiv \frac{x_{1\perp}\cdot x_{2\perp}}{|x_{1\perp}||x_{2\perp}|}\,. 
\end{align} 
Studying this correlator provides the starting point for deriving the defect bootstrap equations.
The bootstrap equations encode the consistency of OPE: the result obtained by first fusing the two bulk operators must coincide with the one obtained by first expanding a bulk operator into defect operators. Since the bulk theory is an ordinary CFT, the corresponding bulk conformal blocks are identical to the well-known expressions in the literature \cite{Billo:2016cpy}. On the other hand, the defect channel exhibits a much richer structure, and the associated defect conformal blocks develop new features characteristic of the logarithmic defect (the discussions on defect conformal blocks of ordinary defect CFT can be found in \cite{Billo:2016cpy}). The main goal of this subsection is to derive the logarithmic defect conformal block explicitly.
\subsubsection*{Specific Gauge Choice}

The two-point functions of logarithmic multiplets contain redundancies that correspond to a change of basis among the primaries that preserve the form of the dilatation operator in \eqref{eq:dilatation} (see \cite[Appendix.\,A]{Hogervorst:2016itc} for a detailed discussion).
To simplify the presentation below, we further make use of this freedom to fix $k_{1}=k_{2}=\cdots =k_{N_{\widehat{\Delta}}-1}=0$ and $k_{\widehat{\mathcal{O}}}\equiv k_{N_{\widehat{\Delta}}}\not=0$ in \eqref{betaexp} which enter in the two-point functions \eqref{eq:d2pt} and \eqref{eq:spin_defect_two} so that,
\begin{align}\label{eq:specific_basis}
    \langle\, \widehat{\mathcal{O}}_{\widehat{\Delta}_1 , i_{1}i_{2}, \cdots i_{s}}^{I_1}(\hat{x})\widehat{\mathcal{O}}_{\widehat{\Delta}_2}^{I_2, j_{1}j_{2}, \cdots j_{s}}(0)\, \rangle =\frac{k_{\widehat{\mathcal{O}}}\cdot \delta_{\widehat{\Delta}_1 , \widehat{\Delta}_2}\mathcal{I}_{i_{1}i_{2}, \cdots i_{s}}^{j_{1}j_{2}, \cdots j_{s}}}{(N_{\wD , s} -I_1 -I_2 +1)!}\frac{(-\log|\hat{x}|^2)^{N_{\wD , s} -I_1 -I_2 +1}}{|\hat{x}|^{2\widehat{\Delta}}}\,, 
\end{align}
and vanishes when $N_{\wD} +1 \leq I_1 +I_2$. In the same basis, the relation \eqref{eq:primary_constraint} is simplified to
\begin{align}\label{eq:simple}
    \lambda_{I}=k_{\widehat{\mathcal{O}}}\,A^{N_{\wD , s}}_{N_{\wD , s}-I+1, 0}\,. 
\end{align}
\subsubsection*{Gram Matrix}
As preparation for the defect conformal block expansion, we introduce the Gram matrix
\begin{align}
    G_{I \vec{a};J\vec{b}}\equiv\braket{\widehat{\mathcal{O}}^{I}_{\widehat{\Delta}, s};\vec{a}|\widehat{\mathcal{O}}^{J}_{\widehat{\Delta}, s};\vec{b}}\,, 
\end{align}
where $\vec{b}=\{b_{1}, b_{2}, \cdots \}$ denotes collectively the vector indices tangential to the defect, and the descendant state $\ket{\widehat{\mathcal{O}}^{J}_{\widehat{\Delta}, s};\vec{b}}$ is written in terms of parallel momenta operators,
\begin{align}
    \ket{\widehat{\mathcal{O}}^{J}_{\widehat{\Delta}, s};\vec{b}}=P_{b_1}P_{b_2}\cdots \ket{\widehat{\mathcal{O}}^{J}_{\widehat{\Delta}, s}}\,.
\end{align}
Also, its BPZ conjugate state can be expressed as
\begin{align}
    \bra{\widehat{\mathcal{O}}^{I}_{\widehat{\Delta}, s};\vec{a}}=\bra{\widehat{\mathcal{O}}^{I}_{\widehat{\Delta}, s}}K_{a_{1}}K_{a_{2}}\cdots\,. 
\end{align}
By using the commutation relation $P_{b}$ and $K_{a}$ and the conformal dimension matrix \eqref{eq:basis}, one can express this Gram matrix as follows:\footnote{Note that the commutation algebra between $P_{b}$ and $K_{a}$ concern not only $D$ but also $M_{ab}$ which is the generator of SO$(p)$ group. However, since we are interested in the defect local operator with trivial SO$(p)$ spin, the contributions from $M_{ab}$ do not appear.}
\begin{align}
    G_{I \vec{a};J\vec{b}}=k_{\widehat{\mathcal{O}}}\delta_{\vec{a}, \vec{b}}V_{IJ}(\partial_{\wD})g_{\vec{a}}(\wD,s )\,, 
\end{align}
where $g_{\vec{a}}(\wD,s)$ depends only on $\wD$ and $s$, and can be fixed by conformal algebra. Also, the matrix $V_{IJ}$ is defined by
\begin{align}
        V_{IJ}(\partial_{\wD}) =\frac{1}{(N_{\wD , s} -I_1 -I_2 +1)!}\times
    \begin{cases}
        &\hspace{-3mm}\partial_{\wD}^{N_{\wD , s} -I_1 -I_2 +1}\,, \quad 2\leq I_{1}+I_{2}\leq N_{\widehat{\Delta} , s}+1\,, \\
        &\hspace{5mm}0 \quad\qquad\, , \quad \text{otherwise}\,. 
    \end{cases}
\end{align}
Moreover, the inverse Gram matrix is given by
\begin{align}\label{eq:inverse_gram}
    \widetilde{G}_{I \vec{a};J\vec{b}}=k^{-1}_{\widehat{\mathcal{O}}}\delta_{\vec{a}, \vec{b}}\widetilde{V}_{IJ}(\partial_{\wD})g^{-1}_{\vec{a}}(\wD,s )\,, 
\end{align}
where $\widetilde{V}_{IJ}$ is defined by
\begin{align}
    \widetilde{V}_{IJ}(\partial_{\wD}) =\frac{1}{(I_1 +I_2-N_{\wD , s} -1)!}\times
    \begin{cases}
        &\hspace{-3mm}\partial_{\wD}^{I_1 +I_2-N_{\wD , s} -1}\,, \quad N_{\widehat{\Delta}, s}+1\leq I_{1}+I_{2}\leq 2N_{\widehat{\Delta}, s}\,, \\
        &\hspace{5mm}0 \quad\qquad\, , \quad \text{otherwise}\,. 
    \end{cases}
\end{align}
One can readily check that the above matrix \eqref{eq:inverse_gram} satisfies the following desired relation,
\begin{align}
    \sum_{K, \vec{c}}G_{I \vec{a};K\vec{c}}\,\widetilde{G}_{K \vec{c};J\vec{b}}=\delta_{I, J}\delta_{\vec{a}, \vec{b}}\,. 
\end{align}
\subsubsection*{Logarithmic Defect Conformal Block Expansion}
With the preparation complete, we are now ready to determine the defect conformal block expansion of the bulk two-point function. The first step is to insert a complete set of defect primary operators:\footnote{Here, we again neglect the contribution from the defect local primaries with SO$(p)$ spin. This is because the correlation function of bulk scalar operator and defect operator with SO$(p)$ spin vanishes.}
\begin{align}
    \mathbf{1}=\sum_{\wD , s}\sum_{I, J=1}^{N_{\wD , s}}\sum_{\vec{a}, \vec{b}}\widetilde{G}_{I \vec{a};J\vec{b}}\ket{\widehat{\mathcal{O}}^{I}_{\widehat{\Delta}, s};\vec{a}}\bra{\widehat{\mathcal{O}}^{J}_{\widehat{\Delta}, s};\vec{b}}
\end{align}
into \eqref{eq:bulk_2pt}. This allows us to decompose the bulk two-point function as follows:
\begin{align}\label{eq:defect_expansion}
    \begin{aligned}
        \langle \mathcal{O}_{\Delta_1}(x_{1})\mathcal{O}_{\Delta_2}(x_{2}) \rangle = \sum_{\wD , s} W^{N_{\wD , s}}_{\widehat{\mathcal{O}}}(x_{1}, x_{2}; \Delta_1 , \Delta_2 )\,, 
    \end{aligned}
\end{align}
where $W^{N_{\wD , s}}_{\widehat{\mathcal{O}}}$ is the conformal partial wave defined by
\begin{align}\label{eq:partial_wave}
    W^{N_{\wD , s}}_{\widehat{\mathcal{O}}}\equiv \sum_{I, J=1}^{N_{\wD , s}}\sum_{\vec{a}, \vec{b}}\widetilde{G}_{I \vec{a};J\vec{b}}\bra{\mathcal{D}}\mathcal{\mathcal{O}}_{\Delta_{1}}(x_{1})\ket{\widehat{\mathcal{O}}^{I}_{\widehat{\Delta}, s};\vec{a}}\bra{\widehat{\mathcal{O}}^{J}_{\widehat{\Delta}, s};\vec{b}}\mathcal{O}_{\Delta_{2}}(x_{2})\ket{\mathcal{D}}\,. 
\end{align}
Here, $\ket{\mathcal{D}}$ represents the defect vacuum, and the radial quantization is performed around a point on the defect. We next show that this conformal partial wave satisfies differential equations. We firstly recall that the residual conformal Lie algebra $\mathfrak{so}(1, p+1)\oplus\mathfrak{so}(d-p)$ has two quadratic Casimir operators $L^{2}$ and $S^2$:
\begin{align}
    L^{2}&\equiv \i D (\i D -p)+\frac{1}{2}M_{ab}M^{ab}+P^{a}K_{a}\,,  \quad 
    S^{2}\equiv \frac{1}{2}M_{ij}M^{ij}\,. 
\end{align}
These Casimir operators act on the logarithmic multiplet as follows,
\begin{align}\label{eq:Casimir_eq}
    \begin{aligned}
    L^{2}\ket{\widehat{\mathcal{O}}^{I}_{\widehat{\Delta}, s};\vec{a}}&=\widehat{C}_{\wD , 0}\ket{\widehat{\mathcal{O}}^{I}_{\widehat{\Delta}, s};\vec{a}}+\partial_{\wD }\widehat{C}_{\wD , 0}\ket{\widehat{\mathcal{O}}^{I+1}_{\widehat{\Delta}, s};\vec{a}}+\frac{1}{2}\partial_{\wD }^{2}\widehat{C}_{\wD , 0}\ket{\widehat{\mathcal{O}}^{I+2}_{\widehat{\Delta}, s};\vec{a}}\,,  \\
    S^{2}\ket{\widehat{\mathcal{O}}^{I}_{\widehat{\Delta}, s};\vec{a}}&=\widehat{C}_{0 , s}\ket{\widehat{\mathcal{O}}^{I}_{\widehat{\Delta}, s};\vec{a}}\,, 
    \end{aligned}
\end{align}
where $\widehat{C}_{\wD , s}=\wD (\wD -p)+s(s+d-p-2)$. In contrast to the case of an ordinary CFT, the Casimir operator for $\mathfrak{so}(1, d+1)$ does not act diagonally on the space of states in an irreducible logarithmic multiplet. Instead, they induce a mixing among states with different Jordan labels $I$ due to the non-diagonal property of conformal dimension matrix $\mathbf{\wD}$. Noting that, for a state of arbitrary rank $N_{\wD, s}$, the operator $(L^{2}-\widehat{C}_{\wD , 0})^{N_{\wD, s}}$ annihilates it, the above Casimir equations \eqref{eq:Casimir_eq} are written as 
\begin{align}
    \begin{aligned}
            (L^{2}-\widehat{C}_{\wD , 0})^{N_{\wD, s}}\ket{\widehat{\mathcal{O}}^{I}_{\widehat{\Delta}, s};\vec{a}}&=0\,, \quad \forall I \,, \\
            (S^{2}-\widehat{C}_{0 , s})\ket{\widehat{\mathcal{O}}^{I}_{\widehat{\Delta}, s};\vec{a}}&=0\,. 
    \end{aligned}
\end{align}
By combining these equations with \eqref{eq:partial_wave}, it turns out that the conformal partial wave $W_{\widehat{\mathcal{O}}}^{N_{\wD , s}}$ satisfies the following differential equations:
\begin{align}\label{eq:differential_eqs}
    \begin{aligned}
        (\mathcal{L}_{1}^{2}-\widehat{C}_{\wD , 0})^{N_{\wD, s}}W_{\widehat{\mathcal{O}}}^{N_{\wD , s}}&=0 \,, \\
        (\mathcal{S}_{1}^{2}-\widehat{C}_{0 , s})W_{\widehat{\mathcal{O}}}^{N_{\wD , s}}&=0\,, 
    \end{aligned}
\end{align}
where $\mathcal{L}_{1}^{2}$ and $\mathcal{S}_{1}^{2}$ are differential representations of Casimir operators $L^2$ and $S^2$ acting on the coordinate $x_{1}$, respectively. In line with \eqref{eq:bulk_2pt}, it is convenient to factor out the conformal factor as follows:
\begin{align}\label{eq:W}
    W_{\widehat{\mathcal{O}}}^{N_{\wD , s}}=\frac{G_{W_{\widehat{\mathcal{O}}}}^{N_{\wD , s}}(\chi)H_{W_{\widehat{\mathcal{O}}}}^{N_{\wD , s}}(\phi)}{|x_{1\perp}|^{\Delta_1}|x_{2\perp}|^{\Delta_2}}\,, \quad \chi\equiv \frac{|\hat{x}_{12}|^{2}+|x_{1\perp}|^2 +|x_{2\perp}|^2}{|x_{1\perp}||x_{2\perp}|} \, .  
\end{align}
Then, the above differential equations \eqref{eq:differential_eqs} become
\begin{align}
        (\widehat{\mathcal{D}}_{L^2})^{N_{\wD, s}}G_{W_{\widehat{\mathcal{O}}}}^{N_{\wD , s}}(\chi)&=0  \,, \label{eq:first_eq} \\
        \widehat{\mathcal{D}}_{S^2} H_{W_{\widehat{\mathcal{O}}}}^{N_{\wD , s}}( \phi)&=0\,, \label{eq:second_eq} 
\end{align}
where $\widehat{\mathcal{D}}_{L^2}$ and $\widehat{\mathcal{D}}_{S^2}$ are differential operators:
\begin{align}
    \begin{aligned}
        \widehat{\mathcal{D}}_{L^2}&\equiv (4-\chi^2 )\frac{\partial^2}{\partial \chi^2} - (p+1)\chi \frac{\partial}{\partial \chi}+\widehat{C}_{\wD , 0}\,, \\
        \widehat{\mathcal{D}}_{S^2}&\equiv (1-\cos^{2}\phi)\frac{\partial^2}{\partial (\cos\phi)^2}+(1-d+p)\cos\phi \frac{\partial}{\partial \cos\phi}+\widehat{C}_{0 , s}\,.
    \end{aligned}
\end{align}
We can immediately solve the second differential equation \eqref{eq:second_eq}, and its physical conformal block is derived as\footnote{In general, not only physical conformal block but also its shadow block contribute to the conformal partial wave expansion. Throughout this paper, we do not consider shadow block contributions.}
\begin{align}\label{eq:defect_block_so(d-p)}
    H_{W_{\widehat{\mathcal{O}}}}^{N_{\wD , s}}(\phi)= {}_{2}F_{1}\left(\frac{d-p+s}{2}-1, -\frac{s}{2};\frac{d-p-1}{2};\sin^{2}\phi\right)\,. 
\end{align}
In particular, when the transverse spin is integer, the above block can be expressed in terms of the Gegenbauer polynomial,\footnote{When the defect codimension is two, it is legitimate to interpret this Gegenbauer expression by taking the limit $d - p = 2 + \epsilon$ with $\epsilon \to 0$. This rule is also applied to the expression of \eqref{eq:as}. \label{footnote:codimension2}
}
\begin{align}\label{eq:gegenbauer}
    H_{W_{\widehat{\mathcal{O}}}}^{N_{\wD , s}}(\phi)=\frac{\Gamma(s+1) \Gamma(d-p-2)}{\Gamma(d-p+s-2)}C_{s}^{(\frac{d-p}{2}-1)}(\cos\phi)
\end{align}
To solve \eqref{eq:first_eq}, let us first consider the following differential equation, which applies in the ordinary DCFT case,
\begin{align}
    \widehat{\mathcal{D}}_{L^2}G_{W_{\widehat{\mathcal{O}}}}(\chi)=0\,, 
\end{align}
and its physical solution is readily computed as \cite{Billo:2016cpy}
\begin{align}\label{eq:so(1,p+1)block}
    \mathsf{G}_{W_{\widehat{\mathcal{O}}}}(\chi)=\chi^{-\wD }{}_{2}F_{1}\left(\frac{\wD +1}{2}, \frac{\wD}{2};\wD-\frac{p-2}{2};\frac{4}{\chi^2}\right)\,. 
\end{align}
In terms of this, the general solution of \eqref{eq:first_eq} can be expressed as follows:
\begin{align}\label{eq:defect_block_so(1,p+1)}
    G_{W_{\widehat{\mathcal{O}}}}^{N_{\wD , s}}(\chi)=\sum_{k=0}^{N_{\wD , s}-1}\frac{C_{k}}{k!}\partial_{\wD}^{k}\mathsf{G}_{W_{\widehat{\mathcal{O}}}}(\chi)\,,
\end{align}
where the logarithmic features are evident from the derivatives in $\widehat\Delta$.

The constants $C_k$ are to be fixed in terms of the bulk-defect two-point functions (equivalently the DOE coefficients). This can be achieved by bringing a bulk local operator close to the defect, i.e. taking $\chi\sim \infty$ while keeping $\cos\phi$ finite. In this limit, the bulk two-point function behaves as
\begin{align}\label{eq:asymp}
W_{\wD}^{N_{\wD , s}}\overset{\chi \sim \infty}{\sim}\frac{1}{|x_{1\perp}|^{\Delta_{1}-\wD}|x_{2\perp}|^{\Delta_{2}-\wD}}H_{W_{\widehat{\mathcal{O}}}}^{N_{\wD , s}}(\phi)\sum_{k=0}^{N_{\wD , s}-1}\frac{C_{k}}{k!}\frac{(-\log\chi)^{k}}{|\hat{x}_{12}|^{2\wD}}\,. 
\end{align}
On the other hand, we can evaluate the bulk two-point function using the logDOE obtained in the previous section:
\begin{align}\label{eq:asymptotic}
    W_{\wD}^{N_{\wD , s}}\overset{\chi \sim \infty}{\sim} &=\frac{a_{s}\, k_{\widehat{\mathcal{O}}}^{-1}}{|x_{1\perp}|^{\Delta_{1}-\wD}|x_{2\perp}|^{\Delta_{2}-\wD}}H_{W_{\widehat{\mathcal{O}}}}^{N}( \phi)\sum_{k=0}^{N-1}\sum_{q=1}^{N-k}\lambda_{N+1-q}\lambda_{k+q}\frac{1}{k!}\frac{(-\log\chi)^{k}}{|\hat{x}_{12}|^{2\wD}}\,, 
\end{align}
where $a_{s}$ is defined by\footnote{See footnote \ref{footnote:codimension2}.} 
\begin{align}\label{eq:as}
    a_{s}\equiv \frac{\Gamma(d-p+s-2)\Gamma(\frac{d-p}{2}-1)}{2^s \Gamma(\frac{d-p}{2}+s-1)\Gamma(d-p-2)}
\end{align}
We relegate its derivation to Appendix \ref{app:asymptotic}. By comparing these two asymptotic formulas \eqref{eq:asymp} and \eqref{eq:asymptotic}, one can relate $C_{k}$ to the bulk-defect coupling:
\begin{align}\label{eq:Ck}
    C_{k}=a_{s}\, k_{\widehat{\mathcal{O}}}^{-1}\sum_{q=1}^{N-k}\lambda_{N+1-q}\lambda_{k+q}\,. 
\end{align}
\subsubsection*{Bulk Conformal Block Expansion}
Finally, let us comment on the bulk conformal block expansion. As mentioned earlier, this discussion proceeds in essentially the same way as in an ordinary defect CFT \cite{Billo:2016cpy} since the logarithmic features of interest here are confined to the defect worldvolume. For completeness and for the reader’s convenience, we record the relevant previous results from \cite{Isachenkov:2018pef,Liendo:2019jpu} below. Following \cite{Liendo:2019jpu}, it is convenient to introduce new cross-ratios $X$ and $\overline{X}$, which are related to $\xi$ and $\phi$ via
\begin{align}\label{eq:new_cross_ratio}
    \frac{X+\overline{X}}{2(X\overline{X})^{1/2}}=\cos\phi\,, \quad \frac{(1-X)(1-\overline{X})}{(X\overline{X})^{1/2}}=\xi\,. 
\end{align}
The bulk-channel conformal block decomposition of the two-point function \eqref{eq:bulk_2pt} reads,
\begin{align}\label{eq:bulk_decomposition}
    F(X, \overline{X})=\xi^{-\frac{\Delta_1 +\Delta_2}{2}}\sum_{{\cal O}_{\Delta, \ell}\in {\cal O}_{\Delta_1}\times{\cal O}_{\Delta_2}}c({{\cal O}_{\Delta_1}, {\cal O}_{\Delta_2}, {\cal O}_{\Delta, \ell}})a_{{\cal O}_{\Delta, \ell}}f^{\text{bulk}}_{\Delta, \ell}(X, \overline{X})\,, 
\end{align}
where $c({\cal O}_{\Delta_1}, {\cal O}_{\Delta_2}, {\cal O}_{\Delta, \ell})$ is the OPE coefficient from ${\cal O}_{\Delta_1}\times{\cal O}_{\Delta_2}$ to ${\cal O}_{\Delta, \ell}$, and $a_{{\cal O}_{\Delta, \ell}}$ is the coefficient in the bulk one-point function of ${\cal O}_{\Delta, \ell}$. The explicit form of each bulk-channel conformal block $f^{\text{bulk}}_{\Delta, \ell}(X, \overline{X})$ is \cite[equation (2.7)]{Liendo:2019jpu}
\begin{align}
    f_{\Delta, \ell}^{\text{bulk}}(X,\overline{X})=f_{\Delta, \ell}^{\text{HS}}(X, \overline{X})+\frac{\Gamma(d+\ell-2)\Gamma(-\ell-\frac{d-2}{2})}{\Gamma(\ell+\frac{d-2}{2})\Gamma(-\ell)}\frac{\Gamma(\frac{\ell+d-p-1}{2})\Gamma(-\frac{\ell-1}{2})}{\Gamma(\frac{\ell+d-1}{2})\Gamma(-\frac{\ell+p-1}{2})}f_{\Delta, 2-d-\ell}^{\text{HS}}(X, \overline{X})\,,
\end{align}
where $f_{\Delta, \ell}^{\text{HS}}(X, \overline{X})$ is the Harish-Chandra function defined by,
\small
\begin{align}
\begin{aligned}
&f_{\Delta, \ell}^{\text{HS}}(X, \overline{X})
= (X\overline{X})^{\Delta_{21}/4}\sum_{n=0}^{\infty} \sum_{m=0}^{\infty}
    h_{n}(\Delta,\ell)\, h_{m}(1-\ell,1-\Delta)
    \frac{\left( \tfrac{\Delta+\ell-\Delta_{21}}{2} \right)_{n-m}}{\left( \tfrac{\Delta+\ell-1}{2} \right)_{n-m}}
    \frac{4^{n-m}}{n!\, m!}  \\
&\times  {}_{4}F_{3}\!\left(
    \begin{array}{c}
        -n,\,-m,\,\tfrac{1-\Delta_{21}}{2},\,\frac{\Delta-\Delta_{21}-\ell-d}{2}+1 \\
        -\tfrac{\Delta+\ell+\Delta_{21}}{2}+1-n,\,\tfrac{\Delta+\ell-\Delta_{21}}{2}-m,\, \tfrac{\Delta-\ell-d+3}{2}
    \end{array}
    ;\,1
\right) (1-X\overline{X})^{\ell-2m}[(1-X)(1-\overline{X})]^{\tfrac{\Delta-\ell}{2}+m+n} \\
&\times 
    {}_{2}F_{1}\!\left(
        \tfrac{\Delta+\ell+\Delta_{21}}{2}-m+n,\,
        \tfrac{\Delta+\ell}{2}-m+n;\,
        \Delta+\ell-2m+2n;\,
        1-X\overline{X}
    \right)\,,
\end{aligned}
\end{align}
\normalsize
with $\Delta_{21}\equiv \Delta_2 -\Delta_1$ and $h_{n}(\Delta, \ell)$ is given by
\begin{align}
    h_{n}(\Delta, \ell)\equiv \frac{\left(\frac{\Delta -1}{2}, \frac{\Delta -p}{2}, \frac{\Delta +\ell+\Delta_{21}}{2}\right)_{n}}{\left(\Delta-\frac{d-2}{2}, \frac{\Delta+\ell+1}{2}\right)_n}\,,
\end{align}
where $(a_1,a_2,\dots,a_m)_n\equiv (a_1)_n(a_2)_n\dots (a_m)_n$ is a shorthand for the product of Pochhammer functions.

\subsubsection*{Bootstrap Constraints}

To summarize, in the above, we have seen that the bulk two-point function \eqref{eq:bulk_2pt} can be expanded in terms of defect conformal block \eqref{eq:defect_expansion} and \eqref{eq:W}. Its precise forms are given by \eqref{eq:defect_block_so(d-p)} and \eqref{eq:defect_block_so(1,p+1)} with \eqref{eq:Ck}.

\begin{figure}[!htb]
    \centering
\begin{tikzpicture}[scale=1.2,
    operator/.style={circle,fill=black,inner sep=1.5pt},
    dashedline/.style={dashed,thick}
]

\node at (-2.5,0.8) {$\displaystyle \sum_{\hat{\Delta},s, I}$};

\draw[dashedline] (-1,0) -- (-1,1.8);
\node[operator] at (-1,1.8) {};
\node at (-1.1,2.1) {$\mathcal{O}_{\Delta_1}$};

\draw[dashedline] (1.0,0) -- (1.0,1.8);
\node[operator] at (1.0,1.8) {};
\node at (0.9,2.1) {$\mathcal{O}_{\Delta_2}$};

\draw[thick] (-1.5,0) -- (1.5,0);
\node at (0,-0.4) {$\widehat{\mathcal{O}}_{\hat{\Delta},s}^{I}$};

\node at (2,1) {$=$};

\node at (2.7, 0.8) {$\displaystyle \sum_{\Delta,\ell}$};

\draw[thick] (3.2,0) -- (6.0,0);

\coordinate (v) at (4.6,1.0);
\draw[dashedline] (4.6,0) -- (v);
\draw[dashedline] (v) -- (5.4,1.6);
\draw[dashedline] (v) -- (3.8,1.6);

\node[operator] at (5.4,1.6) {};
\node[operator] at (3.8,1.6) {};

\node at (3.6,1.9) {$\mathcal{O}_{\Delta_1}$};
\node at (5.6,1.9) {$\mathcal{O}_{\Delta_2}$};
\node at (5.1,0.5) {$\mathcal{O}_{\Delta,\ell}$};
\end{tikzpicture}
    \caption{Schematic picture of logarithmic defect bootstrap equation.}
    \label{fig:bootstrap}
\end{figure}
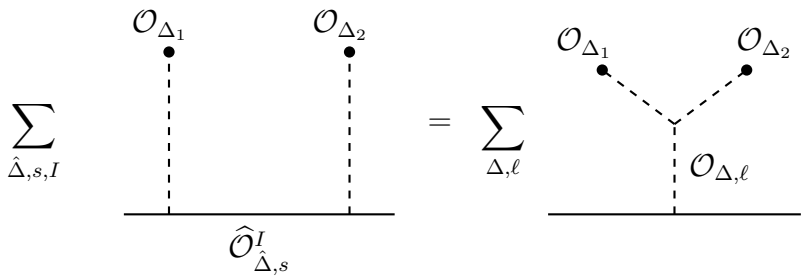

The associativity of OPE in the presence of a logarithmic defect leads to the two-point function bootstrap equation which relates the bulk and the defect channel decompositions of the correlator \cite{Cardy:1984bb,Cardy:1991tv,Lewellen:1991tb,Billo:2016cpy}, as in Figure~\ref{fig:bootstrap}. More explicitly, the logarithmic defect bootstrap equation is written as
\begin{align}\label{bootstrapeqn}
    \sum_{\wD , s}G_{W_{\widehat{\mathcal{O}}}}^{N_{\wD , s}}(\chi)H_{W_{\widehat{\mathcal{O}}}}^{N_{\wD , s}}(\phi)=\xi^{-\frac{\Delta_1 +\Delta_2}{2}}\sum_{{\cal O}_{\Delta, \ell}\in {\cal O}_{\Delta_1}\times{\cal O}_{\Delta_2}}c({{\cal O}_{\Delta_1}, {\cal O}_{\Delta_2}, {\cal O}_{\Delta, \ell}})a_{{\cal O}_{\Delta, \ell}}f^{\text{bulk}}_{\Delta, \ell}(X, \overline{X})\,. 
\end{align}
This equation ties together bulk OPE data and bulk-defect OPE data introduced above and provides a natural arena  to explore constraints on and chart the space of logarithmic defects.

Despite the absence of positivity, the bootstrap equation was analyzed in \cite{Gliozzi:2015qsa} using the determinant method developed in \cite{Gliozzi:2013ysa,Gliozzi:2014jsa}, after truncating the spectrum of exchanged operators. Further applications of this approach can be found in \cite{Gliozzi:2016cmg,Esterlis:2016psv,Li:2017ukc,Li:2017agi,Hikami:2017hwv,Hikami:2018mrf,Leclair:2018trn,Padayasi:2021sik,Niarchos:2023lot,Hu:2025yrs}, where it has been employed in both bulk and defect bootstrap settings. Since this truncation method does not rely on unitarity, it is naturally well-suited for studying systems with logarithmic features. In the present work, we have set up the basic ingredients necessary to pursue further numerical investigations of logarithmic defects along these lines.

In two dimensions, an additional powerful constraint on defects (i.e. boundaries) is provided by the Cardy condition \cite{Cardy:1989ir,Cardy:1991tv}, also known as open–closed duality for the annulus partition function. See e.g. \cite{Friedan:2012jk,Collier:2021ngi,Meineri:2025qwz} where this condition has been analyzed. Notably, the recent work \cite{Meineri:2025qwz} combines the Cardy condition with the two-point function bootstrap equation to yield a positive system in the unitary case, where semidefinite programming can be applied to derive stronger bounds on defect data. In the logarithmic setting, positivity is no longer guaranteed, but it would be interesting to explore possible generalizations of these methods to investigate logarithmic defects in 2d CFTs.

\section{More Details on Logarithmic Conformal Defect in Free Theory}\label{sec:back_to_free}
In the previous section, we have discussed the universal properties of logarithmic conformal defects, i.e. structural features that do not rely on a specific model. In this section, we circle back to our main focus on such defects obtained from subdimensional disorder and discuss how the framework developed above can be applied to the generalized pinning field model \eqref{eq:free_theory2} with random fluctuations. As anticipated in Section~\ref{subsec:log_behaviour}, several defect local operators are expected to behave as logarithmic primaries. A natural question then arises: what is the rank of the corresponding logarithmic multiplets? To address this, in Section~\ref{subsec:multiplet}, we analyze in detail how various primary operators, including the displacement operator and the tilt operator, organize themselves into logarithmic multiplets, sometimes making use of the bulk equations of motion. In Section~\ref{subsec:b_d_data}, we also clarify how the bulk-defect OPE data are determined in this setting. In Section~\ref{subsec:bulk2pt} we present the bulk scalar two-point function together with its explicit conformal block decomposition, emphasizing both the similarities and the differences with respect to the disorder-free case. Finally in Section~\ref{subsec:bulk1pf}, we determine the defect one-point functions for bulk operators in the OPE of the scalar and particularly find that the stress energy one-point function vanishes for this non-topological defect!

\subsection{Zoo of Logarithmic Multiplets in Defect Spectrum}\label{subsec:multiplet}
In this subsection, we examine the spectrum of defect operators generated by subdimensional disorder in the free scalar theory. 
As we will see, the spectrum exhibits a variety of logarithmic structures, which can be directly compared 
with the general considerations developed in the previous section. 
This allows us to pinpoint how the universal features of logarithmic conformal defects are realized 
in the concrete setting of subdimensional disorder given by the generalized pinning field model with randomly fluctuating coupling.
\subsubsection*{Defect operators with $(\wD , s)=(p/2 , 0)$}
We firstly explore the structures of defect operators which are logDOE channels of the bulk scalar field $\sigma$. The logDOE of the bulk operator $\sigma$ is given by\footnote{We omit the descendant contributions, which do not matter in the discussions below.}
\begin{align}
    {\sigma}(x)\supset \sum_{I=1}^{N_{\widehat{\Delta}, s}}\sum_{n=0}^{I-1}\alpha_{I-n}\frac{x_{\perp}^{i_{1}}x_{\perp}^{i_{2}}\cdots x_{\perp}^{i_{s}}}{|x_{\perp}|^{\Delta_{\sigma} - \widehat{\Delta}+s}}\frac{(\log|x_{\perp}|)^{n}}{n!}\,\widehat{\mathcal{O}}^{I}_{\widehat{\Delta}, i_{1}i_{2}\cdots i_s}(\hat{x})\,,
\end{align}
where $\Delta_{\sigma}=(d-2)/2=p/2$ and $\alpha_{I-n}\equiv A^{N_{\widehat{\Delta}, s}}_{I-n , 0}$.
We can extract possible defect channels of $\sigma$ by using the equation of motion \cite{Liendo:2012hy, Billo:2016cpy}. By acting the $d$-dimensional d'lambertian on both hand sides, one can obtain
\begin{align}
    \Box{\sigma}(x)\supset \sum_{\widehat{\Delta}, s}\sum_{I=1}^{N_{\widehat{\Delta}, s}}\sum_{n=0}^{I-1}C_{I-n}\frac{x_{\perp}^{i_{1}}x_{\perp}^{i_{2}}\cdots x_{\perp}^{i_{s}}}{|x_{\perp}|^{\Delta_{\sigma} - \widehat{\Delta}+s}}\frac{(\log|x_{\perp}|)^{n}}{n!}\,\widehat{\mathcal{O}}^{I}_{\widehat{\Delta}, i_{1}i_{2}\cdots i_s}(\hat{x})\,,
\end{align}
where $C_{J}$ $(1\leq J \leq N_{\wD , s})$ is given by
\begin{align}
    C_{J}\equiv\alpha_{J-2}+2(\widehat{\Delta}-\Delta_{\sigma})\alpha_{J-1}+(\widehat{\Delta}-\Delta_{\sigma}+s)(\widehat{\Delta}-\Delta_{\sigma}-s)\alpha_{J}\,. 
\end{align}
The bulk equation of motion then implies that all coefficients $C_{J}$ must vanish:
\begin{align}\label{eq:eom_constraint}
    C_{J}=0\,, \quad 1\leq J \leq N_{\wD , s}\,. 
\end{align}
To make the subsequent statements more precise, let us denote by $\mathcal{V}_{\sigma}$ the module of defect local operators that appear in the DOE of $\sigma$. 
In general, this space can be expressed as
\begin{align}
    \mathcal{V}_{\sigma}=\bigoplus_{\wD , s} \bigoplus_{N_{\wD , s}}\mathcal{V}_{\sigma}^{N_{\wD , s}}\,, 
\end{align}
where $\mathcal{V}_{\sigma}^{N_{\wD , s}}$ is the module associated to the defect primary operator with conformal dimension $\wD$, SO$(d-p)$ spin $s$ and rank $N_{\wD , s}$. In the following, we summarize the results that follow from the bulk equation of motion.
\begin{itemize}
    \item There are no defect logarithmic multiplets in $
    \cV_\sigma$ with non-trivial SO$(d-p)$ spin:\footnote{The displacement multiplet is logarithmic and has rank 2 but does not appear in $\cV_\sigma$. This is consistent with the fact that $\sigma$ has vanishing one-point function with the defect due to the restored global symmetry after disorder averaging.}
    \begin{align}\label{eq:spin_defect}
        \mathcal{V}_{\sigma}^{N_{\wD , s}}=\emptyset \,, \quad N_{\wD , s} \geq 2 \,, \quad s\geq 1\,.   
    \end{align}
    \item Logarithmic scalar multiplets with conformal dimension $p/2$ and rank two appears in $\cV_\sigma$. Moreover, there are no logarithmic scalar multiplets  of higher rank $N_{\wD , 0}>2$ in $\cV_\sigma$:
    \begin{align}
        \mathcal{V}_{\sigma}^{N_{p/2 , 0}=2}\not=\emptyset  \label{eq:scalar_possible}\,,\quad \mathcal{V}_{\sigma}^{N_{\wD , 0}>2} =\emptyset\,. 
    \end{align}
\end{itemize}
The proof of these statements is somewhat technical and is therefore deferred to the Appendix~\ref{app:eom_constraint}. 
Nonetheless, it is clear that the bulk equation of motion imposes non-trivial and strong constraints on the possible defect multiplets. 

We now turn to the task of identifying the defect local operators that belong to the non-trivial logarithmic multiplet 
$\mathcal{V}_{\sigma}^{N_{p/2 , 0}=2}$. The most immediate candidate for a defect local operator with conformal dimension $p/2$ is simply the restriction of the bulk field $\sigma$ to the defect
\begin{align}
    \hat{\sigma}(\hat{x})\equiv\lim_{x_\perp \to 0}\sigma(x)\,.
\end{align}
As we have seen in Section~\ref{subsec:log_behaviour}, the defect two-point function of $\sigma$ exhibits a logarithmic behavior with exponent one,
\begin{align}\label{eq:sigma_sigma_defect}
    \overline{\langle\, \hat{\sigma}(\hat{x})\hat{\sigma}(0)\, \rangle}&=\frac{C_{d}}{|\hat{x}|^{p}}+\frac{C_{d} c^{2}}{2\pi}\frac{\log\mu^{2}|\hat{x}|^{2}}{|\hat{x}|^{p}}\,.
\end{align}
This indicates that $\hat{\sigma}$ does not form an ordinary conformal multiplet but instead belongs to a logarithmic multiplet of rank two. The natural question is then what operator serves as the logarithmic partner of $\hat{\sigma}$. The answer comes from the defect degrees of freedom themselves: the randomly fluctuating local magnetic field $\hat{h}$ precisely plays this role. Indeed, by computing the corresponding defect correlators (at separated points) one finds that they exhibit the structure characteristic of a logarithmic pair as 
\begin{align}\label{eq:sigma_h_defect}
    \overline{\langle\, \hat{\sigma}(\hat{x})\h(0)\, \rangle}=-\frac{C_{d} c^{2}}{|\hat{x}|^{p}}\,, \quad
    \overline{\langle\, \h(\hat{x})\h(0)\, \rangle}=0 \,, \quad (\hat{x}\not= 0)\,. 
\end{align}
thereby confirming that $\hat{\sigma}$ and $\hat{h}$ together generate the non-trivial multiplet $\mathcal{V}_{\sigma}^{N_{p/2,0}=2}$.

Finally, to move to the basis described in Section~\ref{subsec:bulk_two_pt}, 
we perform the following field redefinition:
\begin{align}
    \hat{\sigma}'=\hat{\sigma}+\frac{\h}{2c^2}\,, \quad \h'=\frac{\h}{2\pi}\,.
\end{align}
This brings the two-point functions given above to the form \eqref{eq:specific_basis}:
\begin{align}\label{eq:basis_sigma}
    \overline{\langle\, \hat{\sigma}'(\hat{x})\hat{\sigma}'(0)\, \rangle}=-k_{\{\hat{\sigma}, \h \}}\frac{\log\mu^{2}|\hat{x}|^{2}}{|\hat{x}|^{p}}\,, \quad \overline{\langle\, \hat{\sigma}'(\hat{x})\h'(0)\, \rangle}=\frac{k_{\{\hat{\sigma}, \h \}}}{|\hat{x}|^{p}}\,, \quad \overline{\langle\, \h'(\hat{x})\h'(0)\, \rangle}=0 \, ,
\end{align}
where the coefficient $k_{\{\hat{\sigma}, \h \}}$ is given by
\begin{align}\label{eq:k}
k_{\{\hat{\sigma}, \h \}}=-\frac{C_{d}\,c^2}{2\pi}\,. 
\end{align}
This change of basis will be convenient in the analysis of bulk two-point function in Section~\ref{subsec:bulk2pt}.

\subsubsection*{Defect operators with $(\wD , s)=(np/2 , 0)$}
As a mild generalization of the previous example, we consider defect multiplets with conformal dimension $\widehat\Delta = n p/2$ and transverse spin $s=0$. These provide the simplest nontrivial examples of higher-rank logarithmic multiplets. Clearly, the composite field $\hat{\sigma}^n$ belongs to such a multiplet. Evaluating its defect two-point function, one finds the characteristic logarithmic structure, with the highest exponent of the logarithm equal to $n$:
\begin{align}
    \overline{\langle\, \hat{\sigma}^{n}(\hat{x})\hat{\sigma}^{n}(0)\, \rangle}&=\#\, c^{2n}\frac{(\log|\hat x|)^{n}}{|\hat{x}|^{np}}+\cdots 
\end{align}
For the purposes of this discussion, we omit the explicit numerical coefficients, as they are not essential for the main conceptual point. The crucial observation is that the appearance of $(\log|\hat x|)^n$ indicates that $\hat\sigma^n$ forms a rank-$(n+1)$ logarithmic multiplet. The logarithmic primary operators spanning this multiplet are clearly the composites $\hat\sigma^m \h^{n-m}$ with $0\leq m\leq n$.

\subsubsection*{Displacement Operator Multiplets}
A central and universal ingredient of any conformal defect (non-topological) is the displacement operator. This operator arises because the presence of the defect explicitly breaks the translational invariance of the bulk theory in the directions transverse to the defect. As a result, the Ward-Takahashi identity associated with transverse translations is not preserved in its usual form, but instead acquires an insertion localized on the defect. The operator that generates this insertion is precisely the displacement operator. To be more concrete, we recall that the (improved) energy-momentum tensor for the free theory is given by
    \begin{align}\label{eq:emtensor}
T^{\mu\nu}=\partial^{\mu}\sigma\partial^{\nu}\sigma-\frac{1}{2}\delta^{\mu\nu}\partial^{\rho}\sigma\partial_{\rho}\sigma-\frac{d-2}{4(d-1)}(\partial^{\mu}\partial^{\nu}-\delta^{\mu\nu}\Box)\sigma^{2}\,. 
    \end{align}
By using the equation of motion
\begin{align}
    \Box \sigma(x)=\h(\hat{x})\delta^{(d-p)}(x_{\perp})\,, 
\end{align}
the conservation law of the energy-momentum tensor becomes
\begin{align}\label{eq:dTmunu}
    \partial_{\mu}T^{\mu i}(x)={\h}(\hat{x})\partial^{i}\sigma(x)\delta^{(d-p)}(x_{\perp})\,. 
\end{align}
From this, we can read off the displacement operator $D_{i}$ as
\begin{align}
    D_{i}(\hat{x})\equiv \h(\hat{x})\partial_{i}\hat{\sigma}\,. 
    \label{displace}
\end{align}
The displacement operator transforms as a vector under the transverse rotation group SO$(d-p)$, and it always has fixed conformal dimension $p+1$ from \eqref{eq:dTmunu}. Interestingly, in our disordered system, the displacement operator inherits direct contributions from the randomly fluctuating source $\h(\hat{x})$. This naturally raises the question of whether this universal defect local operator actually organizes into a logarithmic multiplet. To answer this, one should examine the corresponding defect two-point functions, which as we will see, have features that differ from those of conventional defects in curious ways.

A first observation is that the two-point function of the displacement operator vanishes identically at separated points,
\begin{align}\label{eq:displacement_two_pt}
     \overline{\langle\, D_{i}(\hat{x}) D_{j}(0)\, \rangle}=0\,, \quad \hat{x}\not=0\,.  
\end{align}
This signals that, due to the effect of disorder averaging, the displacement operator does not behave as an ordinary CFT primary operator. Instead, it is expected to participate in a logarithmic multiplet. A natural candidate for its logarithmic partner is $\widetilde{D}_{i}(\hat{x}) \equiv \hat{\sigma}(\hat{x}) \partial_{i}\hat{\sigma}(\hat{x})$
which carries the same conformal dimension and SO$(d-p)$ spin as the displacement operator. (This defect local operator has already shown up in Section~\ref{subsec:log_behaviour}.) Computing the relevant correlation functions, one finds
\begin{align}
    \overline{\langle\, D_{i}(\hat{x}) \widetilde{D}_{j}(0)\, \rangle}&=-\frac{p\,C_{d}^{2}\, {c}^{2}\delta_{ij}}{|\hat{x}|^{2p+2}} \,, \\
    \overline{\langle\, \widetilde{D}_{i}(\hat{x}) \widetilde{D}_{j}(0)\, \rangle}&=p\,C_{d}^{2}\delta_{ij}\left(\frac{1}{|\hat{x}|^{2p+2}}+\frac{C_{d}c^2}{2\pi}\frac{\log\mu^{2}|\hat{x}|^{2}}{|\hat{x}|^{2p+2}}\right)\,, 
\end{align}
where the second correlator has already been computed in Section~\ref{subsec:log_behaviour}. These results, together with \eqref{eq:displacement_two_pt} and the general discussion in Section~\ref{sec:systematics}, demonstrate that the displacement operator $D_{i}$ and its partner $\widetilde{D}_{i}$ organize into a rank-2 logarithmic multiplet.

Although in this paper we focus on the free theory as an explicit example, the logarithmic nature of the displacement operator is expected to persist even in interacting theories, as long as the defect is generated by subdimensional disorder.  For instance, it has been shown that in two-dimensional (interacting) CFTs deformed by disorder over the entire space, the energy-momentum tensor acquires the logarithmic structure \cite{cardy2001stresstensorquenchedrandom, Cardy:2013rqg}. More explicitly, in the replica formulation, one considers an $n$-fold replicated theory with $n \in \mathbb{Z}_{+}$. For finite $n$, the replicated theory is expected to be a local CFT, and the two-point function of the energy-momentum tensor exhibits the standard power-law behavior dictated by conformal symmetry. 
However, the quenched disorder average is obtained by analytically continuing to the replica limit $n \to 0$, in which the two-point function of the energy-momentum tensor is known to develop logarithmic dependence \cite{cardy2001stresstensorquenchedrandom, Cardy:2013rqg}. This signals the emergence of a logarithmic partner with the same scaling dimension, such that the energy-momentum tensor forms a rank-2 logarithmic multiplet.
As discussed in Section~\ref{subsec:line_fixed_point}, the replica trick can also be applied to subdimensional disorder, and it is therefore natural to expect that the displacement operator develops a similar logarithmic structure in the replica limit. 

\subsubsection*{Tilt Operator Multiplets}
Whenever a bulk continuous global symmetry is spontaneously broken on the defect, there exist universal defect local operators associated with the broken charges. To introduce the tilt operator explicitly in a example, we consider a natural generalization of the disordered single scalar theory to an $\text{O}(N)$ free scalar theory coupled to a defect via a generalized pinning field:
\begin{align}
S = \frac{1}{2}\int_{\mathbb{R}^{d}} \mathrm{d}^d x (\partial \vec{\sigma})^2 + \int_{\mathbb{R}^{p}} \mathrm{d}^p \hat{x}\h(\hat{x})\vec{n} \cdot \vec{\sigma}(\hat{x})\,,
\end{align}
where $\vec{n}$ is a fixed vector that selects a direction in the internal space, breaking the bulk $\text{O}(N)$ symmetry down to $\text{O}(N-1)$ around $\vec{n}$ on the defect. To achieve defect conformality, we again set the bulk dimension as $d=p+2$. The bulk Noether current  
\begin{align}
J_{\alpha\beta}^{\mu}=\sigma_{\alpha}\partial^{\mu}\sigma_{\beta}-\partial^{\mu}\sigma_{\alpha}\, \sigma_{\beta}\,. 
\end{align}
satisfies the modified conservation equation,
\begin{align}
    \partial_{\mu}J_{\alpha\beta}^{\mu}=\h(\hat{x})(n_{\beta}\sigma_{\alpha}-n_{\alpha}\sigma_{\beta})\delta^{(d-p)}(x_{\perp})\,,
\end{align}
and this defines the tilt operator $t_{\alpha\beta}$ as
\begin{align}
    t_{\alpha\beta}(\hat{x})\equiv \h(\hat{x})(n_{\beta}\hat{\sigma}_{\alpha}(\hat{x})-n_{\alpha}\hat{\sigma}_{\beta}(\hat{x}))\,. 
\end{align}
In the following, we choose the specific vector $\vec{n}=(0, 0, \cdots, 0, 1)$ without loss of generality. Then, the above tilt operators can be written as
\begin{align}
    t_{\hat{\alpha}}(\hat{x})=\h(\hat{x})\hat{\sigma}_{\hat{\alpha}}(\hat{x})\,, \quad \hat{\alpha}=1, \cdots, N-1\,. 
\end{align}
Note that from the conservation equation, the conformal dimension of tilt operators is fixed to be $p$.
In a similar way to the displacement operator, one may then ask whether these tilt operators also participate in logarithmic multiplets. To address this question, we again study the defect two-point function, and the result is clearly
\begin{align}
    \overline{\langle\, t_{\hat{\alpha}}(\hat{x})t_{\hat{\beta}}(0)\, \rangle}=0\,, \quad \hat{x}\not=0\,, 
\end{align}
which implies again that the tilt operator is a nontrivial logarithmic primary. Its logarithmic partner must share the same quantum numbers, namely conformal dimension $p$ and transformation as a vector under the residual O($N-1$) symmetry. The natural candidate is $\tilde{t}_{\alpha}=\hat{\sigma}_{N}\hat{\sigma}_{\alpha}$, and its relevant correlation functions can be easily computed
\begin{align}
    \begin{aligned}
        \overline{\langle\, t_{\hat{\alpha}}(\hat{x})\tilde{t}_{\hat{\beta}}(0)\, \rangle}&=-\frac{C_{d}^{2}c^{2}\delta_{\hat{\alpha}\hat{\beta}}}{|\hat{x}|^{2p}}\,, \\
        \overline{\langle\, \tilde{t}_{\hat{\alpha}}(\hat{x})\tilde{t}_{\hat{\beta}}(0)\, \rangle}&=C_{d}^{2}\delta_{\hat{\alpha}\hat{\beta}}\left(\frac{1}{|\hat{x}|^{2p}}+\frac{C_d c^2}{2\pi}\frac{\log\mu^2 |\hat{x}|^{2}}{|\hat{x}|^{2p}}\right)\,.
    \end{aligned}
\end{align}
These results indicate that, after disorder averaging, the tilt operator $t_{\hat{\alpha}}$ belongs to a rank-2 logarithmic multiplet together with its partner $\tilde{t}_{\hat{\alpha}}$. 

Despite the simplicity of the explicit example presented above, we expect that these special defect multiplets with protected conformal dimensions to become logarithmic in general interacting CFTs with subdimensional disorder. It would be interesting to understand whether the rank-two property of such multiplets found here continue to hold in the interacting examples.

\subsubsection*{Are there ordinary CFT primaries?}
One may wonder whether, after disorder averaging, there still exist ordinary CFT primaries on the defect, namely rank-1 operators. Indeed, such multiplets do exist. They are given by defect local operators with nonzero transverse spin $s$, which take the following form:\footnote{It turns out that only the primaries with transverse spin $s\geq 1$ show up in the defect operator spectrum. See Section~\ref{subsec:bulk2pt} and remark at the end. \label{footnote:non_unitary}}
\begin{align}
    \Psi_{i_{1}, i_{2}, \cdots , i_{s}}(\hat{x})\equiv \partial_{(i_{1}}\cdots \partial_{i_{s})}\hat{\sigma}(\hat{x})\,.
\end{align}
Its two-point function takes the following form:
\begin{align}
    \overline{\langle\, \Psi_{i_{1}, i_{2}, \cdots , i_{s}}(\hat{x})\Psi^{j_{1}, j_{2}, \cdots , j_{s}}(0)\, \rangle}=\frac{2^{s}s!\left(\frac{p}{2}\right)_{\hspace{-1mm}s}C_d}{|\hat{x}|^{p+2s}}\, \mathcal{I}^{j_{1}\cdots j_{s}}_{i_1 \cdots i_s}\,. 
\end{align}
Interestingly, these two-point functions show no dependence on the disorder strength $c$ (i.e. same as without the defect). 
This indicates that the disorder averaging procedure leaves untouched sectors with nonzero transverse spin quantum numbers. This observation is consistent with our earlier conclusion \eqref{eq:spin_defect} from the DOE: channels with nonzero transverse spin can only accommodate ordinary CFT primaries, rather than logarithmic ones. 
This is suggestive of a general feature for subdimensional disorder with scalar operators. We leave the investigation of this feature and generalization to disorder terms involving spinning operators to future studies.

\subsection{Bulk-defect OPE data from Subdimensional Disorder}\label{subsec:b_d_data}
In the previous section, we have seen how various defect local operators in our free theory model organize themselves into logarithmic multiplets. In a defect CFT, besides the usual conformal dimensions and spins, the DOE coefficients $A_{J, 0}^{N_{\wD , s}}$ constitute another set of essential CFT data that characterize the defect. In this section, we focus on explicitly computing the bulk-defect OPE coefficients in our model, with the goal of gaining a deeper understanding of the structure of logarithmic defects.

To derive the DOE coefficients, we firstly need to calculate the bulk-defect two-point function, and obtain the structure constants $\lambda_{I}$ in \eqref{eq:ci}. Then, we can translate this into DOE coefficients $A_{J, 0}^{N_{\wD , s}}$ via the relation \eqref{eq:primary_constraint} with the coefficients $k_{I}$ of defect two-point functions. In what follows, we illustrate how this program works by focusing on the bulk operator $\sigma$. From the discussion in the previous section, we know that the defect local primary operators appearing in the DOE of the bulk field $\sigma$ are precisely the following set:\footnote{See also footnote \ref{footnote:non_unitary}.}
\begin{align}
    \hat{\sigma}(\hat{x}) \,, \quad \h(\hat{x})\,, \quad \Psi_{i_{1}i_{2}\cdots i_{s}}(\hat{x})\,. 
\end{align}
Therefore, to determine the DOE coefficients $A_{J,0}^{N_{\wD,s}}$ associated with $\sigma$, it suffices to compute the bulk-defect two-point functions between $\sigma$ and these defect local operators. In the following, we carry out this analysis in order. 

We firstly analyze the two-point function of $\sigma(x)$ and $\hat{\sigma}(\hat{y})$:
\begin{align}
    \begin{aligned}
    \overline{\langle\sigma(x)\hat{\sigma}(0)\rangle}&=\frac{C_{d}}{|x|^p}+c^2 C_{d}^2 \int \d^{p}\hat{z}\frac{1}{|x-\hat{z}|^p |\hat{z}|^p} \, .
    \end{aligned}
\end{align}
Using the Fourier transformation \eqref{FT} and \eqref{FTmix}, we have
\ie 
\int \d^{p}\hat{z}\frac{1}{|x-\hat{z}|^p |\hat{z}|^p}=
\frac{\pi ^{p/2} \left(\psi \left(\frac{p}{2}\right)-\gamma_{\rm E} +\log  4\right)}{\Gamma \left(\frac{p}{2}\right)|x|^p}
-\frac{4\pi^{p+1}}{\Gamma\left(\frac{p}{2}\right)^2}\mathcal{I}_{1}\,, 
\fe
where $\gamma_{\rm E}$ is the Euler–Mascheroni constant and $\mathcal{I}_{1}$ is the following
\begin{align}\label{eq:I1}
    \begin{aligned}
    \mathcal{I}_{1}=\int \frac{\d^p \hat{p}}{(2\pi)^{p+1}}K_{0}(|\hat{p}||x_{\perp}|)\log|\hat{p}|^{2}e^{\i\hat{p}\cdot \hat x}=\frac{\Gamma(\frac{p}{2})}{2\pi^{\frac{p}{2}+1}|x|^{p}}\left(\tau+\log2+\psi\left(\frac{p}{2}\right)\right)\, . 
    \end{aligned}
\end{align}
Here, $K_{0}$ is the modified Bessel function of the second kind, and $\tau$ is defined in \eqref{eq:tau}. We give the detailed derivation of this integral formula in Appendix \ref{app:integral}. Then, the two-point function is simplified to the following desired form:
\begin{align}\label{eq:sigma_sigma}
    \overline{\langle\sigma(x)\hat{\sigma}(0)\rangle}&=\frac{\widetilde{\lambda}_1+\widetilde{\lambda}_2 \tau}{|x|^p}\,,
\end{align}
where $\widetilde{\lambda}_{1}$ and $\widetilde{\lambda}_{2}$ are given by 
\ie 
\tilde\lambda_1=C_d\left(1-{ c^2\over 4\pi } H_{\frac{p}{2}-1} \right)\,,\quad 
\tilde \lambda_2=-C_d  { c^2\over 2\pi }\,,
\fe
where $H_z\equiv \psi(z+1)+\gamma_{\rm E}$ is the harmonic number continued to the complex plane and $C_d$ is defined in \eqref{Cddef}. 
On the other hand, the two-point function of $\sigma$ and $\h$ can be easily evaluated:
\begin{align}\label{eq:sigma_h}
    \overline{\langle\sigma(x){\h}(0)\rangle}=-\frac{c^{2}C_{d}}{|x|^{p}}\, . 
\end{align}
At this stage, it is convenient to switch to a specific basis \eqref{eq:basis_sigma} in which the two-point functions take a simpler form. 
In this basis, the coefficients $\lambda_1$ and $\lambda_2$ appearing in bulk-defect two-point functions are given by
\begin{align}\label{eq:lambda_sigma}
\lambda_{1}=\frac{C_{d}}{2}\left(1-\frac{c^2}{2\pi}H_{\frac{p}{2}-1}\right)\,, \quad \lambda_{2}=-\frac{C_{d}\, c^2}{2\pi}=k_{\{\hat{\sigma}, \h \}}\,.
\end{align}
Then, the relation \eqref{eq:primary_constraint} and the defect two-point functions \eqref{eq:basis_sigma} can translate $\lambda_{I}$ into the OE coefficients $A_{J, 0}^{N_{p/2 , 0}}$ (see Section~\ref{subsec:logdefect_OPE} and \eqref{eq:simple}):
\begin{align}
    A_{1, 0}^{N_{p/2, 0}}=1\,, \quad A_{2, 0}^{N_{p/2, 0}}=-\frac{\pi}{c^2}+\frac{1}{2}H_{\frac{p}{2}-1}\,. 
\end{align}
Recall that the first subscript $J$ of
$A_{J,\ell}^{N_{\widehat{\Delta}}}$ labels the primary operators in the logarithmic multiplet that enter in the DOE of rank $N_{\widehat{\Delta}}$ and $\ell$ indexes the descendants (with $\ell=0$ corresponding to the primaries).

In contrast, for the defect spinning operator $\Psi_{i_1\cdots i_s}$ discussed in the previous section, the situation is simpler: since it behaves as an ordinary CFT primary, the non-vanishing DOE coefficients are
\ie 
A^{N_{s+p/2, s}}_{1,0}={1\over s!}\,.
\fe 
We have thus completely fixed the logDOE coefficients of the bulk operator $\sigma$ in the presence of subdimensional disorder.

\subsection{Bulk Two-point Function and Deformation from Disorder}\label{subsec:bulk2pt}
Finally, we analyze the bulk two-point function $\sigma$ to illustrate the general discussion in Section~\ref{subsec:bulk_two_pt}.
    \begin{align}\label{eq:twopt_bulk_another}
    \begin{aligned}
        \overline{\langle\, \sigma(x) \sigma(y)\, \rangle}&=\frac{C_{d}}{|x-y|^{p}}+{c}^{2}C_{d}^{2}\int \d^{p}\hat{z}\,\frac{1}{|x-\hat{z}|^{p}|y-\hat{z}|^{p}}\\
        &=\frac{C_{d}}{|x-y|^{p}}+\frac{16\pi^{p+2}}{\Gamma(p/2)^2}{c}^{2}C_{d}^{2}\,{\cal I}_{2}\,, 
    \end{aligned}
\end{align}
where ${\cal I}_{2}$ is the momentum-space integral defined by
\begin{align}\label{eq:integral_I2}
    \begin{aligned}
    {\cal I}_{2}=\int\frac{\d^{p}\hat{p}}{(2\pi)^{p+2}}K_{0}(|\hat{p}||x_{\perp}|)K_{0}(|\hat{p}||y_{\perp}|)e^{\i \hat{p}\cdot (\hat{x}-\hat{y})}\,, 
    \end{aligned}
\end{align}
which follows from the Fourier transformation \eqref{FTmix}. In the following, we will unpack the conformal block decomposition of \eqref{eq:twopt_bulk_another} which includes logarithmic contributions.

The first term on the RHS of \eqref{eq:twopt_bulk_another} is obviously the two-point function with a trivial defect and admits the following conformal block decomposition (as in ordinary defect CFT),
\ie 
  \frac{1}{|x-y|^{p}}=&\,\frac{1}{|x_{\perp}|^{p/2}|y_{\perp}|^{p/2}}\left[\chi^{-\frac{p}{2}}{}_2 F_1 \left(\frac{p}{4}, \frac{p+2}{4}; 1; \frac{4}{\chi^{2}}\right)\right.
  \\
&\left.+\sum_{s=1}^{\infty}\frac{2(\frac{p}{2})_{s}}{s!}\cos(s\phi){}_2 F_1 \left(\frac{p+2s}{4}, \frac{p+2s+2}{4}; s+1; \frac{4}{\chi^{2}}\right)\right]\,,
  \fe 
where $\chi$ is the cross-ratio in \eqref{eq:W}.
   Compared with the general discussion in Section~{\ref{subsec:bulk_two_pt}} for logarithmic defects, the first term in the square bracket corresponds to (a part of) the defect conformal block with $(\widehat{\Delta}, s) = (p/2, 0)$, while the second term corresponds to the spinning defect conformal block with $(\widehat{\Delta}, s) = (p/2+s, s)$ with $s\geq 1$.  
Using the notation of \eqref{eq:defect_block_so(d-p)} and \eqref{eq:so(1,p+1)block}, this tree-level part of \eqref{eq:twopt_bulk_another} can be rewritten as
\begin{align}\label{eq:tree}
    \frac{C_{d}}{|x-y|^{p}}=\frac{1}{|x_{\perp}|^{p/2}|y_{\perp}|^{p/2}}\left(C_{d}\mathsf{G}_{\{\hat{\sigma}, \h\}}(\chi)+\sum_{s=1}^{\infty}H_{\Psi_{s}}^{N_{p/2+s, s}=1}(\phi)G_{\Psi_{s}}^{N_{p/2+s, s}=1}(\chi) \right)\,. 
\end{align}

Next, we turn to the analysis of the second term on the RHS of \eqref{eq:twopt_bulk_another}, which encodes the effects of the subdimensional disorder. To proceed, one needs to evaluate the integral in \eqref{eq:integral_I2}. We defer the details to the Appendix~\ref{app:integral_Is} and only present the final result here,
\begin{align}\label{eq:formula}
    {\cal I}_2 =-\frac{\Gamma(p/2)}{8\pi^{\frac{p}{2}+2}}\frac{1}{|x_{\perp}|^{p/2}|y_{\perp}|^{p/2}}\left[H_{\frac{p}{2}-1}\mathsf{G}_{\wD}(\chi)+\left.\partial_{\wD}\mathsf{G}_{\wD}(\chi)\right]\right|_{\wD=p/2}\,, 
\end{align}
where $\mathsf{G}_{\wD}(\chi)$ is given by \eqref{eq:so(1,p+1)block}. Combined this with the tree-level part \eqref{eq:tree}, the full bulk two-point function can be written as,
\begin{align}\label{free2pfdefexp}
    \overline{\langle\, \sigma(x) \sigma(y)\, \rangle}=\frac{1}{|x_{\perp}|^{p/2}|y_{\perp}|^{p/2}}\left(G_{\{\hat{\sigma}, \h\}}(\chi)+\sum_{s=1}^{\infty}H_{\Psi_{s}}^{N_{p/2+s, s}=1}(\phi)G_{\Psi_{s}}^{N_{p/2+s, s}=1}(\chi) \right)\,, 
\end{align}
where we used \eqref{eq:defect_block_so(1,p+1)}, \eqref{eq:Ck} and \eqref{eq:lambda_sigma}. As desired, the resulting expression matches the structure of the defect conformal block expansion developed in Section~\ref{subsec:bulk_two_pt}. This expression also makes it clear that the disorder dependence (one-parameter family parametrized by the disorder strength $c$) is concentrated in the scalar defect 
channel by comparing to \eqref{subsec:bulk2pt}.

Here we make a brief comment regarding the spectrum of defect spinning operators found from the above analysis. In Section~\ref{subsec:multiplet}, when we impose constraints on the defect operator spectrum associated with bulk operator~$\sigma$ using the equation of motion, the spinning operators with $\wD=p/2-|s|$ are a priori not excluded. This is because unitarity is lost in general on the disordered defect. However, the present analysis of the bulk two-point function makes it clear that such operators do not actually appear in the spectrum. This observation is consistent with the remark made at the end of Section~\ref{subsec:multiplet}, namely that the subdimensional disorder (by a local scalar deformation) does not affect the operators with nonzero transverse spin quantum numbers.

\subsection{Bulk One-point Functions and Crossing}
\label{subsec:bulk1pf}
In the previous section, we have explicitly evaluated the bulk two-point function and verified that it takes the form consistent with the defect conformal block decomposition. In this section, we revisit the same two-point function from the viewpoint of the bulk conformal block expansion (see around \eqref{bootstrapeqn}).

The bulk OPE of two identical free scalar operators $\sigma$ takes the schematic form
\begin{align}\label{bulkOPE}
    \sigma \times \sigma\to \mathbf{1}+\sigma^{2}+\sum_{\ell=1}^{\infty}\Phi_{\mu_1\cdots \mu_{2\ell}}\,,
\end{align}
where $\Phi_{\mu_1\cdots \mu_{2\ell}}(y)$ is bulk local primary operator with even spin $2\ell$ ($\ell\in\mathbb{Z}_{+}$),
\begin{align}
    \Phi_{\mu_1\cdots \mu_{2\ell}}\equiv \sum_{k=0}^{2\ell}c_{2\ell, k}\,\partial_{\mu_1}\cdots \partial_{\mu_k}\sigma\cdot \partial_{\mu_{k+1}}\cdots \partial_{\mu_{2\ell}}\sigma\,.
\end{align}
with coefficients given in \cite{Giombi:2016hkj},
\begin{align}
    c_{2\ell, k}=\frac{4\sqrt{\pi}\Gamma(\frac{p}{2}+2\ell)\Gamma(2\ell+p-1)}{\Gamma(\frac{p-1}{2})}\frac{(-1)^k}{k!\, (2\ell-k)!\, \Gamma(k+\frac{p}{2})\Gamma(2\ell-k+\frac{p}{2})}\,. 
\label{csk}\end{align}
They correspond to the higher even-spin conserved currents in the free theory.

In the following, we will show that only the identity operator and the scalar operator $\sigma^2$ in \eqref{bulkOPE} contribute in the bulk-channel, by directly evaluating the relevant one-point functions in the presence of the subdimensional disorder by Wick contractions and taking into account normal ordering. We will find the result to be consistent with the other form \eqref{eq:twopt_bulk_another}  of the two-point function derived before. 

For the scalar $\sigma^2$, one finds,
\ie  
    \overline{\langle \sigma^2(x)\rangle}=\frac{\pi^{\frac{p}{2}}\Gamma(\frac{p}{2})}{\Gamma(p)}C^{2}_{d}c^2\frac{1}{|x_{\perp}|^{p}}\,. 
\fe 
For the higher spin currents $\Phi_{\m_1\cdots \m_{2\ell}}$, it is convenient to introduce a null polarization vector $\zeta$ (i.e. $\zeta\cdot\zeta=0$) and define,
\begin{align}
    \Phi_{2\ell}(x, \zeta)\equiv \zeta^{\mu_1}\cdots \zeta^{\mu_{2\ell}}\Phi_{\mu_1\cdots \mu_{2\ell}}(x)\,. 
\end{align}
Then the bulk one-point function of $\Phi_{2\ell}(x, \zeta)$ can be evaluated as follows,
\begin{align}
    \begin{aligned}
        \overline{\langle\Phi_{2\ell}(x, \zeta)\rangle}&=C_{d}\,c^2 \lim_{y\to x}\sum_{k=0}^{2\ell}c_{2\ell, k}(\zeta\cdot \partial_{x})^{2\ell-k}(\zeta\cdot \partial_{y})^{k}\int \d^{p}\hat{z}\frac{1}{|x-\hat{z}|^{p}|y-\hat{z}|^{p}}\\
        &=2^{2\ell}C_{d}c^{2}\sum_{k=0}^{2\ell}c_{2\ell , k}\left(\frac{p}{2}\right)_{k}\left(\frac{p}{2}\right)_{2\ell-k}\int \d^{p}\hat{z}\frac{(\zeta\cdot(\hat{z}-x))^{2\ell}}{|x-\hat{z}|^{2p+2\ell}} \\
        &=0\,, 
    \end{aligned}
\end{align}
where in the final line, we have used the following identity satisfied by \eqref{csk}, 
\begin{align}\label{eq:}
\sum_{k=0}^{2\ell}c_{2\ell , k}\left(\frac{p}{2}\right)_{k}\left(\frac{p}{2}\right)_{2\ell-k}=0\,. 
\end{align}

The resulting bulk-channel conformal block decomposition of the scalar two-point function
follows from the general discussion around \eqref{eq:bulk_decomposition} (see notations thereof),
\begin{align}\label{eq:decomposition}
    F(X, \overline{X})=C_{d}\xi^{-p/2}+\frac{\pi^{\frac{p}{2}}\Gamma(\frac{p}{2})}{\Gamma(p)}C^{2}_{d}c^2(X\overline{X})^{p/4}{}_2 F_{1}\left(\frac{p}{2}, \frac{p}{2}; p; 1-X\overline{X}\right)\,.
\end{align}
As a consistency check, we also see the second term on the RHS reproduces the integral ${\cal I}_2$ in \eqref{eq:integral_I2}  (see Appendix \ref{app:integral_bulk_block} for details). 

The above result should be contrasted with what happens in ordinary DCFT, where typically an infinite tower of spinning operators (e.g. double-twist operators) contribute in the bulk-channel conformal block decompositions. 
In particular, the one-point function of the energy-momentum tensor $T^{\mu\nu}$ generally acquires a nontrivial value \cite{Kapustin:2005py, Billo:2016cpy}. However, as demonstrated above, in our free scalar theory with subdimensional disorder, averaged bulk one-point functions of the spinning operator s $\Phi_{2J}$ all vanish, in particular
\begin{align}
    \overline{\langle T^{\mu\nu}(x) \rangle}=0\,. 
\end{align}
In other words, even though the defect is non-topological, the one-point function of the energy-momentum tensor is zero!\footnote{Another indication of the quasi-topological nature of the defect arises from the correlators of the displacement operators $D_i$, which vanish except for contact terms (see \eqref{eq:displacement_two_pt}).

A similar phenomenon occurs for higher-spin symmetries. The corresponding Ward identities take the form
\ie
\partial_{\mu}\Phi^{\mu i_1\dots i_{2J-1}}(x) = c_{2J}{\h}(\hat{x})\partial^{i_1}\cdots \partial^{i_{2J-1}}\sigma(x)\delta^{(d-p)}(x_{\perp}),,
\fe
where $c_{2J}$ is a constant. This generalizes \eqref{eq:dTmunu} to the spin-2 case ($J=1$), and the RHS defines the analog of the displacement operator $D_i$ in \eqref{displace} for higher-spin currents broken by the defect. Like their spin-2 counterparts, these higher-spin displacement operators also belong to rank-2 logarithmic multiplets and exhibit vanishing correlators among themselves. }

\section{Discussion}
\label{sec:disc}

There are several promising directions for future investigation which we will comment below.

\paragraph{Logarithmic defects in interacting CFTs}
A natural extension of our work is to study logarithmic conformal defects arising from disorder in interacting CFTs. For instance, one may generalize the free scalar examples discussed here to the case of random-field disorder in the interacting ${\rm O}(N)$ CFT in $d=4-\epsilon$ (cf. \eqref{eq:freeON}). Since the fundamental scalar has scaling dimension $\Delta_\sigma < 1$, the Harris criterion \eqref{Harrisdef} implies that localized random-field disorder can generate a perturbative fixed point for $p=2$. It would be especially interesting to understand the nature of the resulting conformal surface defect in $d=3$ and to clarify its relation to other surface defects in the ${\rm O}(N)$ CFT \cite{Trepanier:2023tvb,Giombi:2023dqs,Raviv-Moshe:2023yvq,Krishnan:2023cff,Diatlyk:2024ngd}.

The random-field Ising model (RFIM), defined with bulk disorder, is known to exhibit a disordered critical point in dimensions $2<d\leq 6$ \cite{Imry:1975zz} (see also the review \cite{Rychkov:2023rgq}). Near the upper critical dimension, the critical RFIM famously enjoys an emergent Parisi–Sourlas supersymmetry \cite{Parisi:1979ka}, which underlies the phenomenon of dimensional reduction relating the RFIM in $d$ dimensions to the ordinary Ising model in $d-2$ dimensions \cite{Aharony:1976jx} (see also \cite{Kaviraj:2019tbg,Kaviraj:2020pwv,Kaviraj:2021qii,Kaviraj:2022bvd, Nakayama:2024jwq} and the review \cite{Rychkov:2023rgq}).\footnote{See also \cite{Ghosh:2025lzb}, where this correspondence is enriched by incorporating defects.} It would be interesting to explore whether analogous dimensional-reduction phenomena exist for conformal defects with and without disorder in different dimensions of the ${\rm O}(N)$ CFT.\footnote{For recent progress on conformal defects in the ${\rm O}(N)$ CFT (and related models) in the absence of disorder, see e.g. \cite{Billo:2013jda,Gaiotto:2013nva, Soderberg:2017oaa, Carmi:2018qzm,Giombi:2019enr, Herzog:2020lel,Lauria:2020emq,Giombi:2020rmc, Behan:2020nsf,Metlitski:2020cqy,Toldin:2020wbn,Bianchi:2021snj, Padayasi:2021sik,Toldin:2021kun,Cuomo:2022xgw,Gimenez-Grau:2022czc,Giombi:2022vnz,Nishioka:2022qmj,Nishioka:2022odm, Giombi:2022gjj, Bissi:2022bgu, Popov:2022nfq,Pannell:2023pwz, Toldin:2023fny,SoderbergRousu:2023pbe,SoderbergRousu:2023zyj,Cuomo:2023qvp,Harribey:2023xyv, Diatlyk:2024zkk,Diatlyk:2024qpr, Cuomo:2024psk, Giombi:2024qbm, Zhou:2024dbt,Dedushenko:2024nwi, deSabbata:2024xwn,Harribey:2024gjn,Giombi:2025pxx,Lanzetta:2025xfw,Hu:2025yrs, Erramilli:2025pfh,Kravchuk:2025evf}.}

\paragraph{Numerical test of logarithmic behaviors}
It would be interesting to study the expected logarithmic behaviors from numerical simulations in concrete models, for example, to find evidence for new fixed points from subdimensional disorder and to understand how disorder modifies ordinary pinning defects defined by constant coupling.\footnote{We thank Zohar Komargodski for bringing up this suggestion.} The latter is particularly interesting if the coupling is weakly irrelevant according to the Harris criterion. Indeed, in the 2d random bond Ising model, where the disorder is marginally irrelevant, intriguing double-log behaviors were predicted \cite{dotsenko1981phase,dotsenko1982critical} and observed in Monte Carlo simulations \cite{andreichenko1990monte}.
An interesting target with subdimensional disorder would be the random pinning line defect in the 3d Ising model. Here the $p=1$ random-field disorder is very weakly irrelevant since $\Delta_\sigma - 1/2 \approx 0.018 > 0$ and similar logarithmic effects may be testable numerically.

\paragraph{Beyond Gaussian disorder}
In this work, we have restricted attention to Gaussian disorder. An important generalization is to study alternative distributions, such as binary disorder, which realizes the Nishimori line \cite{HNishimori_1980,Nishimori:1981ajd}. This naturally raises the question of possible defect analogs of the Nishimori critical point.

\paragraph{Disorder on nontrivial defects} Another direction is to introduce disorder atop existing nontrivial defects, such as Wilson lines in gauge theories (generally a direct sum of simple Wilson lines). In this case, localized disorder, for example from defect-changing operators, may become relevant under the Harris criterion \eqref{Harrisdef} \cite{Aharony:2022ntz,Aharony:2023amq}.

\paragraph{Constraints on disordered defect RG flows} It is also important to establish general constraints on RG flows involving disordered defects. One natural goal is to formulate a disorder-generalized version of defect monotonicity theorems, extending results from \cite{Affleck:1991tk,Friedan:2003yc,Jensen:2015swa,Casini:2018nym,Kobayashi:2018lil,Wang:2020xkc,Wang:2021mdq,Cuomo:2021rkm,Shachar:2022fqk,Casini:2023kyj}. See \cite{Patil:2025wqp} for recent progress. In Section~\ref{subsec:defect_free_energy}, we have proposed a candidate for the RG monotone in terms of an averaged defect free energy which passes a consistency check on the defect conformal manifold. It would be interesting to study further its properties under nontrivial flows.

\paragraph{Symmetries and anomalies for disordered defects}
Finally, it is worthwhile to clarify the symmetry properties of defects after disorder averaging. Typically, disordered defects preserve bulk symmetries only on average. It would be interesting to investigate the constraints from such average symmetries, as well as their possible anomalies \cite{Ma:2023rji,Antinucci:2023uzq}, on the dynamics of disordered defects. Related ideas have recently been used to solve defect RG flows without disorder (see e.g. \cite{Antinucci:2024izg,Popov:2025cha}).

\section*{Acknowledgement}
We thank Zohar Komargodski for helpful discussions and insightful comments on the draft. This work was initiated while SS was visiting the CCPP at New York University in May 2025, and he gratefully acknowledges the hospitality and support of its members. The work of SS was supported by Grant-in-Aid for JSPS Fellows No.\,23KJ1533. The work of YW was supported in part by the NSF grant PHY-2210420 and by the Simons Junior Faculty Fellows program. 

\appendix
\section{Computations of Beta Functions}\label{app:beta}
In this Appendix, we perform the diagramatic analysis to derive the beta function \eqref{eq:beta_function}, using the Callan-Symanzik equation for the bulk one-point function of $\sigma_{A}\sigma_{B}(x)$,
\begin{align}\label{eq:CS}
    \frac{\d}{\d\log\mu}\langle\sigma_{A}\sigma_{B}(x)\rangle_{\text{rep}}=0\,,
\end{align}
where $\mu$ is a reference energy scale. Using the Wick theorem, we can perturbatively expand this bulk one-point function as a power series of the coupling $m_{AB}$ in the replicated theory,
\begin{align}\label{eq:bulk_one_pt}
    \langle\sigma_{A}\sigma_{B}(x)\rangle_{\text{rep}}=-C_{d}^{2}m_{AB}I_{1}(x)+4C_{d}^{3}\sum_{D}m_{AD}m_{BD}I_{2}(x)+O(m^3)\,,
\end{align}
where $I_{1}$ and $I_{2}$ are the integrals given by
\begin{align}
    \begin{aligned}
        I_{1}(x)&\equiv \int \d^{p}\hat{y}\frac{1}{|x-\hat{y}|^{2p}}\,, \\
        I_{2}(x)&\equiv \int \d^{p}\hat{y}\d^{p}\hat{z}\frac{1}{|x-\hat{y}|^{p}|\hat{y}-\hat{z}|^{p}|x-\hat{z}|^{p}}\,. 
    \end{aligned}
\end{align}
\begin{figure}[!htb]
    \centering
    \begin{tikzpicture}[scale=1.2]
\draw[thick] (-3,0) -- (-1,0); 
\draw[blue, thick] (-2,0) .. controls (-2.5,1.5) and (-1.5,1.5) .. (-2,0); 
\filldraw[black] (-2,1.15) circle (2pt); 
\node at (-2,-0.6) {$I_1$}; 

\draw[thick] (1,0) -- (3,0); 
\draw[blue, thick] (1.2,0) .. controls (1.6,1.8) and (2.4,1.8) .. (2.8,0); 
\draw[blue, thick] (1.2,0) .. controls (2,0.5) .. (2.8,0); 
\filldraw[black] (2,1.35) circle (2pt); 
\node at (2,-0.6) {$I_2$}; 
\end{tikzpicture}
    \caption{Feynman diagrams contributing to the bulk one-point function \eqref{eq:bulk_one_pt}}
    \label{fig:feynman}
\end{figure}
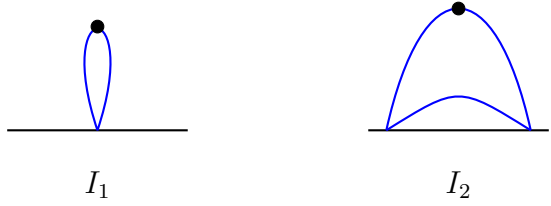
The Feynman diagram corresponding to each integral is drawn in Figure~\ref{fig:feynman}. 
The first integral is the contribution from tree diagram, and can be evaluated using the Schwinger's reparameterization,
\begin{align}
    I_{1}(x)=\frac{\pi^{p/2}\Gamma(p/2)}{\Gamma(p)}\frac{1}{|x_{\perp}|^{p}}\,. 
\end{align}
On the other hand, the second integral $I_{2}$ comes from the loop contribution, and contains logarithmic divergence around $\hat{y}=\hat{z}$. With a proper regularization of this UV divergence, the scheme independent part of $I_{2}$ is written as 
\begin{align}
    I_{2}=\frac{\pi^{p}}{\Gamma(p)}\frac{\log\mu^{2}|x_{\perp}|^2}{|x_{\perp}|^{p}}\,, 
\end{align}
where we introduce the energy scale $\mu$ as a regulator. By plugging these results into \eqref{eq:bulk_one_pt}, we obtain, keeping only contributions up to two-loop, 
\begin{align}
    \frac{\Gamma\left(p\right)C_{d}^{-2}}{\pi^{p/2}\Gamma\left(\frac{p}{2}\right)}\langle\sigma_{A}\sigma_{B}(x)\rangle_{\text{rep}}=\left(c^{2}-2\lambda \delta_{AB}+\frac{8C_{d}\pi^{p/2}}{\Gamma\left(\frac{p}{2}\right)}\left(\frac{n\, c^2}{4}-\lambda c^2 +\lambda^2 \delta_{AB}\right)\log\mu^{2}|x_{\perp}|^2\right)\frac{1}{|x_{\perp}|^{p}}\,.
\end{align}
The Callan-Symanzik equation \eqref{eq:CS} then immediately determines the beta functions as is given in \eqref{eq:beta_function}. 

Let us end by making a brief comment. Clearly, from the viewpoint of conformal symmetry, bulk one-point functions should not exhibit any logarithmic dependence. Indeed, upon setting $n\to0$ and substituting the fixed point values \eqref{eq:line_fixed_point} into the above results, one finds that the logarithmic term disappears, confirming the consistency of the analysis.

\section{Fixing Full LogDOE Structures}\label{app:logDOE}
In Section~\ref{subsec:logdefect_OPE}, we presented the derivation of the coefficients $A_{J, 0}^{N_{\widehat{\Delta}}}$ in \eqref{eq:a}. Here we provide detailed computation of general coefficients $A_{J, \ell}^{N_{\widehat{\Delta}}}$ including the descendant contributions. 
\subsubsection*{Evaluation of \eqref{eq:logdoe_general}}
We note that the derivative part on the RHS in \eqref{eq:logdoe_general} can be written as
\begin{align}\label{eq:logdifferential_formula}
    \Box_{\hat{x}}^{\ell}\frac{(\log |\hat{x}|^2)^m}{|\hat{x}|^{2\widehat{\Delta}}}=\frac{1}{|\hat{x}|^{2\widehat{\Delta}+2\ell}}\sum_{q=0}^{m}a_{\ell , q, m}(\log |\hat{x}|^2)^q\,.
\end{align}
The coefficient $a_{\ell,q,m}$ can be determined as follows. 

First we note that, in the special case with $m=0$ in \eqref{eq:logdifferential_formula},
\ie 
a_{\ell,0,0}=4^{\ell}(\widehat{\Delta})_{\ell}\left(\widehat{\Delta}-\frac{p-2}{2}\right)_{\ell}\,.
\fe
By taking derivative $(-\pa_{\wD})^m$, we find
\ie 
a_{\ell,0,m}=(-\pa_{\wD})^m a_{\ell,0,0}\,.
\fe
Next, we have the following useful recursion relation
\ie 
a_{\ell,n,m}={m\over n} a_{\ell,n-1,m-1}\,,
\label{acrec}
\fe
and thus for $0\leq q\leq m$,
\ie 
a_{\ell,q,m}={m!\over q! (m-q)! }(-\pa_{\wD})^{m-q} a_{\ell,0,0}\,.
\label{acoeff}
\fe

Using the expression \eqref{eq:logdifferential_formula}, one can deform \eqref{eq:logdoe_general} as
\begin{align}
    \begin{aligned}
    \langle\mathcal{O}_{\Delta}(x)\widehat{\mathcal{O}}^{I}_{\widehat{\Delta}}(0)\rangle 
    &=\sum_{\ell = 0}^{\infty}\sum_{k=0}^{N_{\widehat{\Delta}}-I}\sum_{J=1}^{N_{\widehat{\Delta}}-I-k+1}\sum_{n=0}^{k}\frac{(\log|x_{\perp}|)^n}{n!\, (k-n)!}\frac{k_{I+J+k-1}A^{N_{\widehat{\Delta}}}_{J, \ell}}{|x_{\perp}|^{\Delta-\widehat{\Delta}-2\ell}}\,\Box_{\hat{x}}^{\ell}\frac{(-\log |\hat{x}|^2)^{k-n}}{|\hat{x}|^{2\widehat{\Delta}}} \\
    &=\frac{1}{|x_{\perp}|^{\Delta-\widehat{\Delta}}|\hat{x}|^{2\widehat{\Delta}}}\sum_{\ell=0}^{\infty}t^{\ell}\sum_{k=0}^{N_{\widehat{\Delta}}-I}\sum_{J=1}^{N_{\widehat{\Delta}}-I-k+1}k_{I+J+k-1}A^{N_{\widehat{\Delta}}}_{J, \ell}\\
    &\qquad\qquad\qquad\qquad \cdot \sum_{n=0}^{k}\sum_{q=0}^{k-n}\frac{(-1)^{k-n}a_{\ell , q, k-n}}{n!\, (k-n)!}(\log|x_{\perp}|)^n (\log|\hat{x}|^2)^{q} \\
        &=\frac{1}{|x_{\perp}|^{\Delta-\widehat{\Delta}}|\hat{x}|^{2\widehat{\Delta}}}\sum_{\ell=0}^{\infty}t^{\ell}\sum_{k=0}^{N_{\widehat{\Delta}}-I}\sum_{J=1}^{N_{\widehat{\Delta}}-I-k+1}k_{I+J+k-1}A^{N_{\widehat{\Delta}}}_{J, \ell}\\
    &\qquad\qquad\qquad\qquad \cdot \sum_{n=0}^{k}\sum_{q=0}^{k-n}\frac{(-1)^{k-n}a_{\ell , k-n-q, k-n}}{n!\, (k-n)!}(\log|x_{\perp}|)^n (\log|\hat{x}|^2)^{k-n-q} \,, 
    \end{aligned}
\end{align}
where we introduce a new symbol $t\equiv |x_{\perp}|^2/|\hat{x}|^2$, and in the final line we changed the sum variables as
\begin{align}
    \sum_{q=0}^{k-n}f(q)=\sum_{q=0}^{k-n}f(k-n-q)\,. 
\end{align}
Finally, for later convenience, we swap the summations over $n$ and $q$ as
\begin{align}\label{eq:evaluation_logDOE}
    \begin{aligned}
        \langle\mathcal{O}_{\Delta}(x)\widehat{\mathcal{O}}^{I}_{\widehat{\Delta}}(0)\rangle 
    &=\frac{1}{|x_{\perp}|^{\Delta-\widehat{\Delta}}|\hat{x}|^{2\widehat{\Delta}}}\sum_{\ell=0}^{\infty}t^{\ell}\sum_{k=0}^{N_{\widehat{\Delta}}-I}\sum_{J=1}^{N_{\widehat{\Delta}}-I-k+1}k_{I+J+k-1}A^{N_{\widehat{\Delta}}}_{J, \ell}\\
    &\qquad\qquad\qquad\qquad \cdot \sum_{q=0}^{k}\sum_{n=0}^{k-q}\frac{(-1)^{k-n}a_{\ell , k-n-q, k-n}}{n!\, (k-n)!}(\log|x_{\perp}|)^n (\log|\hat{x}|^2)^{k-n-q} \,.
    \end{aligned}
\end{align}
The last summand can be implemented explicitly using \eqref{acoeff},
\ie 
 \sum_{q=0}^{k}\sum_{n=0}^{k-q}\frac{(-1)^{k-n-q} k! }{n!\, (k-n)!}a_{\ell , k-n-q, k-n}(\log|x_{\perp}|)^n (\log|\hat{x}|^2)^{k-n-q}=\sum_{q=0}^{k} a_{\ell,k-q,k} \left(\log\frac{|x_{\perp}|}{|\hat{x}|^2}\right)^{k-q}\,,
\fe 
which takes the expected form from conformal symmetry. In summary, \eqref{eq:evaluation_logDOE} is reduced to
\begin{align}\label{eq:final1}
    \begin{aligned}
        \langle\mathcal{O}_{\Delta}(x)\widehat{\mathcal{O}}^{I}_{\widehat{\Delta}}(0)\rangle 
    &=\frac{1}{|x_{\perp}|^{\Delta-\widehat{\Delta}}|\hat{x}|^{2\widehat{\Delta}}}\sum_{\ell=0}^{\infty}t^{\ell}\sum_{s=0}^{N_{\widehat{\Delta}}-I}\left(\log\frac{|x_{\perp}|}{|\hat{x}|^2}\right)^{\hspace{-1mm}s}\sum_{J=1}^{N_{\widehat{\Delta}}-I-s+1}\sum_{k=s}^{N_{\widehat{\Delta}}-I-J+1}\frac{(-1)^{k+s} k_{I+J+k-1}a_{\ell,s,k}}{k!}A^{N_{\widehat{\Delta}}}_{J, \ell}\,.
    \end{aligned}
\end{align}

\subsubsection*{Evaluation of  \eqref{eq:b_d}}
As mentioned in the main text, we fix $A_{J, \ell}^{N_{\wD}}$ by comparing \eqref{eq:evaluation_logDOE} with the general structure of the bulk-defect two-point function which was given in \eqref{eq:b_d}. Therefore, what we should do next is to express \eqref{eq:b_d} as a power series of $t\equiv |x_{\perp}|^2/|\hat{x}|^2$. For this purpose, we first note the following series expansion formulas:
\begin{align}\label{eq:taylor}
    \begin{aligned}
        \frac{1}{|x|^{2\wD}}&=\frac{1}{|\hat{x}|^{2\wD}}\sum_{m=0}^{\infty}\frac{(-1)^m}{m!}(\wD)_{m}\,t^{m}\,, \\
        \tau^{k} &=\sum_{n=0}^{\infty}\sum_{q=0}^{k}\frac{(-1)^{n}S_{1}(n, q) k!}{n! (k-q)!}\left(\log\frac{|x_{\perp}|}{|\hat{x}|^2}\right)^{k-q}\, t^{n} \,,  
    \end{aligned}
\end{align}
where $S_{1}(n, q)$ is the unsigned Stirling number of first kind, and its generating function is given by
\begin{align}
    (\log(1+t))^{q}=q!\sum_{n=0}^{\infty}\frac{(-1)^{n-q}S_{1}(n, q)}{n!}t^n\,. 
\end{align}
Making use of these Taylor expansions, one can deform \eqref{eq:b_d} as
\begin{align}\label{eq:b_d_2}
    \begin{aligned}
        \langle\,\mathcal{O}_{\Delta}(x)\widehat{\mathcal{O}}^{I}_{\widehat{\Delta}}(0) \, \rangle&=\frac{1}{|x_{\perp}|^{\Delta-\widehat{\Delta}}|x|^{2\widehat{\Delta}}}\sum_{k=0}^{N_{\widehat{\Delta}}-I}\frac{\lambda_{I+k}\tau^{k}}{k!} \\
        &=\frac{1}{|x_{\perp}|^{\Delta-\widehat{\Delta}}|\hat{x}|^{2\widehat{\Delta}}}\sum_{k=0}^{N_{\widehat{\Delta}}-I}\lambda_{I+k}\sum_{q=0}^{k}\sum_{m, n=0}^{\infty}\frac{(-1)^{n+m}(\wD)_{m}S_{1}(n, q)}{m! n! (k-q)!}\left(\log\frac{|x_{\perp}|}{|\hat{x}|^2}\right)^{k-q}\, t^{n+m}\\
        &=\frac{1}{|x_{\perp}|^{\Delta-\widehat{\Delta}}|\hat{x}|^{2\widehat{\Delta}}}\sum_{\ell=0}^{\infty}(-1)^{\ell}t^{\ell}\sum_{k=0}^{N_{\widehat{\Delta}}-I}\lambda_{I+k}\sum_{q=0}^{k}\frac{1}{(k-q)!}\left(\log\frac{|x_{\perp}|}{|\hat{x}|^2}\right)^{k-q}F_{\ell, q} \\
        &=\frac{1}{|x_{\perp}|^{\Delta-\widehat{\Delta}}|\hat{x}|^{2\widehat{\Delta}}}\sum_{\ell=0}^{\infty}(-1)^{\ell}t^{\ell}\sum_{s=0}^{N_{\widehat{\Delta}}-I}\left(\log\frac{|x_{\perp}|}{|\hat{x}|^2}\right)^{\hspace{-1mm}s}\sum_{k=s}^{N_{\widehat{\Delta}}-I}\frac{\lambda_{I+k}}{s!}F_{\ell, k-s}\, ,
    \end{aligned}
\end{align}
where $F_{\ell, q}$ is defined by
\begin{align}
    F_{\ell, q}\equiv\sum_{m=0}^{\ell-q}\frac{(\wD)_{m}S_{1}(\ell -m, q)}{m! (\ell-m)!}\,. 
\end{align}
Noting that the $F_{\ell, q}$ is the coefficient of $y^{q}$ of the following polynomial:
\begin{align}
    \frac{1}{\ell !}\sum_{m=0}^{\ell-q}\binom{\ell}{m}(\wD)_{m}(y)_{\ell-m}=\frac{1}{\ell !}(\wD +y)_{\ell}\,,
\end{align}
and one can express $F_{\ell, q}$ as follows
\begin{align}
    F_{\ell, q}=\frac{1}{\ell! q!} \partial_{\wD }^{q}(\wD )_{\ell}\,. 
\end{align}
Finally, by comparing \eqref{eq:final1} with \eqref{eq:b_d_2}, one can obtain the constraints on $A_{J,\ell}^{N_{\widehat{\Delta}}}$ as
\begin{align}
    \sum_{J=1}^{N_{\widehat{\Delta}}-I_{s}+1}\sum_{k=0}^{N_{\widehat{\Delta}}-I_{s}-J+1}\frac{k_{I_{s}+J+k-1}\partial_{\widehat{\Delta}}^{k}a_{\ell,0,0}}{k!}A^{N_{\widehat{\Delta}}}_{J, \ell}=(-1)^{\ell}\sum_{k=0}^{N_{\widehat{\Delta}}-I_{s}}\lambda_{I_{s}+k}F_{\ell, k}\,, 
\end{align}
where $I_{s}\equiv I+s$, and $I\leq I_{s} \leq N_{\widehat{\Delta}}$. Notably, the dependence on $I$ and $s$ appears only through the combination $I_{s}$, which implies that there are $N_{\widehat{\Delta}}$ independent constraints on~$A_{J, \ell}^{N_{\wD}}$ at each level $\ell$. As a consistency check, by setting $\ell=0$, the above constraint becomes \eqref{eq:primary_constraint}. After some algebras, by using the expression \eqref{eq:primary_constraint},   the above constraint reduces to,
\begin{align}\label{eq:constraint}
    \sum_{k=0}^{J-1}G_{k}A^{N_{\wD}}_{J-k, \ell}=\frac{(-1)^\ell}{\ell!}\sum_{k=0}^{J-1}H_{k}A^{N_{\wD}}_{J-k, 0}\,,
\end{align}
where $G_{k}$ and $H_{k}$ are defined
\begin{align}
    G_{k}\equiv \frac{\partial_{\wD}^k a_{\ell, 0, 0}}{k!}\,, \quad H_{k}\equiv \frac{\partial_{\wD}^k (\wD)_{\ell}}{k!}\,. 
\end{align}
To further simplify these constraints, we introduce the following formal generating functions:
\begin{align}\label{eq:generating_function}
    \begin{aligned}
        A_{\ell}(z)&\equiv \sum_{k=0}^{\infty} A_{k, \ell}^{N_{\wD}}z^{k} \,, \\
        G(z)&\equiv\sum_{k=0}^{\infty}G_{k} z^{k}=4^{\ell}(\widehat{\Delta}+z)_{\ell}\left(\widehat{\Delta}+z-\frac{p-2}{2}\right)_{\ell}\,, \\
        H(z)&\equiv\sum_{k=0}^{\infty}H_{k} z^{k}=(\widehat{\Delta}+z)_{\ell} \,.
    \end{aligned}
\end{align}
Then, we should notice that, in \eqref{eq:constraint}, the left and right hand sides correspond to the $z^{J}$ coefficients of the polynomials $G(z)A_{\ell}(z)$ and $H(z)A_{0}(z)$, respectively. This identification allows us to rewrite the above constraint in a drastically simpler form:
\begin{align}
    G(z)A_{\ell}(z)=\frac{(-1)^\ell}{\ell!}H(z)A_{0}(z)\,. 
\end{align}
Therefore, by using the explicit forms of $G(z)$ and $H(z)$ in \eqref{eq:generating_function}, one can obtain the generating function of all logDOE coefficients:
\begin{align}
    A_{\ell}(z)=\frac{(-4)^{-\ell}}{\ell! \left(\widehat{\Delta}+z-\frac{p-2}{2}\right)_{\ell}}A_{0}(z)\,. 
\end{align}
From this expression, we arrive at the explicit form of $A_{J, \ell}^{N_{\wD }}$:
\begin{align}
    A_{J, \ell}^{N_{\wD }}=\frac{(-4)^{-\ell}}{\ell! \left(\widehat{\Delta}+z-\frac{p-2}{2}\right)_{\ell}}\sum_{m=1}^{J}(-1)^{J+m}A_{m, 0}^{N_{\wD }}\sum_{0\leq k_{1}< \cdots < k_{J-m}\leq \ell -1}\prod_{r=1}^{J-m}\left(\wD -\frac{p-2}{2}+k_{r}\right)^{\hspace{-1mm}-1}\,. 
\end{align}

\section{Derivation of \eqref{eq:asymptotic}}\label{app:asymptotic}
In this Appendix, we derive the asymptotic formula \eqref{eq:asymptotic}. For simplicity, we denote $N_{\wD , s}$ simply by $N$ below. Firstly, by using the expression of logDOE \eqref{eq:ansatz}, we obtain
\begin{align}\label{eq:w}
    \begin{aligned}
        W_{\wD}^{N}&\overset{\chi \sim \infty}{\sim}\frac{k_{\widehat{\mathcal{O}}}}{|x_{1\perp}|^{\Delta_{1}-\wD}|x_{2\perp}|^{\Delta_{2}-\wD}}\frac{x^{(i_{1}}_{1\perp}\cdots x^{i_{s})}_{1\perp}x_{2\perp i_{1}}\cdots x_{2\perp i_{s}}}{|x_{1\perp}|^{s}|x_{2\perp}|^{s}}\frac{1}{|\hat{x}_{12}|^{2\wD}} \\
        &\qquad \cdot \sum_{I_{1}=1}^{N}\sum_{I_{2}=1}^{N-I_{1}+1}\sum_{n=0}^{I_{1}-1}\sum_{m=0}^{I_{2}-1}A_{I_{1}-n , 0}A_{I_{2}-m , 0}\frac{(\log|x_{1\perp}|)^{n}(\log|x_{2\perp}|)^{m}(-\log|\hat{x}_{12}|^2)^{N-I_{1}-I_{2}+1}}{n!\, m!\, (N-I_{1}-I_{2}+1)!} \\
        &=\frac{a_{s}\, k_{\widehat{\mathcal{O}}}}{|x_{1\perp}|^{\Delta_{1}-\wD}|x_{2\perp}|^{\Delta_{2}-\wD}}H_{W_{\widehat{\mathcal{O}}}}^{N}( \phi)\frac{1}{|\hat{x}_{12}|^{2\wD}}\\
        &\qquad \cdot \sum_{q=1}^{N}\sum_{r=1}^{N+1-q}A^{N}_{q, 0}A^{N}_{r, 0}\sum_{I_{1}=1}^{N+1-r}\sum_{I_{2}=r}^{N-I_{1}+1}\frac{(\log|x_{1\perp}|)^{I_{1}-q}(\log|x_{2\perp}|)^{I_{2}-r}(-\log|\hat{x}_{12}|^2)^{N-I_{1}-I_{2}+1}}{(I_{1}-q)!\, (I_{2}-r)!\, (N-I_{1}-I_{2}+1)!}\,. 
    \end{aligned}
\end{align}
In the second line, we change the order of summations and used the following identity:
\begin{align}
    \begin{aligned}
        \frac{x^{(i_{1}}_{1\perp}\cdots x^{i_{s})}_{1\perp}x_{2\perp i_{1}}\cdots x_{2\perp i_{s}}}{|x_{1\perp}|^{s}|x_{2\perp}|^{s}}&=\frac{\Gamma(s+1)\Gamma(\frac{d-p}{2}-1)}{2^s \Gamma(\frac{d-p}{2}+s-1)}C_{s}^{(\frac{d-p}{2}-1)}(\cos\phi)\\
        &=a_{s}H_{W_{\widehat{\mathcal{O}}}}^{N_{\wD , s}}( \phi)\,, 
    \end{aligned}
\end{align}
where $H_{W_{\widehat{\mathcal{O}}}}^{N_{\wD , s}}( \phi)$ is given by \eqref{eq:gegenbauer}, and $a_{s}$ is defined in \eqref{eq:as}. Moreover, by using the multinomial theorem:
\begin{align}
    \sum_{a+b+c=L}\frac{s^a t^b u^c}{a!b!c!}=\frac{(s+t+u)^{L}}{L!}\,, 
\end{align}
the expression \eqref{eq:w} is simplified to
\begin{align}
    \begin{aligned}
        W_{\wD}^{N}&\overset{\chi \sim \infty}{\sim}\frac{a_{s}\, k_{\widehat{\mathcal{O}}}}{|x_{1\perp}|^{\Delta_{1}-\wD}|x_{2\perp}|^{\Delta_{2}-\wD}}H_{W_{\widehat{\mathcal{O}}}}^{N}( \phi)\sum_{q=1}^{N}\sum_{r=1}^{N+1-q}A^{N}_{q, 0}A^{N}_{r, 0}\frac{1}{(N+1-q-r)!}\frac{(-\log\chi)^{N+1-q-r}}{|\hat{x}_{12}|^{2\wD}} \\
        &=\frac{a_{s}\, k_{\widehat{\mathcal{O}}}}{|x_{1\perp}|^{\Delta_{1}-\wD}|x_{2\perp}|^{\Delta_{2}-\wD}}H_{W_{\widehat{\mathcal{O}}}}^{N}( \phi)\sum_{k=0}^{N-1}\sum_{q=1}^{N-k}A_{q, 0}^{N}A_{N+1-k-q, 0}^{N}\frac{1}{k!}\frac{(-\log\chi)^{k}}{|\hat{x}_{12}|^{2\wD}} \\
        &=\frac{a_{s}\, k_{\widehat{\mathcal{O}}}^{-1}}{|x_{1\perp}|^{\Delta_{1}-\wD}|x_{2\perp}|^{\Delta_{2}-\wD}}H_{W_{\widehat{\mathcal{O}}}}^{N}( \phi)\sum_{k=0}^{N-1}\sum_{q=1}^{N-k}\lambda_{N+1-q}\lambda_{k+q}\frac{1}{k!}\frac{(-\log\chi)^{k}}{|\hat{x}_{12}|^{2\wD}}\,, 
    \end{aligned}
\end{align}
where in the final line, we used the relation \eqref{eq:simple}. By comparing this expression with \eqref{eq:asymp}, one can fix the coefficient $C_{k}$ as \eqref{eq:Ck}.

\section{Constraints from Bulk Equation of Motion}\label{app:eom_constraint}
In this Appendix, we prove the statements described \eqref{eq:spin_defect} and \eqref{eq:scalar_possible} in order.
\subsection{Proof of \eqref{eq:spin_defect}}
We start with giving the proof of the statement \eqref{eq:spin_defect}. This can be shown by contradiction. Suppose that the associated module is non-trivial, i.e. $\mathcal{V}_{\sigma}^{N_{\wD , s}}\not=\emptyset\,, N_{\wD , s} \geq 2  \text{ and } s\geq 1$. The constraint \eqref{eq:eom_constraint} at $J=1$ reads
\begin{align}
    (\widehat{\Delta}-\Delta_{\sigma}+s)(\widehat{\Delta}-\Delta_{\sigma}-s)\alpha_{1}=0\,.
\end{align}
Our assumption implies that there is non-zero solution for $\alpha_{1}$, hence the above equation restricts the defect conformal dimensions as $\wD = \Delta_{\sigma}\pm s$. However, with these dimensions, one can show that there is no non-zero solution compatible with other constraints with $J\not =1$. Indeed, the constraint \eqref{eq:eom_constraint} with $J=2$ reads $s\,\alpha_{1}=0$, which forces $\alpha_{1}=0$ for $s\geq 1$. This completes the proof for \eqref{eq:spin_defect}. Also, the above proof is enough to show the statement \eqref{eq:scalar_possible} by setting $s=0$.
\subsection{Proof of \eqref{eq:scalar_possible}}
We next move on to the proof for \eqref{eq:scalar_possible}. For simplicity, we give the proof of $\mathcal{V}_{\sigma}^{N_{\wD , 0}=3}=\emptyset$, although the generalization to generic rank $N_{\widehat{\Delta}, 0}$ is quite straightforward. In the case of rank three, the constraints associated to equation of motion read
\begin{align}
    (\widehat{\Delta}-\Delta_{\sigma})^{2}\alpha_{1}&=0 \label{eq:1st_rank3}\,,   \\
    2(\widehat{\Delta}-\Delta_{\sigma})\alpha_{1}+(\widehat{\Delta}-\Delta_{\sigma})^{2}\alpha_{2}&=0\,, \\
    \alpha_{1}+2(\widehat{\Delta}-\Delta_{\sigma})\alpha_{2}+(\widehat{\Delta}-\Delta_{\sigma})^{2}\alpha_{3}&=0\,. \label{eq:2nd_rank3}
\end{align}
We show that $\alpha_{1}$ must be trivial again by the contradiction. Suppose that $\alpha_{1}\not=0$, the first constraint \eqref{eq:1st_rank3} implies again that the defect conformal dimensions are restricted as $\wD=\Delta_{\sigma}$. Then, the final constraint  \eqref{eq:2nd_rank3} forces $\alpha_{1}$ to be zero, which contradicts our assumption. This completes the proof.

\section{Derivation of Selected Integrals}\label{app:integral}

We first list two integrals that are useful in performing the Fourier transforms:
\ie 
\int \frac{\d^{n}k}{(2\pi)^n}\frac{e^{\i k\cdot x}}{|k|^{2\alpha}}&=\frac{\Gamma(\frac{n}{2}-\alpha)}{\pi^{n/2}4^{\alpha}\Gamma(\alpha)}\frac{1}{|x|^{n-2\alpha}}\,,
\label{FT}
\fe
and the following generalization that involves mixed bulk-defect coordinates,
\ie 
{2\pi^{p\over 2}\over \Gamma({\A\over 2})}\int \frac{\d^{p}\hat k}{(2\pi)^p} e^{\i \hat k\cdot \hat x} |k|^{\A-p \over 2} K_{p-\A\over 2}(|\hat k| |x_\perp|)= {|2x_\perp|^{\A-p\over 2}\over (\hat x^2+x_\perp^2)^{\A\over 2}}\,,
\label{FTmix}
\fe
where $K_\n(x)$ is the Bessel $K$-function.

\subsection{Integral Formula \eqref{eq:I1}}

Here we present the derivation of \eqref{eq:I1} by solving the following integral:
\begin{align} 
{\cal I}_{1}= \int \frac{\d^p \hat{p}}{(2\pi)^{p+1}}K_{0}(|\hat{p}||x_{\perp}|)\log|\hat{p}|^{2}e^{\i\hat{p}\cdot \hat x}\, . 
\end{align}
In the following, we will make use of the Mellin representation of the Bessel $K$-function
\begin{align}\label{eq:mellin}
K_{0}(2y)={1\over 8\pi \i}\int_{-\i\infty+\epsilon}^{\i\infty+\epsilon} \d s\, y^{-s} \Gamma\left(s\over 2\right)^2\,,
\end{align}
with $\epsilon>0$ and the following two integral identities involving the Bessel $J$-function:
\begin{align}\label{eq:spherical_integral}
\int \d \Omega_p \,e^{\i \hat p \cdot \hat x}
={(2\pi)^{p\over 2}  \over (\rho |\hat x|)^{p-2\over 2}}J_{p-2\over 2} (\rho |\hat x|)
\end{align}
where the integral is over the unit sphere $S^{p-1}$ and $\rho\equiv |\hat p|$, and from \cite{2014637}
\begin{align} 
\int_0^\infty \d\rho\, \rho^{\m+{1\over 2}} \log \rho J_\n(\rho r)={2^{\m -{1\over 2}}\over r^{\m +{3\over 2}}}{\Gamma\left({\n+\m\over 2}+{3\over 4}\right)\over \Gamma\left({\n-\m\over 2}+{1\over 4}\right)}\left(
\psi\left({\n+\m\over 2}+{3\over 4}\right)+\psi\left({\n-\m\over 2}+{1\over 4}\right)-\log {r^2\over 4}
\right)\,,
\end{align}
where $\psi(z)\equiv \Gamma'(z)/\Gamma(z)$ is the digamma function.

In particular, with $\m={p-1\over 2}-s$ and $\n={p-2\over2}$ in the above identities, we have
\begin{align}\label{eq:contor_integral} 
{\cal I}_{1}
={1\over |\hat{x}|^p}{1\over \pi^{p/2+1}}
{1\over 16\pi \i}\int_{-\i\infty+\epsilon}^{\i\infty+\epsilon} \d s\,  \left(|\hat{x}| \over |x_\perp|\right)^s  \Gamma\left(s\over 2\right) \Gamma\left(p-s\over 2\right)
\left(\psi\left(s\over 2\right)+\psi\left(p-s\over 2\right)-\log {|\hat{x}|^2\over 4}\right)\,.
\end{align}

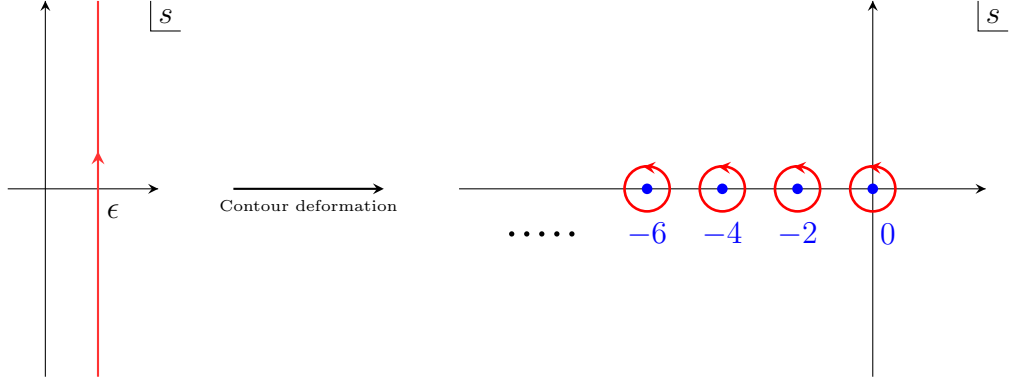
\begin{figure}[!htb]
    \centering
    \begin{tikzpicture}[>=stealth]

\begin{scope}[shift={(-6,0)}]
  \draw [decoration={markings, mark=at position 1 with {\arrow[scale=1.2]{>}}}, postaction={decorate}] (-0.5,0) -- (1.5,0);
  \draw [decoration={markings, mark=at position 1 with {\arrow[scale=1.2]{>}}}, postaction={decorate}] (0,-2.5) -- (0,2.5);
  \draw (1.4, 2.5) -- (1.4, 2.1) -- (1.8,2.1);
  \node at (1.6,2.3) {$s$};

  \draw [red!80, thick,decoration={markings, mark=at position 0.6 with {\arrow[scale=1.2]{>}}}, postaction={decorate}] (0.7,-2.5) -- (0.7,2.5);
  \node at (0.9,-0.3) {$\epsilon$};
\end{scope}

\draw[->, thick] (-3.5,0) -- (-1.5,0) node[midway, below] {{\tiny Contour deformation}};;

\begin{scope}[shift={(5,0)}]
  \draw [decoration={markings, mark=at position 1 with {\arrow[scale=1.2]{>}}}, postaction={decorate}] (-5.5,0) -- (1.5,0);
  \draw [decoration={markings, mark=at position 1 with {\arrow[scale=1.2]{>}}}, postaction={decorate}] (0,-2.5) -- (0,2.5);
    \draw (1.4, 2.5) -- (1.4, 2.1) -- (1.8,2.1);
  \node at (1.6,2.3) {$s$};

  \foreach \x/\label in {0, -1, -2, -3} {
    \fill[blue] (\x,0) circle (2pt);
  }

  \foreach \x in {0,-1,-2,-3} {
  \draw[red, line width=1,->] (\x,0.3) arc(-270:100:0.3);
  }

  \node[blue] at (0.2,-0.6) {$0$};
  \node[blue] at (-1,-0.6) {$-2$};
  \node[blue] at (-2,-0.6) {$-4$};
  \node[blue] at (-3,-0.6) {$-6$};
  
  \foreach \x in {-4,-4.2,-4.4,-4.6, -4.8}
  \fill (\x,-0.6) circle (1pt);
\end{scope}
\end{tikzpicture}
    \caption{Contour deformation of the integral \eqref{eq:contor_integral}. }
    \label{fig:contor}
\end{figure}
To evaluate the integral, we deform the contour to the left (assuming $|\hat x|>|x_\perp|$) and pick up the residues at the poles $s=0, -2, -4, \cdots$ (see also Figure~\ref{fig:contor}):
\begin{align}
{\cal I}_{1}=
{1\over |\hat{x}|^p}{1\over \pi^{p/2+1}}
{1\over 8}
\sum_{n=0}^\infty 
{2(-1)^n\over n!}\left (|x_\perp| \over |\hat{x}|\right)^{2n} \Gamma\left(n+{p\over 2}\right)\left(
\log {4 x_\perp^2\over |\hat{x}|^4}+ \psi \left(n+{p\over 2}\right)
\right)\,.
\end{align} 
Performing the summation, this gives what we want to show,
\begin{align}
{\cal I}_{1}={\Gamma\left({p\over 2}\right) \over 2 \pi^{{p \over 2}+1}|x|^{p}}   \left(
\log {|x_\perp| \over |x|^{2}}+\log2+ \psi \left({p\over 2}\right)
\right)\, .
\label{besselint}
\end{align}

\subsection{Integral Formula \eqref{eq:formula}}\label{app:integral_Is}
Finally, we evaluate the following integral,
\begin{align}
\mathcal{I}_{2}=\int\frac{\d^{p}\hat{p}}{(2\pi)^{p+2}}K_{0}(|\hat{p}||x_{\perp}|)K_{0}(|\hat{p}||y_{\perp}|)e^{\i \hat{p}\cdot (\hat{x}-\hat{y})}\,. 
\end{align}
We again use the Mellin representation of the Bessel-$K$ function \eqref{eq:mellin}, 
\begin{align}
    {\cal I}_{2}=\frac{1}{(8\pi\i)^2}\int_{-\i \infty+\epsilon}^{\i \infty+\epsilon}\d s\int_{-\i \infty+\epsilon}^{\i \infty+\epsilon}\d t \frac{2^{s+t}}{|x_{\perp}|^s |y_{\perp}|^t}\Gamma^{2}\left(\frac{s}{2}\right)\Gamma^{2}\left(\frac{t}{2}\right)\int \frac{\d^{p}\hat{p}}{(2\pi)^{p+2}}\frac{e^{\i \hat{p}\cdot(\hat{x}-\hat{y})}}{|\hat{p}|^{s+t}} \,. 
\end{align}
Firstly, we perform the integral over $\hat{p}$ using \eqref{eq:spherical_integral} and the following formula,
\begin{align}
    \int_{0}^{\infty}\d \rho \rho^{\mu}J_{\nu}(a \rho)=\frac{2^\mu}{a^{\mu+1}}\frac{\Gamma(\frac{1+\mu+\nu}{2})}{\Gamma(\frac{1+\nu-\mu}{2})} \,,
\end{align}
and the result is
\begin{align}
        {\cal I}_{2}=-\frac{1}{256\pi^{p/2 +4}}\frac{1}{|\hat{x}-\hat{y}|^p}\int_{-\i \infty+\epsilon}^{\i \infty+\epsilon}\d s\int_{-\i \infty+\epsilon}^{\i \infty+\epsilon}\d t \, \xi_{1}^{-s}\xi_{2}^{-t}\, \frac{\Gamma^{2}\left(\frac{s}{2}\right)\Gamma^{2}\left(\frac{t}{2}\right)\Gamma(\frac{p-s-t}{2})}{\Gamma(\frac{s+t}{2})}\,. 
\end{align}
Here, $\xi_1$ and $\xi_2$ are the cross-ratios defined by
\begin{align}
    \xi_{1}\equiv \frac{|x_{\perp}|}{|\hat{x}-\hat{y}|} \,, \quad \xi_{2}\equiv \frac{|y_{\perp}|}{|\hat{x}-\hat{y}|}\,. 
\end{align}
We are interested in the defect block expansion, namely with $0<\xi_{i}<1\, (i=1,2)$.

We next evaluate the following integral,
\begin{align}\label{eq:tildeK}
    \widetilde{{\cal I}}_{2}\equiv\int_{-\i \infty+\epsilon}^{\i \infty+\epsilon}\d s\int_{-\i \infty+\epsilon}^{\i \infty+\epsilon}\d t \, \xi_{1}^{-s}\xi_{2}^{-t}\, \frac{\Gamma^{2}\left(\frac{s}{2}\right)\Gamma^{2}\left(\frac{t}{2}\right)\Gamma(\frac{p-s-t}{2})}{\Gamma(\frac{s+t}{2})} \,,
\end{align}
by deforming the contour to the left and using the residue theorem. As in Figure \ref{fig:contor}, this picks up the poles at $t=0, -2, -4, \cdots$ and gives,
\begin{align}
    \begin{aligned}
    &\int_{-\i \infty+\epsilon}^{\i \infty+\epsilon}\d t \, \xi_{2}^{-t}\frac{\Gamma^{2}\left(\frac{t}{2}\right)\Gamma(\frac{p-s-t}{2})}{\Gamma(\frac{s+t}{2})} \\
    &\ = 4\pi \i \sum_{n=0}^{\infty}\xi_{2}^{2n}\frac{\Gamma(n+\frac{p-s}{2})}{(n!)^2 \Gamma(-n+\frac{s}{2})}\left(2\psi(n+1)-\psi\left(n+\frac{p-s}{2}\right)-\psi\left(-n+\frac{s}{2}\right)-2\log\xi_{2}\right) \,.
    \end{aligned}
\end{align}
We next deform the contour associated to $s$-integral to the left, and pick up the poles at $s=0, -2, -4, \cdots$. This enables us to express the integral $\widetilde{{\cal I}}_2$ as a double sum,
\begin{align}
    \begin{aligned}
    &\widetilde{{\cal I}}_2=-32\pi^2 \sum_{n, m=0}^{\infty}\frac{(-\xi_{1}^2 )^m (-\xi_{2}^2 )^n}{(m+n)!}\binom{m+n}{m}^{2}\Gamma\left(m+n+\frac{p}{2}\right) \\
    &\qquad \times \left(\psi(n+1)+\psi(m+1)-\psi(m+n+1)-\psi\left(m+n+\frac{p}{2}\right)-\log \xi_1 \xi_2\right) \,. 
    \end{aligned}
\end{align}
In summary so far, we obtain the following expression of ${\cal I}_2$
\begin{align}
\begin{aligned}
\mathcal{I}_{2}
=\frac{1}{8\pi^{\frac{p}{2}+2}}\frac{1}{|x_{\perp}|^{p/2}|y_{\perp}|^{p/2}}\left[{\cal K}_1-{\cal K}_2 -\log(1+\xi_1^2 +\xi_2^2){\cal K}_3 +\log\chi\, {\cal K}_3 \right]\,, 
\end{aligned}
\end{align}
where ${\cal K}_1$, ${\cal K}_2$ and ${\cal K}_3$ are given by
\begin{align}\label{eq:loop}
    \begin{aligned}
        {\cal K}_1&\equiv \chi^{-\frac{p}{2}}(1+\xi_{1}^2 +\xi_{2}^2)^{\frac{p}{2}}\sum_{n, m=0}^{\infty}\frac{(-\xi_{1}^2 )^m (-\xi_{2}^2 )^n}{(m+n)!}\binom{m+n}{m}^{2}\Gamma\left(m+n+\frac{p}{2}\right) \\
    &\qquad \cdot \left(\psi(n+1)+\psi(m+1)-\psi(m+n+1)\right)\,, \\
    {\cal K}_2&\equiv \chi^{-\frac{p}{2}}(1+\xi_{1}^2 +\xi_{2}^2)^{\frac{p}{2}}\sum_{n, m=0}^{\infty}\frac{(-\xi_{1}^2 )^m (-\xi_{2}^2 )^n}{(m+n)!}\binom{m+n}{m}^{2}\Gamma\left(m+n+\frac{p}{2}\right)\psi\left(m+n+\frac{p}{2}\right)\,, \\
    {\cal K}_3&\equiv \chi^{-\frac{p}{2}}(1+\xi_{1}^2 +\xi_{2}^2)^{\frac{p}{2}}\sum_{n, m=0}^{\infty}\frac{(-\xi_{1}^2 )^m (-\xi_{2}^2 )^n}{(m+n)!}\binom{m+n}{m}^{2}\Gamma\left(m+n+\frac{p}{2}\right)
    \end{aligned}
\end{align}
Note that $\chi$ is written in terms of $\xi_1$ and $\xi_2$ as
\begin{align}
    \chi=\frac{1+\xi_1^2 +\xi_2^2}{\xi_1 \xi_2}\,. 
\end{align}
Introducing a new quantity ${\cal K}(\Delta)$ as
\begin{align}
    {\cal K}(\Delta)\equiv \chi^{-\frac{p}{2}}(1+\xi_{1}^2 +\xi_{2}^2)^{\frac{p}{2}}\sum_{n, m=0}^{\infty}\frac{(-\xi_{1}^2 )^m (-\xi_{2}^2 )^n}{(m+n)!}\binom{m+n}{m}^{2}\Gamma\left(m+n+\Delta \right)\,, 
\end{align}
and the double sums ${\cal K}_2$ and ${\cal K}_3$ can be written in terms of ${\cal K}$ as
\begin{align}
    {\cal K}_2 = \partial_{\Delta}{\cal K}\left(\frac{p}{2}\right)\,, \quad {\cal K}_3 = {\cal K}\left(\frac{p}{2}\right)\,. 
\end{align}
This motivate us to firstly evaluate ${\cal K}(\Delta)$,
\begin{align}\label{eq:F1}
    \begin{aligned}
        {\cal K}(\wD)
        &=\chi^{-\frac{p}{2}}(1+\xi_{1}^2 +\xi_{2}^2)^{\frac{p}{2}}\sum_{\ell=0}^{\infty}\frac{\Gamma(\ell+\wD)(\xi_{2}^2-\xi_{1}^2)^{\ell}}{\ell!}P_{\ell}\left(\frac{\xi_{1}^2 + \xi_{2}^2}{\xi_{1}^2 - \xi_{2}^2}\right)\\
        &=\chi^{-\frac{p}{2}}(1+\xi_{1}^2 +\xi_{2}^2)^{\frac{p}{2}}\int_{0}^{\infty}\d t\, t^{\wD -1}e^{-(1+\xi_{1}^2 + \xi_{2}^2)t}I_{0}(2\xi_{1}\xi_{2}t) \\
    &=\Gamma\left(\wD\right)(1+\xi_{1}^2 +\xi_{2}^2)^{\frac{p}{2}-\wD}\chi^{-\frac{p}{2}}{}_2 F_{1}\left(\frac{\wD}{2}, \frac{\wD +1}{2}; 1; \frac{4}{\chi^{2}}\right)\,.
    \end{aligned}
\end{align}
From this, one can derive ${\cal K}_2$ and ${\cal K}_3$ as follows:
\begin{align}
    \begin{aligned}
        {\cal K}_2 &=\Gamma\left(\frac{p}{2}\right)\left[\left(\psi\left(\frac{p}{2}\right)-\log(1+\xi_{1}^2 +\xi_{2}^2)\right)\mathsf{G}_{\{\hat{\sigma}, \h\}}(\chi)+\chi^{-\frac{p}{2}}\partial_{\wD}\left. {}_2 F_{1}\left(\frac{\wD}{2}, \frac{\wD +1}{2}; 1; \frac{4}{\chi^{2}}\right)\right|_{\wD = p/2} \right] \\
        {\cal K}_3 &=\Gamma\left(\frac{p}{2}\right)\chi^{-\frac{p}{2}}{}_2 F_{1}\left(\frac{p}{4}, \frac{p+2}{4}; 1; \frac{4}{\chi^{2}}\right)=\Gamma\left(\frac{p}{2}\right)\mathsf{G}_{\{\hat{\sigma}, \h\}}(\chi)\,. 
    \end{aligned}
\end{align}
By plugging these results into \eqref{eq:loop}, one can obtain
\begin{align}
        {\cal I}_2 =-\frac{\Gamma(p/2)}{8\pi^{\frac{p}{2}+2}}\frac{1}{|x_{\perp}|^{p/2}|y_{\perp}|^{p/2}}\left[H_{\frac{p}{2}-1}\mathsf{G}_{\wD}(\chi)+\left.\partial_{\wD}\mathsf{G}_{\wD}(\chi)\right|_{\wD=p/2}-{\cal R}\right]\,,
\end{align}
where ${\cal R}$ is given by
\begin{align}
    \mathcal{R}=\frac{1}{\Gamma(p/2)}{\cal K}_1 +\gamma_{\text{E}}\mathsf{G}_{\{\hat{\sigma}, \h\}}(\chi)+\chi^{-\frac{p}{2}}\frac{\partial}{\partial a}\left.{}_2 F_{1}\left(\frac{p}{4}, \frac{p+2}{4}; a; \frac{4}{\chi^{2}}\right)\right|_{a = 1}\,. 
\end{align}
By expanding $\mathcal{R}$ in powers of $\xi_1$ and $\xi_2$ and examining the coefficients, one finds that $\mathcal{R} = 0$. The desired integral formula \eqref{eq:formula} then follows.

\section{Bulk Conformal Block Expansion in Free Scalar Theory}\label{app:integral_bulk_block}
In this Appendix, we derive the bulk-channel conformal block expansion \eqref{eq:decomposition}  by directly evaluating the integral $\cI_2$ that appears in the bulk two-point function \eqref{eq:twopt_bulk_another}. In fact we will solve the following more general integral,
\begin{align}
    I_{\alpha, \beta}\equiv\int \d^{p}\hat{z}\,\frac{1}{|x-\hat{z}|^{2\alpha}|y-\hat{z}|^{2\beta}}\,,
    \label{integralIab}
\end{align}
which may be useful for more general DCFT setups.
Later we will specialize to the case $\A=\B={p\over 2}$, which is relevant for \eqref{eq:twopt_bulk_another} that describes the free scalar theory with subdimensional disorder. 

By using the Schwinger reparameterization, the integral \eqref{integralIab} becomes
\begin{align}
    \begin{aligned}
        I_{\alpha, \beta}&=\frac{1}{\Gamma(\alpha)\Gamma(\beta)}\int \d^{p}\hat{z}\int_{0}^{\infty}\d u\, u^{\alpha -1}e^{-u|x-\hat{z}|^2}\cdot \int_{0}^{\infty}\d v \, v^{\beta -1}e^{-v|y-\hat{z}|^2} \\
        &=\frac{\pi^{p/2}}{\Gamma(\alpha)\Gamma(\beta)}\int_{0}^{\infty}\d u\d v \, u^{\alpha -1}\, v^{\beta -1}\, (u+v)^{-p/2}e^{-\frac{uv}{u+v}|\hat{x}-\hat{y}|^2} e^{-u|x_{\perp}|^2 -v|y_{\perp}|^2} \,, 
    \end{aligned}
\end{align}
where, in the final equality, we have performed the integral over $\hat{z}$. Next we expand the last exponential in terms of the power of $u$ and $v$,
\begin{align}
     I_{\alpha, \beta}&=\frac{\pi^{p/2}}{\Gamma(\alpha)\Gamma(\beta)}\sum_{n, m=0}^{\infty}\frac{(-|x_{\perp}|^{2})^{n}(-|y_{\perp}|^{2})^{m}}{n!\, m!}\int_{0}^{\infty}\d u\d v \, u^{n+\alpha -1}\, v^{m+\beta -1}\, (u+v)^{-p/2}e^{-\frac{uv}{u+v}|\hat{x}-\hat{y}|^2}\,. 
\end{align}
By changing the integral variables as
\begin{align}
    u=y\,t\,,\qquad v=y\,(1-t)\,,
\end{align}
under which the integration measure turns into
\begin{align}
    \int_{0}^{\infty}\d u\d v \mapsto \int_0^\infty\, y\, \d y\,\int_0^1\d t\,,
\end{align}
the integral becomes
\begin{align}
    \begin{aligned}
     I_{\alpha, \beta}&=\frac{\pi^{p/2}}{\Gamma(\alpha)\Gamma(\beta)}\sum_{n, m=0}^{\infty}\frac{(-|x_{\perp}|^{2})^{n}(-|y_{\perp}|^{2})^{m}}{n!\, m!}\int_{0}^{1}\d t\, t^{n+\alpha-1}(1-t)^{m+\beta-1} \\
     &\qquad \qquad \qquad \qquad\qquad \times\int_{0}^{\infty}\d y \, y^{n+m+\alpha+\beta-\frac{p}{2}-1}e^{-y t (1-t)|\hat{x}-\hat{y}|^2}\,. 
     \end{aligned}
\end{align}
After another change of variable by,
\begin{align}
    y=\frac{\tilde{y}}{t(1-t)|\hat{x}-\hat{y}|^2}\,,
\end{align}
the above integral can be written as,
\begin{align}
    \begin{aligned}
         I_{\alpha, \beta}&=\frac{\pi^{p/2}F_{\alpha, \beta}}{\Gamma(\alpha)\Gamma(\beta)|\hat{x}-\hat{y}|^{2\alpha+2\beta-p}}\,, 
    \end{aligned}
\end{align}
where $F_{\alpha, \beta}$ is defined by
\begin{align}\label{eq:ab}
    \begin{aligned}
            F_{\alpha, \beta}&\equiv \sum_{n, m=0}^{\infty}\frac{\Gamma(n+m+\alpha+\beta-\frac{p}{2})}{n!\, m!}A^{n}\, B^{m}\int_{0}^{1}\d t\, t^{\frac{p}{2}-\beta-m-1}(1-t)^{\frac{p}{2}-\alpha-n-1}\,, \\ 
            A&\equiv-\frac{|x_{\perp}|^{2}}{|\hat{x}-\hat{y}|^2}\,, \quad B\equiv-\frac{|y_{\perp}|^{2}}{|\hat{x}-\hat{y}|^2}\,.
    \end{aligned}
\end{align}
We can evaluate $F_{\alpha, \beta}$ as
\begin{align}
    \begin{aligned}
    F_{\alpha, \beta}&\equiv \sum_{k=0}^{\infty}\sum_{n=0}^{k}\frac{\Gamma(k+\alpha+\beta-\frac{p}{2})}{n!\, (k-n)!}A^{n}\, B^{k-n}\int_{0}^{1}\d t\, t^{\frac{p}{2}-\beta-k+n-1}(1-t)^{\frac{p}{2}-\alpha-n-1}\\ 
    &=\sum_{k=0}^{\infty}\frac{\Gamma\left(k+\alpha+\beta-\frac{p}{2}\right)}{k!}\int_{0}^{1}\d t\, t^{\frac{p}{2}-\beta-k-1}(1-t)^{\frac{p}{2}-\alpha-1} \left(B+\frac{A t}{1-t}\right)^{k}\\
    &=\Gamma\left(\alpha+\beta-\frac{p}{2}\right)\int_{0}^{1}\d t\, t^{\frac{p}{2}-\beta-1}(1-t)^{\frac{p}{2}-\alpha-1} \left(1-\frac{A}{1-t}-\frac{B}{t}\right)^{\frac{p}{2}-\alpha-\beta}\\
    &=\Gamma\left(\alpha+\beta-\frac{p}{2}\right)(-B)^{\frac{p}{2}-\alpha-\beta}\int_{0}^{1}\d t\, t^{\alpha-1}(1-t)^{\beta-1} (1-t x')^{\frac{p}{2}-\alpha-\beta} (1-ty')^{\frac{p}{2}-\alpha-\beta} \,, 
    \end{aligned}
\end{align}
where $x'$ and $y'$ satisfy
\begin{align}
    x'y'=\frac{1}{B}\,, \quad x'+y'=-\frac{A-B-1}{B}\,. 
\end{align}
Then, by using the following integral formula \cite[eq.(B.22)]{Billo:2016cpy},
\begin{align}
    \int_{0}^{1}\d t\, t^{w-1}(1-t)^{r-1}(1-tx')^{-\rho}(1-ty')^{-\sigma}=B(w, r)F_{1}(w,\rho, \sigma, w+r; x', y')\,, 
\end{align}
where $F_{1}$ is the Appel function, $F_{\alpha, \beta}$ can be written as
\begin{align}
    \begin{aligned}
        F_{\alpha, \beta}=B\left(\alpha, \beta\right)\Gamma\left(\alpha+\beta-\frac{p}{2}\right)(-B)^{\frac{p}{2}-\alpha-\beta}F_{1}\left(\alpha, \alpha+\beta-\frac{p}{2}, \alpha+\beta-\frac{p}{2}, \alpha+\beta;x';y'\right)\,. 
    \end{aligned}
\end{align}
Particularly, when $w+r=\rho+\sigma$, the Appel function is reduced to the hypergeometric function via the following identity,
\begin{align}
    F_{1}(w,\rho, \sigma, \rho+\sigma; x'; y')=(1-y')^{w}{}_{2}F_{1}\left(w,\rho, \rho+\sigma; \frac{x'+y'}{1-y'}\right)\,. 
\end{align}
In our case, $\alpha=\beta=p/2$, so $F_{\alpha, \beta}$ can be simplified to
\begin{align}
    F_{p/2, p/2}=B\left(\frac{p}{2}, \frac{p}{2}\right)\Gamma\left(\frac{p}{2}\right)\left(\frac{1-y'}{-B}\right)^{p/2}{}_{2}F_{1}\left(\frac{p}{2},\frac{p}{2}, p; \frac{x'+y'}{1-y'}\right)\,.
\end{align}
To sum up, the bulk two-point function of $\sigma$ is given by
\begin{align}\label{eq:bulk_decomposition_sigma}
    \overline{\langle\, \sigma(x) \sigma(y)\, \rangle}&=\frac{C_{d}}{|x-y|^{p}}+\frac{\pi^{\frac{p}{2}}\Gamma(\frac{p}{2})}{\Gamma(p)}C_{d}^{2}{c}^{2}\frac{1}{|\hat{x}-\hat{y}|^{p}}\left(\frac{1-y'}{-B}\right)^{p/2}{}_{2}F_{1}\left(\frac{p}{2},\frac{p}{2}, p; \frac{x'+y'}{1-y'}\right) \,. 
\end{align}
Finally, by using the following identities,
\begin{align}
    \frac{|x_{\perp}||y_{\perp}|}{|\hat{x}-\hat{y}|^2}\left(\frac{1-y'}{-B}\right)=(X\overline{X})^{1/2}\,, \quad \frac{x' + y'}{1-y'}=1-X\overline{X}\,, 
\end{align}
one can show that the expression \eqref{eq:bulk_decomposition_sigma} perfectly matches the desired form \eqref{eq:decomposition} from the bulk-channel conformal block expansion.  

\bibliographystyle{JHEP}
\bibliography{disorder}
 \end{document}